\documentclass[aip,reprint,author-year,nofootinbib]{revtex4-1}

\usepackage{graphicx}
\usepackage{amsmath,amssymb}             
\usepackage{stmaryrd}					
\usepackage[normalem]{ulem}
\usepackage{color}
\usepackage[dvipsnames]{xcolor}
\usepackage{color,hyperref}
\definecolor{darkblue}{rgb}{0.0,0.0,0.4}
\definecolor{red}{rgb}{1.0,0.0,0.0}
\definecolor{green}{rgb}{0.0,0.5,0.0}
\hypersetup{colorlinks,breaklinks,
            linkcolor=darkblue,urlcolor=darkblue,
            anchorcolor=darkblue,citecolor=RoyalBlue}

\usepackage{gensymb} 
\usepackage[mathscr]{euscript}
\usepackage{upgreek}
\DeclareFontFamily{OT1}{pzc}{}
\DeclareFontShape{OT1}{pzc}{m}{it}{<-> s * [1.10] pzcmi7t}{}
\DeclareMathAlphabet{\mathpzc}{OT1}{pzc}{m}{it}
\usepackage{stmaryrd}

\def\apjl{Astrophys. J. Lett.}
\def\apj{Astrophys. J.}
\def\aj{Astronom. J.}

\def\mnras{Mon. Not. R. Astron. Soc.}

\def\prd{Phys. Rev. D}

\def\physrep{Physics Reports}
\def\nat{Nature}

\def\aap{Astron. \& Astrophys.}

\def\jcap{Journal of Cosmology and Astroparticle Physics}

\newcommand{\comment}[1]{}

\defcitealias{Jasche2015BORGSDSS}{\textsc{borg sdss}}
\newcommand{\borg}{\textsc{borg}}
\newcommand{\cola}{\textsc{cola}}
\newcommand{\tweb}{\textsc{T-web}}
\newcommand{\diva}{\textsc{diva}}
\newcommand{\origami}{\textsc{origami}}
\newcommand{\class}{\textsc{class}}

\begin{document}


\title{Comparing cosmic web classifiers using information theory}


\author{Florent Leclercq}
\email{florent.leclercq@polytechnique.org}
\affiliation{Institute of Cosmology and Gravitation (ICG), University of Portsmouth,\\ Dennis Sciama Building, Burnaby Road, Portsmouth PO1 3FX, United Kingdom}

\author{Guilhem Lavaux}
\affiliation{Institut d'Astrophysique de Paris (IAP), UMR 7095, CNRS -- UPMC Universit\'e Paris 6, Sorbonne Universit\'es, 98bis boulevard Arago, F-75014 Paris, France}
\affiliation{Institut Lagrange de Paris (ILP), Sorbonne Universit\'es,\\ 98bis boulevard Arago, F-75014 Paris, France}

\author{Jens Jasche}
\affiliation{Excellence Cluster Universe, Technische Universit\"at M\"unchen,\\ Boltzmannstrasse 2, D-85748 Garching, Germany}

\author{Benjamin Wandelt}
\affiliation{Institut d'Astrophysique de Paris (IAP), UMR 7095, CNRS -- UPMC Universit\'e Paris 6, Sorbonne Universit\'es, 98bis boulevard Arago, F-75014 Paris, France}
\affiliation{Institut Lagrange de Paris (ILP), Sorbonne Universit\'es,\\ 98bis boulevard Arago, F-75014 Paris, France}
\affiliation{Department of Physics, University of Illinois at Urbana-Champaign,\\ 1110 West Green Street, Urbana, IL~61801, USA}
\affiliation{Department of Astronomy, University of Illinois at Urbana-Champaign,\\ 1002 West Green Street, Urbana, IL~61801, USA}


\date{\today}

\begin{abstract}
\noindent We introduce a decision scheme for optimally choosing a classifier, which segments the cosmic web into different structure types (voids, sheets, filaments, and clusters). Our framework, based on information theory, accounts for the design aims of different classes of possible applications: (i)~parameter inference, (ii)~model selection, and (iii)~prediction of new observations. As an illustration, we use cosmographic maps of web-types in the Sloan Digital Sky Survey to assess the relative performance of the classifiers {\tweb}, {\diva} and {\origami} for: (i)~analyzing the morphology of the cosmic web, (ii)~discriminating dark energy models, and (iii)~predicting galaxy colors. Our study substantiates a data-supported connection between cosmic web analysis and information theory, and paves the path towards principled design of analysis procedures for the next generation of galaxy surveys. We have made the cosmic web maps, galaxy catalog, and analysis scripts used in this work publicly available.
\end{abstract}


\maketitle



\section{Introduction}

The large-scale distribution of matter in the Universe is not uniform, but forms a complex pattern known as the cosmic web \citep{Bond1996,vandeWeygaertBond2008}. As it retains memory about structure formation processes, it contains a rich variety of astrophysical and cosmological information. Applications of mapping the cosmic web include in particular: correlating with galaxy properties \citep[e.g.][]{Blanton2005a}; studying the effect of the large-scale structure (LSS) on light propagation through cosmic expansion, dust extinction, and absorption by the intergalactic medium \citep[e.g.][]{Planck2015ISW}; testing general relativity \citep{Falck2015a}; probing dark matter annihilation along caustics \citep{VogelsbergerWhite2011}; and looking for ``bullet cluster'' objects \citep{Harvey2015}. Properties of cosmic web elements can also be viewed as statistical summaries of the large-scale structure, and serve as alternatives to correlation functions in order to learn about cosmological parameters \citetext{for recent results, see e.g. \citealp{deHaan2016} using clusters; \citealp{Hamaus2016} using voids}.

Important tools for cosmic web analysis are classifiers, i.e. algorithms that dissect the entire large-scale structure into one of its structural elements. In contrast to structure finders that focus on one component at a time (typically clusters, filaments, or voids), they allow an analysis of the connection between cosmic web components, identified in the same framework. The richly-structured morphology of the cosmic web is simultaneously sensitive to the original phases of the field, the local density and velocity, and the growth history. Classifiers reduce this complex information to the common concepts of voids, sheets, filaments, and clusters. Many such algorithms have been proposed over the last decade, exploiting different physical information to perform the classification: the eigenvalues of the tidal tensor \citetext{the {\tweb}, \citealp{Hahn2007a}, and its extensions, \citealp{Forero-Romero2009}}; the eigenvalues of the velocity shear tensor \citetext{the \textsc{V-web}, \citealp{Hoffman2012}, and its particle-based formulation, \citealp{FisherFaltenbacherJohnson2016}}; the eigenvalues of the shear of the Lagrangian displacement field \citetext{{\diva}, \citealp{Lavaux2010}}; the number of orthogonal axes along which stream-crossing occurs \citetext{{\origami}, \citealp{Falck2012}}. Different classifiers provide different insights into cosmic web morphology. The aim of this paper is to offer a principled way of choosing among possible classifiers, depending on the application of interest.

As outlined by \citet{Leclercq2015ST} and further demonstrated in this work, the need for information theory in cosmic web analysis uniquely emerges from the uncertainties inherent to actual observations, as opposed to the unique answer provided by any one simulation. Indeed, building a complete cosmographic description of the real Universe from galaxy positions requires high-dimensional, non-linear probabilistic methods. As a response, Bayesian large-scale structure inference \citep{Lahav1994,Zaroubi2002,Erdovgdu2004,Kitaura2008,JascheKitaura2010,Jasche2010b,Jasche2013BORG,Kitaura2013,Wang2013,Wang2014a} offers a methodical approach. It has been shown recently that resulting reconstructions can be used for the detection of cosmic web elements \citetext{halos, \citealp{MersonJascheAbdallaEtAl2016}; voids, \citealp{Leclercq2015DMVOIDS}} and for the application of cosmic web classifiers \citep{Nuza2014,Leclercq2015ST,Leclercq2016}.

In previous work, we studied the dynamic cosmic web of the nearby Universe, relying on the analysis of the Sloan Digital Sky Survey (SDSS) main galaxy sample with the {\borg} algorithm \citep{Jasche2015BORGSDSS} and using different classifiers. Specifically, in \citet{Leclercq2015ST}, we used the {\tweb} definition; and in \citet{Leclercq2016} we used {\diva} and {\origami}. A legitimate question, unanswered yet, regards the relative merits of our different cosmic web maps, which is linked to the relative performance of the different classifiers, depending on the desired use. These possible applications are of three broad classes: (i) optimal parameter inference, (ii) model comparison, and (iii) prediction of future observations. In each case, the optimal choice of a classifier is naturally expressed as a Bayesian experimental design problem. Beyond the Gaussian assumption and Fisher matrix forecasts \citep[which are known to suffer from severe shortcomings, see][]{Wolz2012}, information-theoretic approaches to the design of experiments and analysis procedures, especially for the second and third kind of problems, remain largely unused in cosmology \citep[see however][concerning the optimal design of cosmological surveys for parameter estimation]{Bassett2005}. Nevertheless, interestingly, problems that share strong mathematical similarity have been studied in the bioinformatics literature \citep[e.g.][]{Vanlier2012,Vanlier2014}.

One of the most commonly used and versatile Bayesian design criteria is to maximize the mutual information between the data and some quantity of interest. Mutual information is an information-theoretic notion based on entropy that reflects how much of the uncertainty in one random variable is reduced by knowing about the other. In \citet{Leclercq2015DT}, we discussed an optimal decision-making criterion for segmenting the cosmic web into different structure types on the basis of their respective probabilities and the strength of data constraints. In the present paper, we use classifier utilities and the concept of mutual information to extend the decision problem to the space of classifiers. We illustrate this methodological discussion with three cosmological problems of the types mentioned above: (i) optimization of the information content of cosmic web maps, (ii) discrimination of dark energy models, and (iii) prediction of galaxy colors. In doing so, we quantify the relative performance of the {\tweb}, {\diva} and {\origami} for each of these applications.

After discussing information mapping in the cosmic web in section \ref{sec:Mapping information in the cosmic web}, we introduce utilities for cosmic web classifiers in section \ref{sec:Classifier utilities}. We discuss our results and give our conclusions in section \ref{sec:Conclusions}. The relevant notions of information theory and of Bayesian experimental design are respectively reviewed in appendices \ref{apx:Information theory} and \ref{apx:Bayesian experimental design}.

\section{Mapping information in the cosmic web}
\label{sec:Mapping information in the cosmic web}

The goal of this section is to introduce probabilistic maps of the cosmic web and assess their information content. We briefly review Bayesian large-scale structure analysis in section \ref{sec:Bayesian large-scale structure analysis}. We then discuss probabilistic classifications of the cosmic web in section \ref{sec:Classifications} and introduce the relevant information-theoretic notions in section \ref{sec:Information-theoretic comparison of classifiers}. 

\subsection{Bayesian large-scale structure analysis}
\label{sec:Bayesian large-scale structure analysis}

The cosmic web maps used in this work have been built upon results previously obtained by the application of {\borg} \citep[Bayesian Origin Reconstruction from Galaxies,][]{Jasche2013BORG} to the SDSS main galaxy sample \citep{Jasche2015BORGSDSS}. {\borg} is a Bayesian large-scale structure inference code that reconstructs the primordial density fluctuations and produces physical reconstructions of the dark matter distribution that underlies observed galaxies, by assimilating the survey data into a cosmological structure formation model. To do so, it samples a complex posterior distribution in a multi-million dimensional parameter space (corresponding to the voxels of the discretized domain) by means of the Hamiltonian Monte Carlo algorithm \citep{Duane1987}.

For each move in parameter space, the code does several evaluations of the data model, which involves second-order Lagrangian perturbation theory \citep[see e.g.][]{Bernardeau2002} to describe large-scale structure formation between initial density fields (at a scale factor $a=10^{-3}$) and the present day (at $a=1$). In this fashion, the code jointly accounts for the shape of the three-dimensional matter field and its formation history, in the linear and mildly non-linear regimes. Besides large-scale structure formation, {\borg} accounts for uncertainties coming from luminosity-dependent galaxy biases and observational effects such as selection functions, the survey mask, and shot noise. The distribution of galaxies is modeled as an inhomogeneous Poisson process on top of evolved, biased density fields. For a more extensive discussion of the {\borg} data model, the reader is referred to chapter 4 in \citet{LeclercqThesis}.

Starting from samples of inferred initial conditions, which contain the data constraints, we perform a non-linear filtering step \citep[see chapter 7 in][]{LeclercqThesis}. This is achieved by evolving samples forward in time with second-order Lagrangian perturbation theory (2LPT) to the redshift of $z=69$, then running a constrained simulation with the {\cola} method \citep{Tassev2013} from $z=69$ to $z=0$. 

When producing the maps used in this work \citep{Leclercq2015ST,Leclercq2016}, we used a set of 1,097 non-linear {\borg}-{\cola} samples. Their initial conditions are defined on a $750$~Mpc/$h$ (comoving) cubic grid of $256^3$ voxels. The evolved realizations contain $512^3$ particles and have been obtained with $30$ {\cola} timesteps. Whenever it is necessary, particles are binned to the grid using the cloud-in-cell scheme.

\subsection{Classifications}
\label{sec:Classifications}

\begin{table*}\centering
\begin{tabular}{llll}
\hline\hline
Classifier $\xi$: & {\tweb} & {\diva} & {\origami}\\
\hline
Type & Eulerian & Lagrangian & Lagrangian\\
\hline
Structure type\\
\hline
Void & $\mu_1,\mu_2,\mu_3 < 0$ & $\lambda_1,\lambda_2,\lambda_3 < 0$ & no-stream crossing\\
Sheet & $\mu_1,\mu_2 < 0$ and $\mu_3 > 0$ $\quad$ & $\lambda_1,\lambda_2 < 0$ and $\lambda_3 > 0$ $\quad$ & stream-crossing along one axis\\
Filament & $\mu_1 < 0$ and $\mu_2,\mu_3 > 0$ & $\lambda_1 < 0$ and $\lambda_2,\lambda_3 > 0$ & stream-crossing along two orthogonal axes\\
Cluster & $\mu_1,\mu_2,\mu_3 > 0$ & $\lambda_1,\lambda_2,\lambda_3 > 0$ & stream-crossing along three orthogonal axes\\
\hline\hline
\end{tabular}
\caption{Rules for classification of structure types according to the {\tweb}, {\diva}, and {\origami} procedures.}
\label{tb:rules}
\end{table*}

This paper focuses on the possibility to classify the cosmic web into four different structure types: voids, sheets, filaments, and clusters. Any of the algorithms cited in the introduction can be used on our set of constrained realizations. However, for the purpose of this paper, we will compare the results of three classifiers:
\begin{itemize}
\item the {\tweb} \citep{Hahn2007a},
\item {\diva} \citep{Lavaux2010},
\item and {\origami} \citep{Falck2012}.
\end{itemize}

With the {\tweb}, structures are classified according to the sign of the eigenvalues $\mu_1(\vec{x}) \leq \mu_2(\vec{x}) \leq \mu_3(\vec{x})$ of the tidal field tensor $\mathscr{T}$, the Hessian of the rescaled gravitational potential $\Phi$:
\begin{equation}
\mathscr{T}_{ij} \equiv \mathrm{H}(\Phi)_{ij} = \frac{\partial^2 \Phi}{\partial \vec{x}_i \partial \vec{x}_j},
\end{equation}
where $\Phi$ obeys the reduced Poisson equation
\begin{equation}
\Delta \Phi(\vec{x}) = \delta(\vec{x}),
\end{equation}
$\delta$ being the local density contrast. A voxel belongs to a cluster, a filament, a sheet or a void, if, respectively, three, two, one or zero of the $\mu_i$ are positive. The {\tweb} is a Eulerian procedure, in the sense that it operates at the level of voxels of the discretized domain. It can be applied at any time, but does not use the time-evolution of structures to classify them. 

In contrast, Lagrangian classifiers rely on the displacement field $\vec{\Psi}(\vec{q})$, which maps the initial position of particles $\vec{q}$ to their final position $\vec{x}(\vec{q})$ \citep[see e.g.][]{Bernardeau2002}:
\begin{equation}
\vec{x}(\vec{q}) \equiv \vec{q} + \vec{\Psi}(\vec{q}).
\label{eq:mapping}
\end{equation}
Such classifiers provide a description of the cosmic web at the level of the initial grid of particles.

Instead of the tidal field tensor $\mathscr{T}$, {\diva} uses the shear of the displacement field $\mathscr{R}$, defined by
\begin{equation}
\mathscr{R}_{\ell m} \equiv \frac{\partial \vec{\Psi}_\ell}{\partial \vec{q}_m} .
\end{equation}
Denoting by $\lambda_1(\vec{q}) \leq \lambda_2(\vec{q}) \leq \lambda_3(\vec{q})$ the eigenvalues of $\mathscr{R}$, a particle's structure type is defined as before by counting the number of positive $\lambda_i$ (instead of $\mu_i$). Note that at first order in Lagrangian perturbation theory (the Zel'dovich approximation), $\mathscr{T}$ and $\mathscr{R}$ are proportional, so the {\tweb} and {\diva} yield the same classification of the cosmic web. Differences only arise at higher order.

An alternative way to classify particles is to consider the evolution of the matter streams they belong to. During gravitational collapse, ``shell-crossing'' happens when different streams pass through a single location. {\origami} defines structure types according to the number of orthogonal axes along which a Lagrangian patch undergoes shell-crossing. Specifically, void, sheet, filament, and cluster particles are defined as particles that have been crossed along zero, one, two, or three orthogonal axes, respectively. The {\tweb}, {\diva} and {\origami} rules for cosmic web classification are summarized in table \ref{tb:rules}.

In Bayesian large-scale structure inference, uncertainties are quantified by the variation of density fields among constrained samples. As shown in previous work \citep{Jasche2010a,Leclercq2015ST,Lavaux2016BORG2MPP,Leclercq2016}, uncertainties can be self-consistently propagated to structure type classification as follows. Let us denote by $\xi$ one of the classifiers. By applying $\xi$ to a specific large-scale structure realization, we obtain a unique answer in the form of four scalar fields that obey the following conditions for any $\vec{\sigma}_p$:
\begin{equation}
\mathrm{T}_i(\vec{\sigma}_p|\xi) \in \{0,1\} \; \mathrm{for} \; i \in \llbracket 0,3 \rrbracket \quad \mathrm{and} \quad \sum_{i=0}^{3} \mathrm{T}_i(\vec{\sigma}_p|\xi) = 1
\end{equation}
where $\mathrm{T}_0=$ void, $\mathrm{T}_1=$ sheet, $\mathrm{T}_2=$ filament, $\mathrm{T}_3=$ cluster, and where $\vec{\sigma}_p$ is $\vec{x}_k$ (the location of a voxel) if $\xi$ is a Eulerian classifier, or $\vec{q}_\ell$ (the location of a particle on the initial grid) if $\xi$ is a Lagrangian classifier. By applying $\xi$ to the complete set of constrained realizations and counting the relative frequencies of structure types at each spatial coordinate $\vec{\sigma}_p$, we obtain a posterior probability mass function (pmf) in the form of four scalar fields $\mathcal{P}(\mathrm{T}_i(\vec{\sigma}_p)|d,\xi) \equiv \mathcal{T}_i(\vec{\sigma}_p|\xi)$ that take their values in the range $[0,1]$ and sum up to one at each $\vec{\sigma}_p$:
\begin{equation}
\mathcal{T}_i(\vec{\sigma}_p|\xi) \in [0,1] \; \mathrm{for} \; i \in \llbracket 0,3 \rrbracket \quad \mathrm{and} \quad \sum_{i=0}^{3} \mathcal{T}_i(\vec{\sigma}_p|\xi) = 1.
\end{equation}

The corresponding prior probabilities $\mathcal{P}(\mathrm{T}_i|\xi)$ can be estimated by applying the same procedure to a set of unconstrained realizations produced using the same setup as for constrained samples. We found that these probabilities are well approximated by Gaussians whose means and standard deviations are given in table \ref{tb:prior_initial} for the primordial large-scale structure and \ref{tb:prior_final} for the late-time large-scale structure.

\begin{figure*}
\begin{center}
{\tweb}
\includegraphics[width=\textwidth]{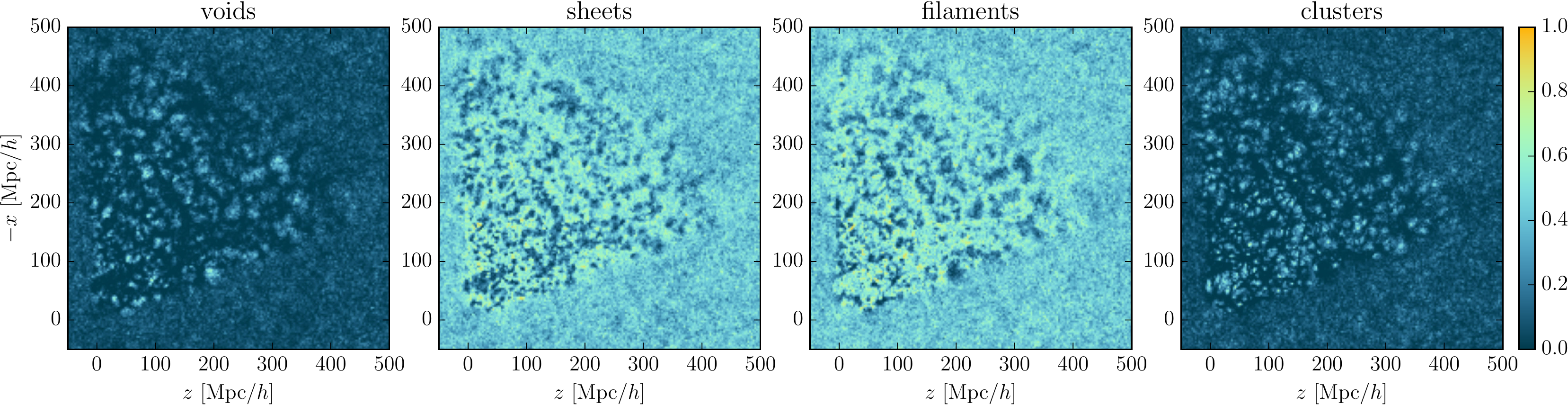}
{\diva}
\includegraphics[width=\textwidth]{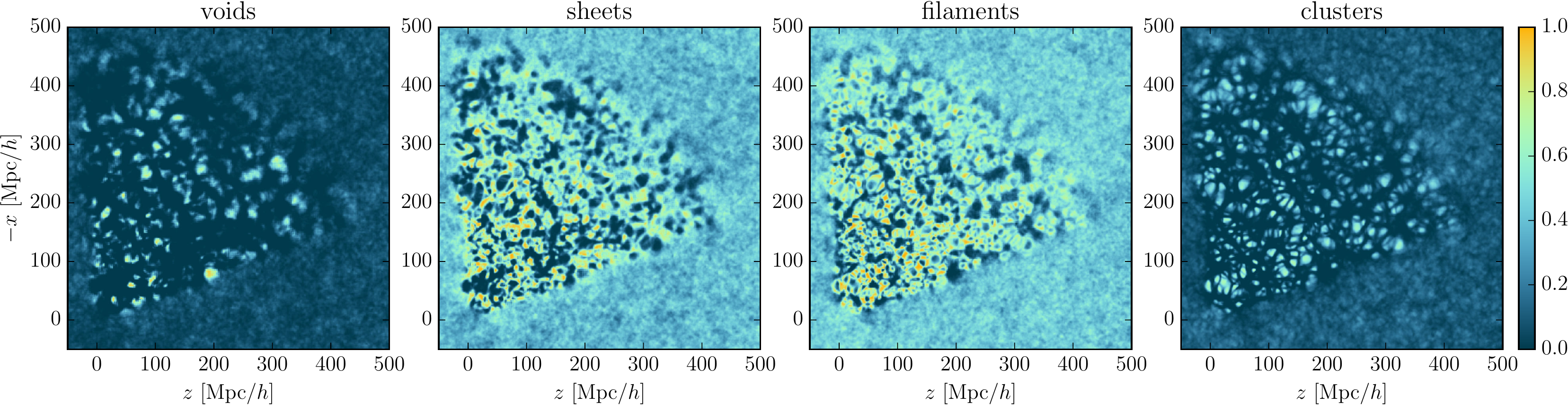}
{\origami}
\includegraphics[width=\textwidth]{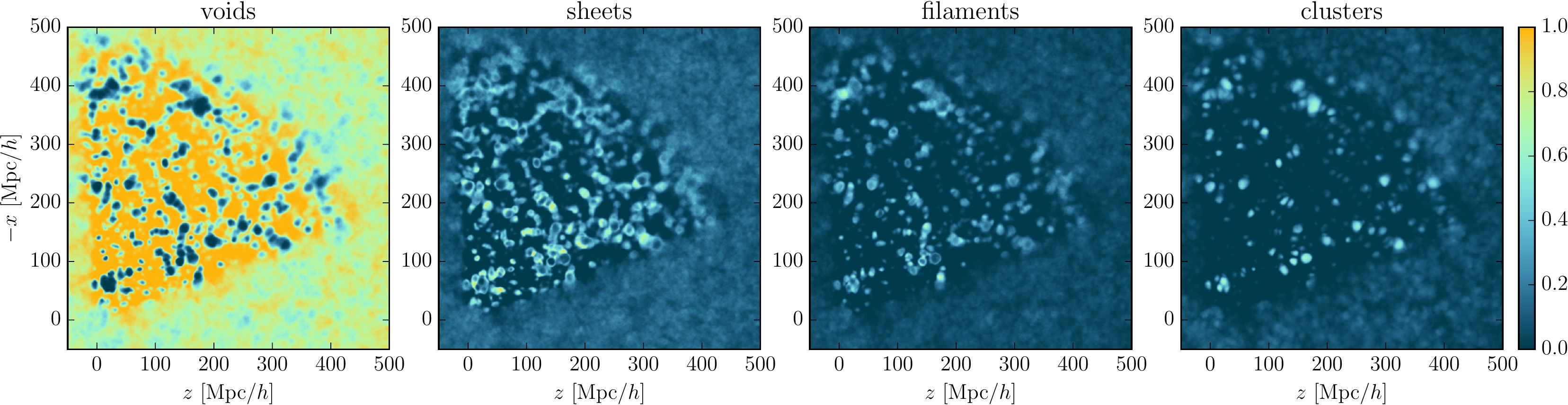}
\caption{Slices through the posterior probabilities for different structure types (from left to right: void, sheet, filament, and cluster), in the primordial large-scale structure in the Sloan volume ($a=10^{-3}$). These four three-dimensional probabilities sum up to one at each location. From top to bottom, structure types are defined using the {\tweb}, {\diva} and {\origami}. The first row \citep[{\tweb}, reproduced from][]{Leclercq2015ST} shows $256^3$-voxel grids; the second and third row ({\diva} and {\origami}, reproduced from \citealp{Leclercq2016}) show Lagrangian grids of $512^3$ particles.\label{fig:pdf_initial}}
\end{center}
\end{figure*}

With Eulerian classifiers, a classification of the primordial large-scale structure is obtained when the $\vec{x}_k$ are voxels of the grid on which the initial density field is defined \citep[see section IV in][]{Leclercq2015ST}. With Lagrangian classifiers, it is directly obtained by looking at the initial grid of particles \citep[see section II in][]{Leclercq2016}. The web-type posterior maps for the primordial large-scale structure in the SDSS volume are shown in figure \ref{fig:pdf_initial}. 

\begin{figure*}
\begin{center}
{\tweb}
\includegraphics[width=\textwidth]{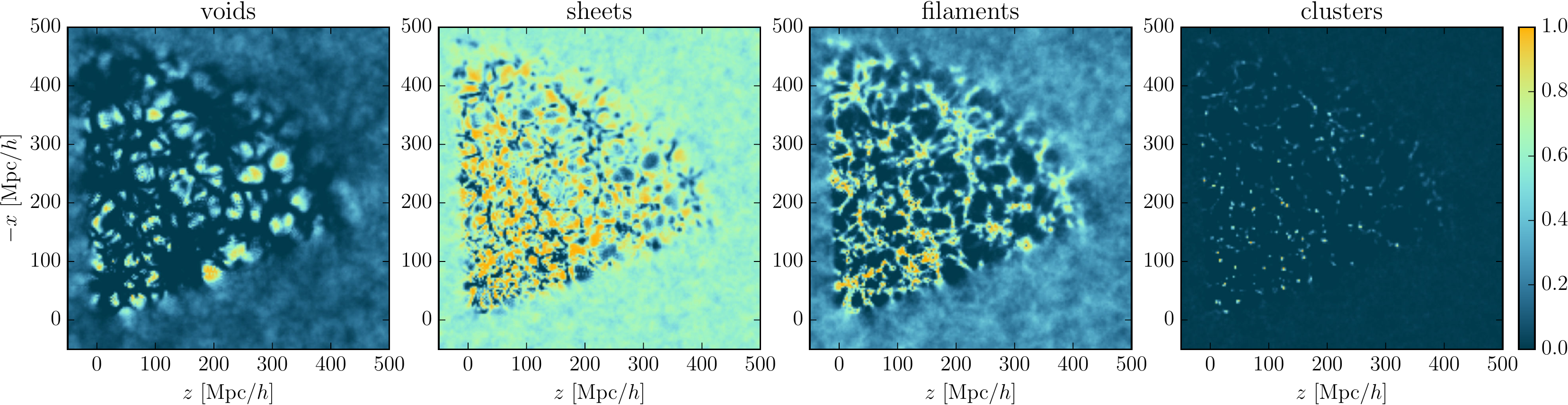}
{\diva}
\includegraphics[width=\textwidth]{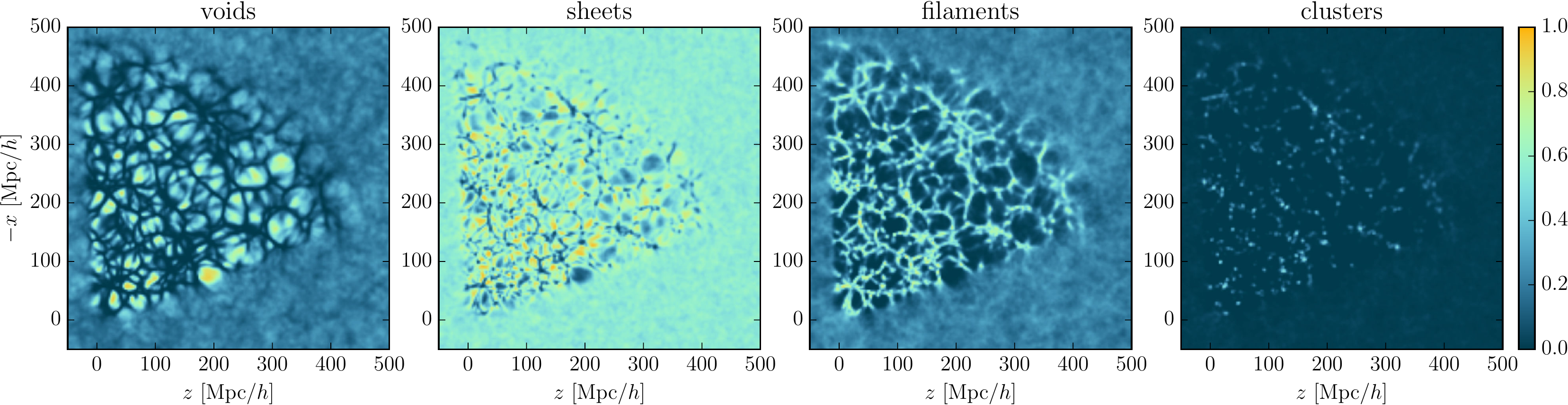}
{\origami}
\includegraphics[width=\textwidth]{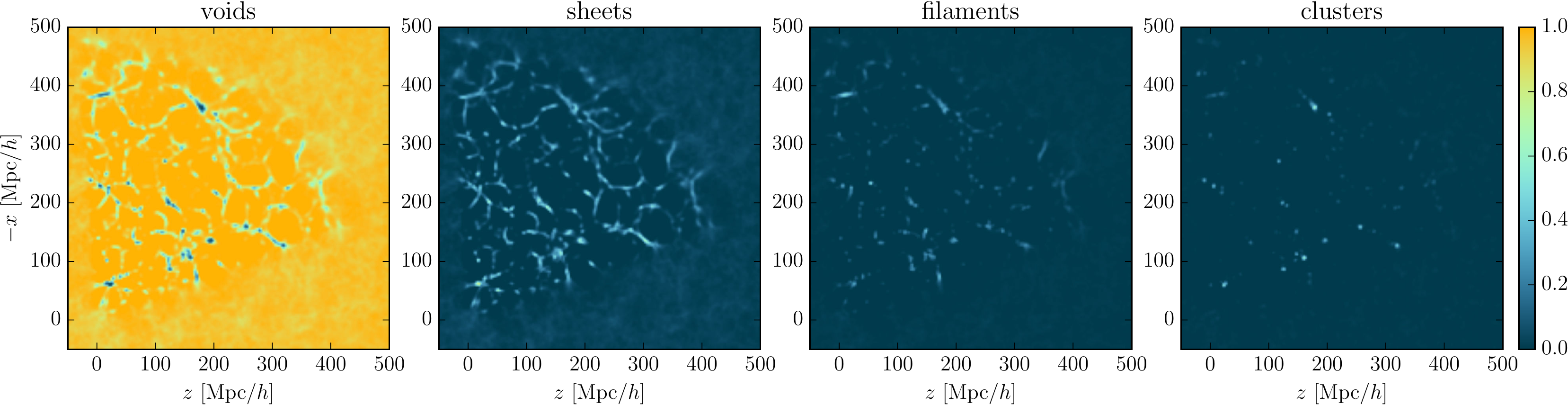}
\caption{Slices through the posterior probabilities for different structure types (from left to right: void, sheet, filament, and cluster), in the late-time large-scale structure in the Sloan volume ($a=1$). These four three-dimensional probabilities sum up to one on a voxel basis. From top to bottom, structure types are defined using the {\tweb}, {\diva} and {\origami}. The first row is reproduced from \citet{Leclercq2015ST}, the second and third rows from \citet{Leclercq2016}.\label{fig:pdf_final}}
\end{center}
\end{figure*}

In \citet{Leclercq2016}, we also showed how to translate the result of Lagrangian classifiers from particles' positions $\vec{q}_\ell$ to Eulerian voxels $\vec{x}_k$, so as to obtain a description of the late-time large-scale structure: particles transport their Lagrangian structure type along their trajectory, and are binned to the grid at their final Eulerian position. In figure \ref{fig:pdf_final}, we show the web-type posterior for evolved structures in the SDSS. We focus on these maps in the rest of this paper.

\begin{table*}\centering
\begin{tabular}{lcccccc}
\hline\hline
Classifier $\xi$: & \multicolumn{2}{c}{\tweb} & \multicolumn{2}{c}{\diva} & \multicolumn{2}{c}{\origami}\\
\hline
Structure type & $\mu_{\mathcal{P}(\mathrm{T}_i|\xi)}$ & $\sigma_{\mathcal{P}(\mathrm{T}_i|\xi)}$ & $\mu_{\mathcal{P}(\mathrm{T}_i|\xi)}$ & $\sigma_{\mathcal{P}(\mathrm{T}_i|\xi)}$ & $\mu_{\mathcal{P}(\mathrm{T}_i|\xi)}$ & $\sigma_{\mathcal{P}(\mathrm{T}_i|\xi)}$\\
\hline
\multicolumn{7}{c}{Primordial large-scale structure ($a=10^{-3}$)} \\
Void & $0.07979$ & $5.4875 \times 10^{-5}$ & $0.06288$ & $7.5158 \times 10^{-5}$ & $0.59079$ & $7.3765 \times 10^{-4}$ \\
Sheet & $0.42022$ & $1.0240 \times 10^{-4}$ & $0.38229$ & $1.7662 \times 10^{-4}$ & $0.16487$ & $3.7328 \times 10^{-4}$ \\
Filament & $0.42022$ & $1.0412 \times 10^{-4}$ & $0.42716$ & $1.4050 \times 10^{-4}$ & $0.09775$ & $2.2532 \times 10^{-4}$ \\
Cluster & $0.07978$ & $5.6337 \times 10^{-5}$ & $0.12767$ & $1.7990 \times 10^{-4}$ & $0.14659$ & $5.0061 \times 10^{-4}$ \\
\hline\hline
\end{tabular}
\caption{Prior probabilities assigned by the {\tweb}, {\diva} and {\origami} to the different structures types, in the primordial large-scale structure ($a=10^{-3}$), i.e. in the initial density field for the {\tweb}, and on the Lagrangian grid of particles for {\diva} and {\origami}.}
\label{tb:prior_initial}
\end{table*}

\begin{table*}\centering
\begin{tabular}{lcccccc}
\hline\hline
Classifier $\xi$: & \multicolumn{2}{c}{\tweb} & \multicolumn{2}{c}{\diva} & \multicolumn{2}{c}{\origami}\\
\hline
Structure type & $\mu_{\mathcal{P}(\mathrm{T}_i|\xi)}$ & $\sigma_{\mathcal{P}(\mathrm{T}_i|\xi)}$ & $\mu_{\mathcal{P}(\mathrm{T}_i|\xi)}$ & $\sigma_{\mathcal{P}(\mathrm{T}_i|\xi)}$ & $\mu_{\mathcal{P}(\mathrm{T}_i|\xi)}$ & $\sigma_{\mathcal{P}(\mathrm{T}_i|\xi)}$\\
\hline
\multicolumn{7}{c}{Late-time large-scale structure ($a=1$)} \\
Void & $0.14261$ & $6.1681 \times 10^{-4}$ & $0.20216$ & $4.7733 \times 10^{-4}$ & $0.89459$ & $4.4745 \times 10^{-4}$ \\
Sheet & $0.59561$ & $6.3275 \times 10^{-4}$ & $0.54845$ & $2.5827 \times 10^{-4}$ & $0.06727$ & $3.0459 \times 10^{-4}$ \\
Filament & $0.24980$ & $5.5637 \times 10^{-4}$ & $0.22587$ & $3.6287 \times 10^{-4}$ & $0.02249$ & $1.0619 \times 10^{-4}$ \\
Cluster & $0.01198$ & $5.8793 \times 10^{-5}$ & $0.02352$ & $6.8724 \times 10^{-5}$ & $0.01565$ & $7.9767 \times 10^{-5}$ \\
\hline\hline
\end{tabular}
\caption{Prior probabilities assigned by the {\tweb}, {\diva} and {\origami} to the different structures types, in the late-time large-scale structure ($a=1$).}
\label{tb:prior_final}
\end{table*}

\subsection{Information-theoretic comparison of classifiers}
\label{sec:Information-theoretic comparison of classifiers}

The posterior probability maps for each classifier show complex and distinct features, coming both from the quantification of observational uncertainty and from the various physical criteria used to define structures. It is therefore important to use appropriate tools to characterize their information content and agreement. As discussed in \citet{Leclercq2015ST}, information theory offers a natural language to address these questions. In this framework, the uncertainty content of a pmf $\mathcal{P}$ is the Shannon entropy \citep{Shannon1948}, $H[\mathcal{P}]$ (in shannons, Sh); the information gain due to the data is the relative entropy or Kullback-Leibler divergence \citep{Kullback1951} of the posterior $\mathcal{P}$ from the prior $\uppi$, $\mathcal{D}_\mathrm{KL}[\mathcal{P}||\uppi]$; finally, the similarity between two pmfs $\mathcal{P}$ and $\mathcal{Q}$ is measured by the Jensen-Shannon divergence \citep{Lin1991}, $D_\mathrm{JS}\left[ \mathcal{P}\!:\!\mathcal{Q} \right]$ (see appendix \ref{apx:Information theory}). For our analysis, these quantity read generically
\begin{widetext}
\begin{gather}
H\left[ \mathcal{P}(\mathrm{T}(\vec{x}_k)|d,\xi) \right] \equiv - \sum_{i=0}^{3} \mathcal{P}(\mathrm{T}_i(\vec{x}_k)|d,\xi) \log_2 \mathcal{P}(\mathrm{T}_i(\vec{x}_k)|d,\xi) ,\label{eq:definition_entropy}\\
D_\mathrm{KL}\left[ \mathcal{P}(\mathrm{T}(\vec{x}_k)|d,\xi) \middle\| \mathcal{P}(\mathrm{T}|\xi) \right] \equiv \sum_{i=0}^{3} \mathcal{P}(\mathrm{T}_i(\vec{x}_k)|d,\xi) \log_2\left(\frac{\mathcal{P}(\mathrm{T}_i(\vec{x}_k)|d,\xi)}{\mathcal{P}(\mathrm{T}_i|\xi)}\right) ,\label{eq:definition_DKL}\\
D_\mathrm{JS}\left[ \mathcal{P}(\mathrm{T}(\vec{x}_k)|d,\xi_\alpha)\!:\!\mathcal{P}(\mathrm{T}(\vec{x}_k)|d,\xi_\beta) \right] \equiv H\left[ \frac{\mathcal{P}(\mathrm{T}(\vec{x}_k)|d,\xi_\alpha)+\mathcal{P}(\mathrm{T}(\vec{x}_k)|d,\xi_\beta)}{2} \right] - \frac{H\left[ \mathcal{P}(\mathrm{T}(\vec{x}_k)|d,\xi_\alpha) \right] + H\left[ \mathcal{P}(\mathrm{T}(\vec{x}_k)|d,\xi_\beta) \right]}{2}
\label{eq:definition_DJS}
\end{gather}
\end{widetext}
where the space of structure types is $\{\mathrm{T}_0~=$ void, $\mathrm{T}_1~=$ sheet, $\mathrm{T}_2~=$ filament, $\mathrm{T}_3~=$ cluster$\}$ and the space of classifiers is $\{\xi_0 =$ {\tweb}, $\xi_1 =$ {\diva}, $\xi_2 =$ {\origami}$\}$.

\begin{figure*}
\begin{center}
\includegraphics[width=\textwidth]{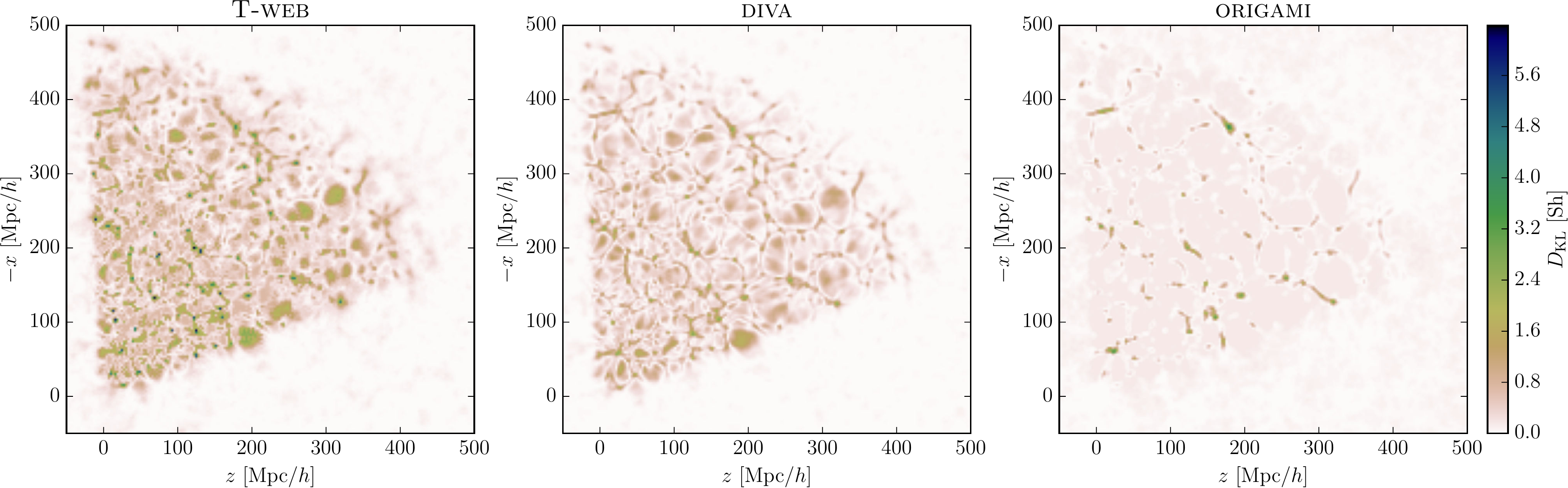}
\caption{Slices through the Kullback-Leibler divergence of the web-type posterior from the prior. This quantity, defined by equation \eqref{eq:definition_DKL}, represents the information gained on structure type classification by looking at SDSS galaxies. It corresponds to the joint utility for parameter inference of the SDSS data set $\tilde{d}$ and the classifier $\xi$ ($U_1(\tilde{d},\xi)$, see equation \eqref{eq:joint_utility_inference}). From left to right, structures are defined using the {\tweb}, {\diva} and {\origami}.\label{fig:comparison_pdf_kl}}
\end{center}
\end{figure*}

\begin{figure*}
\begin{center}
\includegraphics[width=\textwidth]{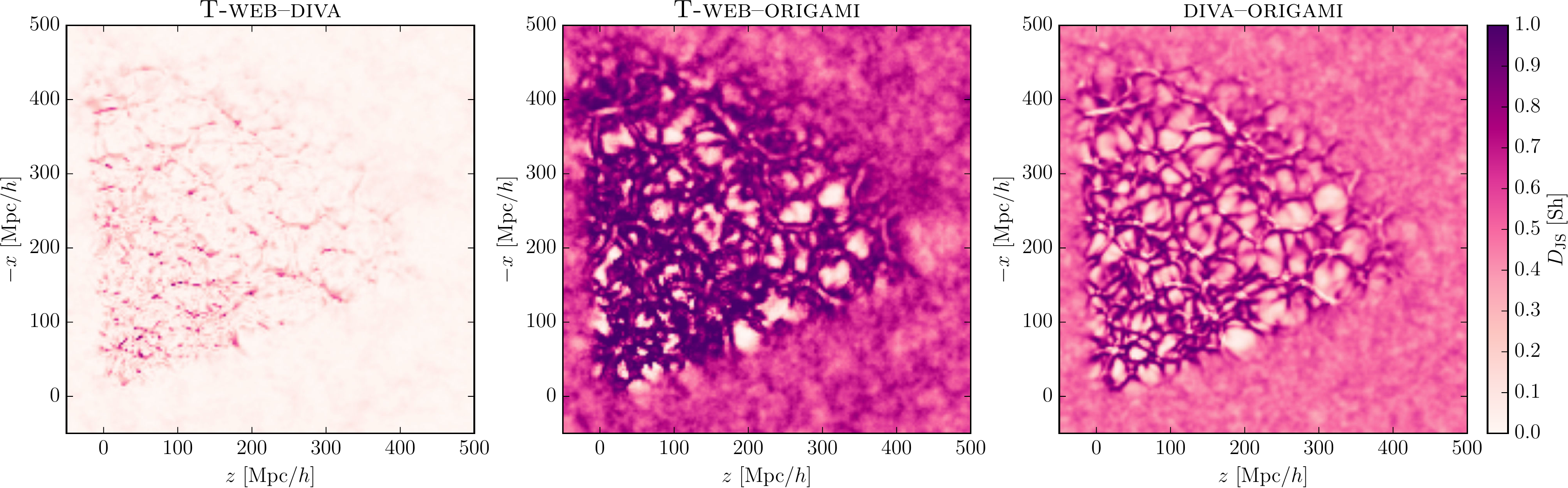}
\caption{Slices through the Jensen-Shannon divergence between pairs of web-type posteriors, as indicated above the panels. The Jensen-Shannon divergence, defined by equation \eqref{eq:definition_DJS}, is a symmetric measure of the disagreement (between $0$ and $1$ Sh) between the different classifiers.\label{fig:comparison_pdf_js}}
\end{center}
\end{figure*}

Slices through the voxel-wise Kullback-Leibler divergence of web-type posteriors from their respective priors, for different classifiers, are shown in figure \ref{fig:comparison_pdf_kl}. As expected, the information gain is close to zero out of the survey boundaries. There, the information gain fluctuates around $\sim 0.03$ Sh ({\tweb}), $\sim 0.02$ Sh ({\diva}), $\sim 0.05$ Sh ({\origami}). These values are small, but positive. This artifact is due to the limited number of samples used in our analysis: because of the finite length of the Markov Chain, the sampled representation of the posterior has not yet fully converged to the true posterior, and therefore it can show artificial information gain with respect to the prior \citep[see also the discussion in][]{Leclercq2015DT}. In observed regions, SDSS galaxies are informative on the underlying cosmic web at the level of several shannons in most of the volume for the {\tweb} and {\diva}, this information being more evenly distributed in the {\diva} map. With {\origami}, the information gain can reach $\sim 3$ Sh in shell-crossed structures, but in most of the volume, filled with voids, it cannot exceed the value of $0.1614$ Sh (which corresponds to the certain inference of a void, i.e. $\mathcal{P}(\mathrm{T}_0(\vec{x}_k)|d,\xi_2) = 1$, with a prior value of $\mathcal{P}(\mathrm{T}_0 |\xi_2) = 0.89459$). 

In figure \ref{fig:comparison_pdf_js}, we show slices through the Jensen-Shannon divergence of pairs of web-type posteriors. These maps confirm and precisely quantify the visual impression, obtained with figure \ref{fig:pdf_final}, that the {\tweb} and {\diva} classifications do not differ much and are far from the {\origami} result.

\section{Classifier utilities}
\label{sec:Classifier utilities}

This section describes how to set up utility functions in the space of classifiers. This space can contain any of the algorithms mentioned in the introduction. Formally, all the implementation details (e.g. threshold, smoothing scale, method-internal parameters) should also be considered as yielding different classifiers. For simplicity, we limit the space of classifiers to $\{\xi_0 =$ {\tweb}, $\xi_1 =$ {\diva}, $\xi_2 =$ {\origami}$\}$, using the detailed setups described in \citet{Leclercq2015ST} and \citet{Leclercq2016}. In particular, we adopted natural choices for the method-internal parameters suggested by their authors and by our {\borg} analysis,\footnote{In particular, {\tweb} classifications are defined at a comoving Eulerian scale of $\sim~3$~Mpc/$h$ (corresponding to a grid of $256^3$ voxels on a cube of $750$ Mpc/$h$ side length), and {\diva} and {\origami} classifications are defined at a comoving Lagrangian scale of $\sim~1.5$~Mpc/$h$ (corresponding to a regular lattice of $512^3$ particles in the same cube).} but did not further explore these choices in this study. 

In analogy with the formalism of Bayesian experimental design (see appendix \ref{apx:Bayesian experimental design}), we introduce the utility of a classifier as $U(\xi) = \left\langle U(d,\xi) \right\rangle_{\mathcal{P}(d|\xi)}$, where $U(d,\xi)$ is the joint utility of a data set $d$ and a classifier $\xi$. The decision problem will consist in maximizing the utility $U(\xi)$.

As noted in the introduction, the choice of a classifier should be specific to the application of interest. For example, a classifier which efficiently estimates the shape of the cosmic web may not extract the relevant information for discriminating among cosmological models, or may not be useful for predicting future observations. In this section, we introduce some Bayesian utility functions for various experimental goals. We illustrate each situation with a physical question and in each case, we give an estimate of the relative performance of the different classifiers considered in this paper.

\subsection{Utility for parameter inference: cosmic web analysis}
\label{sec:Utility for parameter inference: cosmic web analysis}

In Bayesian experimental design, common utility functions that aim at optimal parameter inference are information-based (see section \ref{apx:Parameter inference utility functions}). Following this idea, we propose that the optimal classifier for cosmic web analysis should simply maximize the expected information gain, i.e. the utility $U_1(\xi) \equiv \left\langle U_1(d,\xi) \right\rangle_{\mathcal{P}(d|\xi)}$, with 
\begin{eqnarray}
U_1(d,\xi)(\vec{x}_k) & \equiv & D_\mathrm{KL}\left[\mathcal{P}(\mathrm{T}(\vec{x}_k)|d,\xi) || \mathcal{P}(\mathrm{T}|\xi)\right] \label{eq:joint_utility_inference} \\
& = & \sum_{i=0}^3 \mathcal{P}(\mathrm{T}_i(\vec{x}_k)|d,\xi) \log_2\left( \frac{\mathcal{P}(\mathrm{T}_i(\vec{x}_k)|d,\xi)}{\mathcal{P}(\mathrm{T}_i|\xi)} \right) . \nonumber
\end{eqnarray}
Note that in this case, $U_1(d,\xi)(\vec{x}_k)$ depends on the location (for each data set, $U_1(d,\xi)$ is a three-dimensional map of the large-scale structure), but $U_1(\xi)$ should not depend on the location once the expectation over all possible data realizations is taken.

Using property \eqref{eq:utility_inference_mutual_info}, we obtain
\begin{equation}
U_1(\xi) = I[\mathrm{T}\!:\!d|\xi],
\label{eq:utility_inference}
\end{equation}
the mutual information between the inferred parameters (the web-types) and the data. This utility will therefore maximize the information content of the inferred cosmic web map.

Figure \ref{fig:comparison_pdf_kl} shows the joint utility $U_1(\tilde{d},\xi)$ for different classifiers and for one particular data set $\tilde{d}$, namely the SDSS galaxies used in our {\borg} analysis \citep[see section 2 in][]{Jasche2015BORGSDSS}. In order to estimate $U_1(\xi)$, one should in principle consider the expectation of such maps over all possible data sets. This task involves building many synthetic galaxy catalogs mimicking the SDSS and performing on them a {\borg} analysis followed by different cosmic web classifications. Considering computational time requirements, such an endeavor is unattainable. Instead, we propose to estimate $U_1(\xi)$ by considering $U_1(\tilde{d},\xi)(\vec{x}_k)$ at different locations. This idea is analog to the hypothesis of ergodicity: if the SDSS is a fair sample of the Universe, then the ensemble average and the sample average of any quantity coincide. For cosmic web analysis, this means supposing that the SDSS contains a large enough variety of voids, sheets, filaments, and clusters so that all possible configurations of such structures are represented fairly.

\begin{table}\centering
\begin{tabular}{lccc}
\hline\hline
Utility & {\tweb} & {\diva} & {\origami}\\
\hline
$\widehat{U_1}(\xi)$ [Sh] & 0.4573 & 0.2664 & 0.1347\\
$\widehat{U_1'}(\xi)$ [Sh$^{-1}$] & 36.28 & 55.09 & 20.92\\
$\widehat{U_2}(\xi)$ [$10^{-3}$ Sh] & 5.53 & 2.22 & 3.24\\
$\widehat{U_2'}(\xi)$ [Sh$^{-1}$] & 1454.2 & 1782.9 & 861.06\\
$\widehat{U_3}(\xi)$ [Sh] & 0.0152 & 0.0101 & 0.0143\\
\hline\hline
\end{tabular}
\caption{Estimation of the utility of different classifiers (the {\tweb}, {\diva} and {\origami}) for different optimization problems: parameter inference (cosmic web analysis, $\widehat{U_1}$), insensitivity to artifacts for parameter inference (cosmic web analysis, $\widehat{U_1'}$), model comparison (dark energy equation of state, $\widehat{U_2}$), insensitivity to artifacts for model comparison (dark energy equation of state, $\widehat{U_2'}$), prediction of additional observations (galaxy colors, $\widehat{U_3}$).}
\label{tb:utilities}
\end{table}

Formally, we introduce the following estimator:
\begin{equation}
U_1(\xi) \approx \widehat{U_1}(\xi) \equiv \frac{1}{N_\mathrm{obs}} \sum_{\vec{x}_k} U_1(\tilde{d},\xi)(\vec{x}_k) .
\end{equation}
where the summation runs over voxels of the observed regions, characterized by the three-dimensional survey response operator being positive \citep[see][]{Jasche2015BORGSDSS}. There are $N_\mathrm{obs}=3,148,504$ such voxels out of $N_\mathrm{tot}~=~256^3~=~16,777,216$. The results, given in table \ref{tb:utilities}, indicate that for cosmic web inference, preference should be given, in this order, to the {\tweb}, {\diva}, then {\origami}. This ordering is mostly due to the very high information gain in {\tweb} clusters, and to the strong prior preference of {\origami} for voids, which limits its information gain -- as noted in section \ref{sec:Information-theoretic comparison of classifiers}. 

A disadvantage of using information gain as the utility function is its sensitivity to artifacts. This is a general feature of all information-theoretic quantities that are maximized in case of maximal randomness (such as entropy): they are not only sensitive to ``interesting'' patterns, but also to ``incidental'' information. In our case, classifiers have different sensitivities to artifacts in our cosmic web reconstructions, of various origin: noise in the data, approximate physical modeling, limited number of samples, etc. In order to assess the ``risk'' taken by different classifiers when producing the final cosmic web map, one needs to quantify the average number of ``false positives''. To do so, we propose to use the information gain in unobserved regions as a proxy for the sensitivity to artifacts, and to minimize its expectation value. Therefore, we introduce the utility $U_1'(\xi) = \left\langle U_1'(d,\xi) \right\rangle_{\mathcal{P}(d|\xi)}$, where 
\begin{equation}
U_1'(d,\xi)(\vec{x}_k) \equiv D_\mathrm{KL}\left[\mathcal{P}(\mathrm{T}(\vec{x}_k)|d,\xi) || \mathcal{P}(\mathrm{T}|\xi)\right]^{-1}
\end{equation}
and $\vec{x}_k$ has not been observed when the data set $d$ has been taken.

For the SDSS, with a similar argument as before, $U_1'$ is estimated by the inverse of the average artificial information gain in unconstrained regions, i.e.
\begin{equation}
U_1'(\xi) \approx \widehat{U_1'}(\xi) \equiv \left[ \frac{1}{N_\mathrm{unobs}} \sum_{\vec{x}_k } U_1(\tilde{d},\xi)(\vec{x}_k) \right]^{-1}
\end{equation}
where the summation now runs on unobserved voxels (i.e. where the survey response operator is zero), and where $N_\mathrm{unobs}\equiv N_\mathrm{tot}-N_\mathrm{obs}$. Numerical values, given in table \ref{tb:utilities}, show that, from this point of view, {\diva} outperforms the {\tweb} and {\origami}.

Considering simultaneously $U_1$ and $U_1'$, a user can make a decision based on a quantitative criterion that weights the utility of different classifiers, accounting for the user's preferred trade-off between information gain and sensitivity to artifacts.

\subsection{Utility for model selection: dark energy equation of state}
\label{sec:Utility for model selection: dark energy equation of state}

\begin{figure*}
\begin{center}
\includegraphics[width=\textwidth]{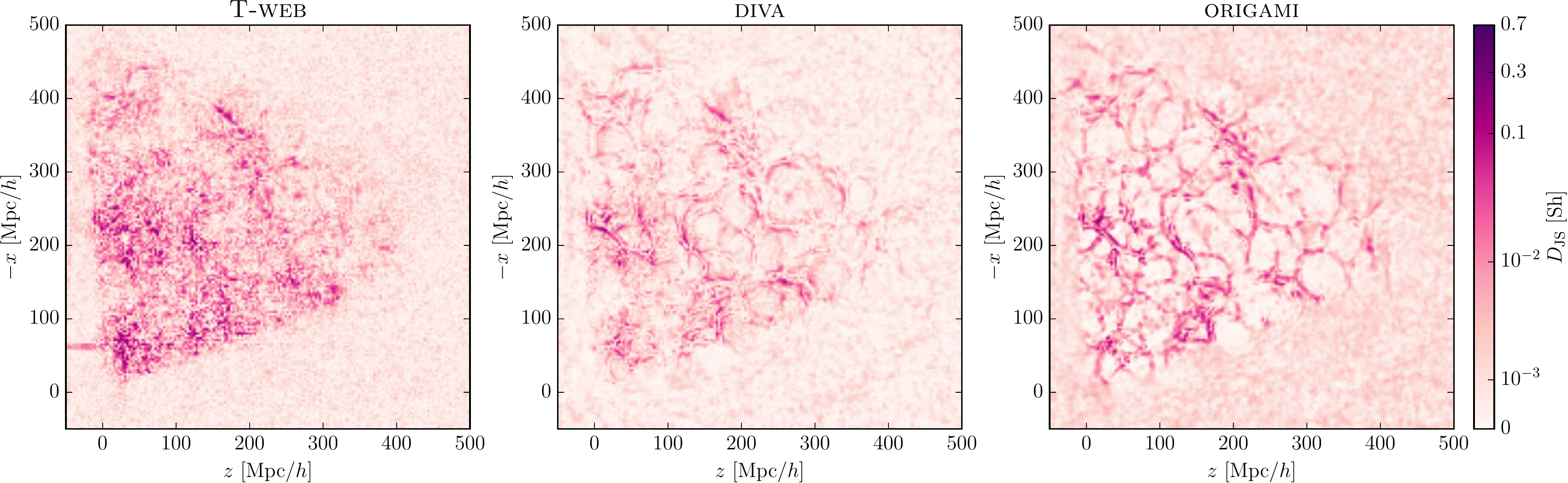}
\caption{Slices through the Jensen-Shannon divergence between the three cosmic web-type posteriors $\mathcal{P}(\mathrm{T}(\vec{x}_k)|d,\mathcal{M}_w,\xi)$, for the dark energy equation of state $w \in \{ -0.9, -1, -1.1\}$ and the classifier $\xi$ indicated above the panels (from left to right: the {\tweb}, {\diva}, and {\origami}). This quantity corresponds to the joint utility for model selection of the SDSS data set $\tilde{d}$ and of the considered classifier $\xi$ ($U_2(\tilde{d},\xi)$, see equation \eqref{eq:joint_utility_model_selection}). The color scale has been stretched around zero using the mapping $x \mapsto \mathrm{argsinh}(10^3 \, x)$. \label{fig:comparison_pdf_js_DE}}
\end{center}
\end{figure*}

Model selection is an important experimental design problem which has generated some research interest (see section \ref{apx:Model selection utility functions}). In a Bayesian context, model selection is typically based on the Bayes factor, which measures the amount of evidence that the data provide for one model over another. In cosmology, competing models can be for example the standard $\Lambda$CDM paradigm and one of its extensions. Our aim in this section is to choose the cosmic web classifier that selects the best features to discriminate between such models.

It has recently been shown that the Jensen-Shannon divergence between posterior predictive distributions can be used as an approximate predictor for the change in the Bayes factor \citep{Vanlier2014}. Following this idea, we propose a model selection utility for classifiers as $U_2(\xi) \equiv \left\langle U_2(d,\xi) \right\rangle_{\mathcal{P}(d|\xi)}$, where 
\begin{equation}
U_2(d,\xi)(\vec{x}_k) \equiv D_\mathrm{JS}\left[\mathcal{P}(\mathrm{T}(\vec{x}_k)|d,\mathcal{M}_1)\!:\!\mathcal{P}(\mathrm{T}(\vec{x}_k)|d,\mathcal{M}_2) |\xi\right] ,
\label{eq:utility_model_selection_DJS}
\end{equation}
where $\mathcal{M}_1$ and $\mathcal{M}_2$ are two competing cosmological models.

In the following, we exemplify for three cosmological models in which the dark energy component has different equations of state: $\mathcal{M}_{-0.9}$, $\mathcal{M}_{-1}$, $\mathcal{M}_{-1.1}$ corresponding respectively to $w$CDM with $w=-0.9$, $\Lambda$CDM ($w=-1$) and $w$CDM with $w=-1.1$. For simplicity, we note $\mathcal{P}_w(d) \equiv \mathcal{P}(\mathrm{T}(\vec{x}_k)|d,\mathcal{M}_w)$ for $w \in \{-0.9, -1, -1.1\}$. Different values for the equation of state of dark energy will mean different expansion history and growth of structure in the Universe, which affects the late-time morphology of the cosmic web. We aim here at finding the classifier which best separates the predictions of different models. Following equation \eqref{eq:utility_model_selection_DJS}, the joint utility of a data set $d$ and a classifier $\xi$ is the Jensen-Shannon divergence between the three probabilities
\begin{equation}
U_2(d,\xi)(\vec{x}_k) = D_\mathrm{JS} [\mathcal{P}_{-0.9}(d)\!:\!\mathcal{P}_{-1}(d)\!:\!\mathcal{P}_{-1.1}(d) | \xi ],
\label{eq:joint_utility_model_selection}
\end{equation}
where we need the generalized definition of $D_\mathrm{JS}$ (equation \eqref{eq:def_DJS_generalized}). Given property \eqref{eq:utility_DJS_mixture}, we have
\begin{equation}
U_2(\xi) = I[ \mathcal{M}\!:\!\mathcal{R}(d) | \xi],
\label{eq:utility_model_selection}
\end{equation}
the mutual information between $\mathcal{M}$ and $\mathcal{R}(d)$, respectively the model indicator and the mixture of distributions $\mathcal{P}_{-0.9}(d),\mathcal{P}_{-1}(d),\mathcal{P}_{-1.1}(d)$:
\begin{equation}
\mathcal{R}(d) \equiv \frac{1}{3} \left[ \mathcal{P}_{-0.9}(d) + \mathcal{P}_{-1}(d) + \mathcal{P}_{-1.1}(d) \right].
\end{equation}

For the SDSS data set $\tilde{d}$, the probabilities $\mathcal{P}_{-1}(\tilde{d})$ have been already inferred and discussed within standard $\Lambda$CDM cosmology (see section \ref{sec:Classifications} and figure \ref{fig:pdf_final}). To evaluate $\mathcal{P}_{-0.9}(\tilde{d})$ and $\mathcal{P}_{-1.1}(\tilde{d})$, we ran a set of constrained simulations within $w$CDM cosmology, corresponding to our existing set.\footnote{This treatment is approximate, since the calculation of $\mathcal{P}_{-0.9}(\tilde{d})$ and $\mathcal{P}_{-1.1}(\tilde{d})$ should in principle involve inference of the initial conditions with a modified version of {\borg}, accounting for $w~\neq~-1$. Considering computational requirements, we leave this exact study for future work.} More precisely, we started from the set of {\borg}-inferred initial phases (obtained by dividing the initial density realizations by the square root of the fiducial power spectrum, in Fourier space) and rescaled the Fourier modes so as to reproduce the linear matter power spectrum for our set of cosmological parameters and for the correct value of $w$. These power spectra have been obtained with the cosmological Boltzmann code \textsc{class} \citep{Blas2011}. The resulting initial conditions have been evolved with 2LPT to the redshift $z=69$ and with 30 {\cola} timesteps from $z=69$ to $z=0$. During the evolution, we fixed the dark energy equation of state to $w=-0.9$ or $w=-1.1$. Finally, we performed cosmic web analysis as before to get $\mathcal{P}_{-0.9}(\tilde{d})$ and $\mathcal{P}_{-1.1}(\tilde{d})$ for each classifier. 

Figure \ref{fig:comparison_pdf_js_DE} shows the Jensen-Shannon divergence between $\mathcal{P}_{-0.9}(\tilde{d})$, $\mathcal{P}_{-1}(\tilde{d})$ and $\mathcal{P}_{-1.1}(\tilde{d})$; i.e. the joint utility $U_2(\tilde{d},\xi)$ of the SDSS data set and each of our three classifiers (see equation \eqref{eq:joint_utility_model_selection}). There, one can clearly notice that Lagrangian classifiers ({\diva} and {\origami}) pick out more structure than the {\tweb}. In particular, we find that the surroundings of voids are especially sensitive regions to separate the predictions of different dark energy models. This can be easily interpreted: as the cosmic web is affected by dark energy throughout its growth, the Lagrangian displacement field (used by {\diva} and {\origami}) keeps a better memory of the expansion history of the Universe than the final Eulerian position of particles (used by the {\tweb}).

Since we have only one data set at hand, it is possible, as in section \ref{sec:Utility for parameter inference: cosmic web analysis}, to use an estimator for $U_2$:
\begin{equation}
U_2(\xi) \approx \widehat{U_2}(\xi) \equiv \frac{1}{N_\mathrm{obs}} \sum_{\vec{x}_k} U_2(\tilde{d},\xi)(\vec{x}_k) ,
\end{equation}
and for the corresponding ``risk'' taken by classifiers when separating different models, $U_2'$:
\begin{equation}
U_2'(\xi) \approx \widehat{U_2'}(\xi) \equiv \left[ \frac{1}{N_\mathrm{unobs}} \sum_{\vec{x}_k } U_2(\tilde{d},\xi)(\vec{x}_k) \right]^{-1} .
\end{equation}
Numerical results are given in table \ref{tb:utilities}. Noticeably, this crude estimator for $U_2$ favors the {\tweb} versus {\diva} and {\origami}: though the Jensen-Shannon divergence between the different pmfs is more evenly distributed with the {\tweb} (see figure \ref{fig:comparison_pdf_js_DE}), its average value within the entire volume is the highest.

\subsection{Utility for predictions: galaxy colors}
\label{sec:Utility for predictions: galaxy colors}

In the context of optimizing the predictive power of experiments, the expected information gain from the prior to the posterior predictive distributions is a useful utility function. As discussed in section \ref{apx:Utilities for prediction of future observations}, it is also the mutual information between predicted and upcoming observations, conditional on the experimental design.

Let us denote by $c$ future observations, or observations already available but that have not been used so far. We introduce the utility of a classifier $\xi$ to perform predictions as $U_3(\xi) \equiv \left\langle U_3(d,\mathrm{T},\xi) \right\rangle_{\mathcal{P}(d,\mathrm{T}|\xi)}$, where the joint utility of a data set $d$, a classification $\mathrm{T}$ and a classifier $\xi$ is the information gain on $c$, i.e.
\begin{equation}
U_3(d,\mathrm{T},\xi) \equiv D_\mathrm{KL}[\mathcal{P}(c|d,\mathrm{T},\xi)||\mathcal{P}(c|\xi)] .
\label{eq:joint_utility_prediction}
\end{equation}

\begin{figure*}
\begin{center}
\includegraphics[width=\textwidth]{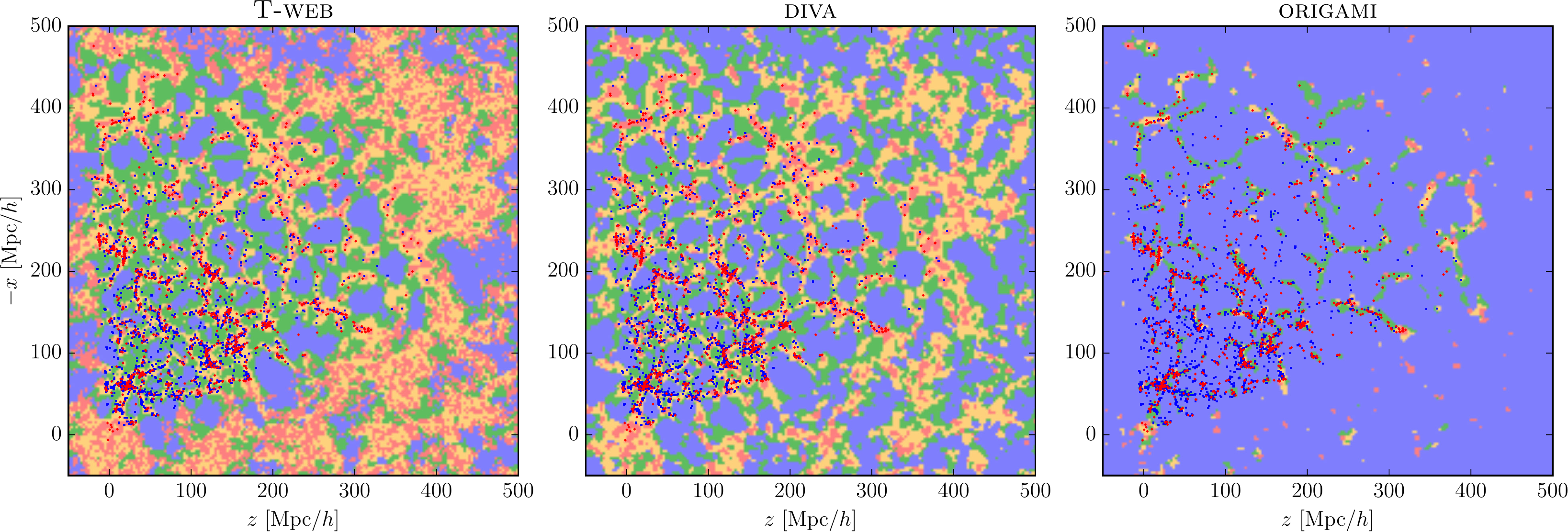}
\caption{Slices through maps of structure types in the late-time large-scale structure as observed by the SDSS. The color coding is blue for voids, green for sheets, yellow for filaments, and red for clusters. From left to right, structures are defined using the {\tweb}, {\diva} and {\origami}. These maps are based on the posterior probabilities shown in figure \ref{fig:pdf_final}, using the Bayesian decision rule of \citet{Leclercq2015DT} for $\alpha=1.0$ (the \textit{fair game} situation, in which a decision is made everywhere). Blue galaxies are overplotted as blue squares and red galaxies as red diamonds.\label{fig:comparison_decision_maps}}
\end{center}
\end{figure*}

In the following, we exemplify for the prediction of a property of galaxies that has been used neither in our {\borg} inference nor cosmic web analyses: their color. More specifically, we started from the objects used in the {\borg} SDSS DR7 run \citep[see][section 2]{Jasche2015BORGSDSS}. We queried the SDSS database in order to keep only those identified as galaxies after spectral analysis ($\mathtt{specClass}=2$) and to get for each of them the $r$-band apparent magnitude ($\mathtt{modelMag\_r}$) and the $(g-r)$ color ($\mathtt{grModelColor}$). From the apparent magnitude $r$ and the redshift $z$, we computed the $r$-band absolute magnitude $M_{0.1\,r}$. Absolute magnitudes receive appropriate $K$-correction to their $z=0.1$ value using the code of \citet{Blanton2003a,Blanton2007} and $E$-correction using the luminosity evolution model of \citet{Blanton2003}. Each galaxy is then given a color label following the criterion of \citealt{Li2006} (formula 7 and table 4): it is ``red'' if its $(g-r)$ color satisfies
\begin{equation}
(g-r) \geq -0.788-0.078 \times M_{0.1\,r}, 
\end{equation}
and ``blue'' otherwise. Therefore, in the following, the color $c$ is seen as a two-valued random variable in the set $\left\lbrace c_\mathrm{I}=\mathrm{blue}, c_\mathrm{II}=\mathrm{red} \right\rbrace$. The following step is to determine in which web-type environment each galaxy lives, given the different classifiers. To do so, we adopted the criterion of \citet{Leclercq2015DT} for optimal decision-making, combined with the probabilities presented in section \ref{sec:Classifications}. Since we want to commit to a structure type for each galaxy, we adopted $\alpha=1$ in the notations of \citet{Leclercq2015DT} (the \textit{fair game} situation). This choice ensures that a decision is made for each voxel of the cosmic web map and results in the ``speculative maps'' of the large-scale structure, shown in figure \ref{fig:comparison_decision_maps}. We then assigned to each galaxy the structure type of its voxel using the Nearest-Grid-Point scheme.

\begin{table*}\centering
\begin{tabular}{ccccccc}
\hline\hline
ra & dec & $z$ & \tweb & {\diva} & {\origami} & color\\
\hline
$189.41567183$ & $-0.82251765$ & $0.072441$ & $3$ & $2$ & $1$ & I\\
$178.47971762$ & $-0.79366005$ & $0.132380$ & $2$ & $2$ & $1$ & II\\
$211.60459472$ & $0.89288053$ & $0.047731$ & $3$ & $2$ & $0$ & I\\
$132.11854651$ & $0.27207873$ & $0.051950$ & $2$ & $3$ & $1$ & II\\
$174.52633299$ & $43.94979005$ & $0.052975$ & $1$ & $0$ & $0$ & I\\
$222.71912116$ & $38.99904573$ & $0.056868$ & $0$ & $0$ & $0$ & I\\
$201.00744331$ & $13.95337526$ & $0.023865$ & $3$ & $3$ & $3$ & II\\
$209.12213650$ & $12.81801605$ & $0.027113$ & $2$ & $1$ & $0$ & I\\
\hline\hline
\end{tabular}
\caption{Some rows of the galaxy catalog used in section \ref{sec:Utility for predictions: galaxy colors}. The columns are: right ascension and declination (in degrees, J2000.0 equatorial coordinates); redshift; web-type environment as defined by the {\tweb}, {\diva} and {\origami} ($\mathrm{T}_0 =$ void, $\mathrm{T}_1~=~$ sheet, $\mathrm{T}_2 =$ filament, $\mathrm{T}_3 = $ cluster); galaxy color label ($c_\mathrm{I} = $ blue, $c_\mathrm{II} =$ red). The optimal choice of a classifier can be seen as a machine learning problem: in this training set, which classification is the most relevant to predict galaxy color?}
\label{tb:galaxy_catalog}
\end{table*}

\begin{figure*}
\begin{center}
\includegraphics[width=0.85\textwidth]{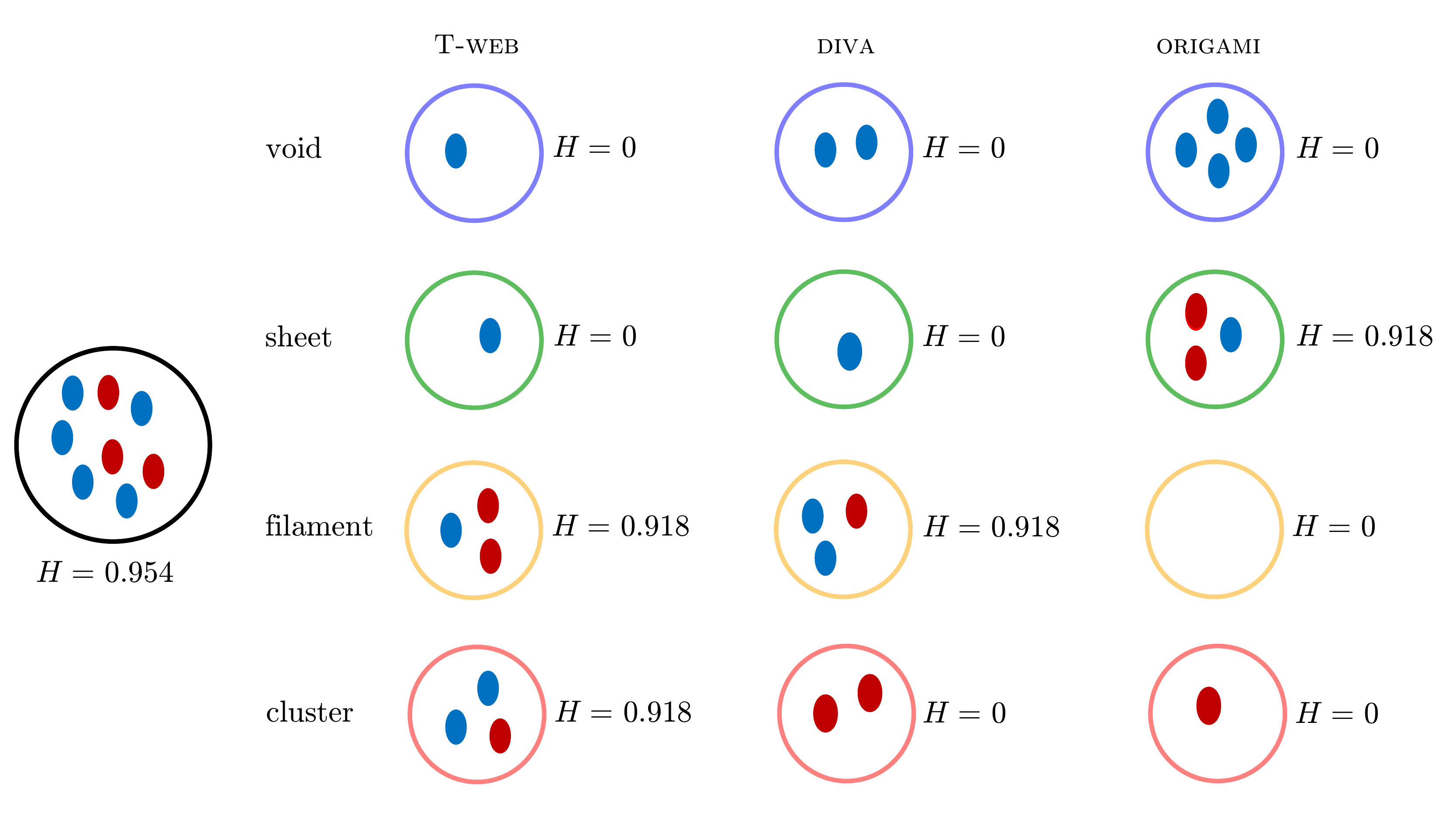}
\caption{Schematic illustration of the procedure to compute the utility of classifiers for predicting galaxy colors (equations \eqref{eq:utility_colors}--\eqref{eq:entropy_child}), using the subcatalog given in table \ref{tb:galaxy_catalog}. In this example, the parent entropy is $H=-3/8 \log_2(3/8) -5/8 \log_2(5/8) \approx 0.954$~Sh. The child entropy in each structure type is computed similarly and reported above. For each classifier, the utility is the parent entropy minus the weighted average of children entropies. In this case, we get $\widehat{U_3}(\xi_0)=0.266$~Sh for the {\tweb}, $\widehat{U_3}(\xi_1)=0.610$~Sh for {\diva} and $\widehat{U_3}(\xi_2)=0.610$~Sh for {\origami}.\label{fig:illustration_classifiers}}
\end{center}
\end{figure*}

Summing up the discussion above, we built a catalog containing, for each galaxy: a color $c_a$ ($a\in \left\lbrace \mathrm{I},\mathrm{II} \right\rbrace$) and three structure types $\mathrm{T}_i$ ($i\in \llbracket 0,3 \rrbracket$) -- one for each of the classifiers $\xi_\alpha$ ($\alpha \in \llbracket 0,2 \rrbracket$). Some rows of this catalog are given in table \ref{tb:galaxy_catalog} as examples. This catalog contains $N_\mathrm{gal}=367,157$ galaxies. 

In many respects, the question of choosing the best classifier for predicting galaxy colors can now be viewed as a supervised machine learning problem. Each of the input galaxies is assigned a set of \textit{attributes}, the web-type in which it lives according to each classifier, and a \textit{class label}, its color (see table \ref{tb:galaxy_catalog}). The design problem of choosing the most efficient classifier for predicting galaxy properties is analog to the machine learning problem of determining the most relevant attribute for discriminating among the classes to be learned \citep[for a cosmological example, see e.g.][]{Hoyle2015}.

Following equation \eqref{eq:joint_utility_prediction}, the utility of a classifier for predicting galaxy colors, that we seek to maximize, is
\begin{equation}
U_3(\xi) = \left\langle D_\mathrm{KL}[\mathcal{P}(c|\mathrm{T},\xi)||\mathcal{P}(c)] \right\rangle_{\mathcal{P}(\mathrm{T}|\xi)}
\end{equation}
where we have used that $\mathcal{P}(c|\xi)= \mathcal{P}(c)$ (before looking at the data, galaxy colors do not depend on the chosen classifier), and the simplifying assumption that $\mathcal{P}(c|d,\mathrm{T},\xi) = \mathcal{P}(c|\mathrm{T},\xi)$ (galaxy colors do not further depend on the data once their web-type environment is specified). It follows that the utility is the mutual information between the classification and the new observations (see section \ref{apx:Utilities for prediction of future observations}):
\begin{eqnarray}
U_3(\xi) & = & I[\mathrm{T}\!:\!c|\xi] \label{eq:utility_colors}\\
& = & H[\mathcal{P}(c)] - H[\mathcal{P}(c|\mathrm{T},\xi)] \nonumber\\
& = & H[\mathcal{P}(c)] - \sum_{i=0}^3 \mathcal{P}(\mathrm{T}=\mathrm{T}_i|\xi) H[\mathcal{P}(c|\mathrm{T}=\mathrm{T}_i,\xi)] \nonumber
\end{eqnarray}
The weighting coefficients to be used in the last line represent the probability that a \textit{galaxy} lives in web-type $\mathrm{T}_i$, given classifier $\xi$, irrespective of its color. They are approximated as (see also equation \eqref{eq:machine_learning})
\begin{eqnarray}
\mathcal{P}(\mathrm{T} = \mathrm{T}_i|\xi) & \approx & \frac{|\left\lbrace c | \mathrm{T}=\mathrm{T}_i,\xi \right\rbrace|}{N_\mathrm{gal}} \label{eq:weighting_colors}\\
& = & \sum_{a \in \left\lbrace \mathrm{I},\mathrm{II} \right\rbrace} \frac{N_{c_\mathrm{a}}(\mathrm{T}=\mathrm{T}_i|\xi)}{N_\mathrm{gal}} \nonumber\\
& = & \frac{N_\mathrm{blue}(\mathrm{T}=\mathrm{T}_i|\xi) + N_\mathrm{red}(\mathrm{T}=\mathrm{T}_i|\xi)}{N_\mathrm{gal}}, \nonumber
\end{eqnarray}
i.e. the fraction of galaxies (blue or red) that live in web-type $\mathrm{T}_i$. Note that this is different from $\mathcal{P}(\mathrm{T}_i|\xi)$, the prior probability for a given \textit{voxel} to belong to a structure of type $\mathrm{T}_i$. This difference accounts in particular for the fact that galaxies live preferentially in the most complex structures of the cosmic web.

The first term in equation \eqref{eq:utility_colors} is the ``parent'' entropy
\begin{eqnarray}
H[\mathcal{P}(c)] & = & \sum_{a \in \left\lbrace \mathrm{I},\mathrm{II} \right\rbrace} \frac{N_{c_\mathrm{a}}}{N_\mathrm{gal}} \log_2\left( \frac{N_{c_\mathrm{a}}}{N_\mathrm{gal}} \right) \label{eq:entropy_parent}\\
& = & \frac{N_\mathrm{blue}}{N_\mathrm{gal}} \log_2\left( \frac{N_\mathrm{blue}}{N_\mathrm{gal}} \right) + \frac{N_\mathrm{red}}{N_\mathrm{gal}} \log_2\left( \frac{N_\mathrm{red}}{N_\mathrm{gal}} \right). \nonumber
\end{eqnarray}
Similarly, for each classifier and each structure type, the ``child'' entropy $H[\mathcal{P}(c|\mathrm{T}=\mathrm{T}_i,\xi)]$ is estimated as
\begin{equation}
\sum_{a \in \left\lbrace \mathrm{I},\mathrm{II} \right\rbrace} \frac{N_{c_\mathrm{a}}(\mathrm{T}=\mathrm{T}_i|\xi)}{N_\mathrm{gal}} \log_2\left( \frac{N_{c_\mathrm{a}}(\mathrm{T}=\mathrm{T}_i|\xi)}{N_\mathrm{gal}} \right) .
\label{eq:entropy_child}
\end{equation}
Eventually, equations \eqref{eq:utility_colors}, \eqref{eq:weighting_colors}, \eqref{eq:entropy_parent} and \eqref{eq:entropy_child} yield $\widehat{U_3}(\xi)$, an estimator of $U_3$ for each of the classifiers. A schematic illustration of the entire procedure is given in figure \ref{fig:illustration_classifiers}.

\begin{table*}\centering
\begin{tabular}{lcccccccc}
\hline\hline
& \multicolumn{4}{c}{blue} & \multicolumn{4}{c}{red} \\
\hline
all & \multicolumn{4}{c}{194,503} & \multicolumn{4}{c}{172,654} \\
& \multicolumn{4}{c}{(53.0\%)} & \multicolumn{4}{c}{(47.0\%)} \\
\hline\hline
& \multicolumn{4}{c}{blue} & \multicolumn{4}{c}{red} \\
& void & sheet & filament & cluster & void & sheet & filament & cluster \\
\hline
{\tweb} & 19,150 & 19,290 & 63,318 & 92,745 & 18,456 & 8,370 & 42,678 & 103,150\\
& (50.9 \%) & (69.7 \%) & (59.7 \%) & (47.3 \%) & (49.1 \%) & (30.3 \%) & (40.3 \%) & (52.7 \%)\\
{\diva} & 26,358 & 27,051 & 67,515 & 73,579 & 22,436 & 15,289 & 51,402 & 83,527\\
& (54.0 \%) & (63.9 \%) & (56.8 \%) & (46.8 \%) & (46.0 \%) & (36.1 \%) & (43.2 \%) & (53.2 \%)\\
{\origami} & 82,805 & 69,583 & 29,748 & 12,367 & 55,362 & 60,775 & 36,589 & 19,928\\
& (59.9 \%) & (53.4 \%) & (44.8 \%) & (38.3 \%) & (40.1 \%) & (46.6 \%) & (55.2 \%) & (61.7 \%)\\
\hline\hline
\end{tabular}
\caption{Number and percentage of blue and red galaxies as a function of their web-type environment. In the first row, the number of blue and red galaxies in the entire catalog are reported. For the other rows, the number corresponds to the number of blue/red galaxies in the web-type indicated by the column, given the classifier indicated by the row; the percentage corresponds to the fraction of galaxy living in this web-type that are blue/red.}
\label{tb:galaxy_colors}
\end{table*}

\begin{figure*}
\begin{center}
\includegraphics[width=0.85\textwidth]{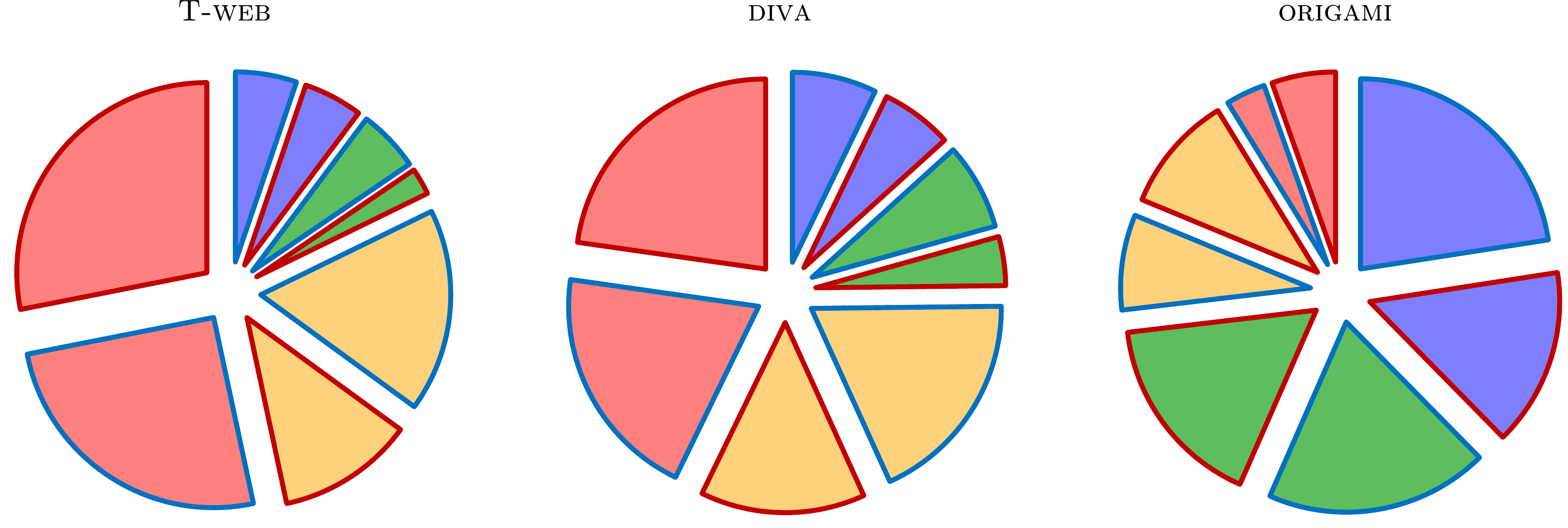}
\caption{Pie diagrams illustrating the data of table \ref{tb:galaxy_colors}, with the area of sectors proportional to the corresponding number of galaxies. The color of sectors represent the structure type (blue for voids, green for sheets, yellow for filaments, and red for clusters), and the borders represent the galaxy color ($c_\mathrm{I} =$ blue, $c_\mathrm{II} =$ red).\label{fig:illustration_pie}}
\end{center}
\end{figure*}

The result of our analysis is presented in table \ref{tb:galaxy_colors} and illustrated by the pie charts of figure \ref{fig:illustration_pie}. In the table, the first row represents the number of blue and red galaxies of our catalog, irrespective of their web-type environment. It permits to estimate the parent entropy $H[\mathcal{P}(c)] = 0.9974 \mathrm{~Sh}$. We also quote the number of blue and red galaxies that live in voids, sheets, filaments, and clusters, as defined by the {\tweb} (second row), by {\diva} (third row) and by {\origami} (fourth row). Resulting numbers for estimated utilities $\widehat{U_3}(\xi)$ are given in the last row of table \ref{tb:utilities}.

Several physical comments can be made at this point. All classifiers agree on a general trend, which can be observed in table \ref{tb:galaxy_colors}: red galaxies live preferentially in clusters, while blue galaxies live preferentially in sheets and voids. This is in agreement with earlier results \citep[e.g][]{Hogg2003,Patiri2006,Alpaslan2015}. It is interesting to note, that though almost half of the galaxies are found in a void according to {\origami}, its utility stays comparable to that of the other classifiers. In our setup, the {\tweb} and {\origami} have similar performance at predicting galaxy colors, and outperform {\diva}. This could be due to the weaker sensitivity of {\diva} classifications to the local density, which is known to correlate with galaxy colors \citep{Hogg2003}. It is also notable that for all classifiers, the information gained on galaxy colors once their web-type is known is rather small -- of the order of $10^{-2}$~Sh, to be compared, for example, to the $\sim$ 1 Sh gained on cosmological models from cosmic microwave background experiments \citep{Seehars2014,Martin2016}. From a machine learning perspective, this result means that none of the attributes that we considered are really relevant to learn the class label. This suggests that galaxy colors are only loosely related to the physical information exploited by the {\tweb}, {\diva} and {\origami} (tidal field, shear of the Lagrangian displacement field and number of particle crossings, respectively). It further highlights the necessity of developing targeted cosmic web classifiers for cross-correlating with galaxy properties. Unsupervised machine learning from an extended space of attributes, including galaxy properties and their cosmic environment at different scales \mbox{\citep[e.g.][]{FrankJascheEnslin2016}} could allow progress for the design of such classifiers.

\section{Summary and Conclusions}
\label{sec:Conclusions}

Following \citet{Leclercq2015ST}, this study discusses the data-supported connection between cosmic web analysis and information theory. It is a project exploiting the cosmic web maps of \citet{Leclercq2015ST} and \citet{Leclercq2016}, which are part of the rich variety of chrono-cosmographic results produced by the application of the Bayesian inference engine {\borg} \citep{Jasche2013BORG} to the Sloan Digital Sky Survey main sample galaxies \citep{Jasche2015BORGSDSS}. Using information-theoretic concepts, in section \ref{sec:Mapping information in the cosmic web}, we measure and characterize the extent to which the SDSS is informative about the morphological features of the underlying cosmic web, as defined by the {\tweb} \citep{Hahn2007a}, {\diva} \citep{Lavaux2010}, and {\origami} \citep{Falck2012}.

In section \ref{sec:Classifier utilities}, this paper quantitatively addresses the question of choosing a cosmic web classifier, depending on the desired use. To do so, we extend the decision-theory framework of \citet{Leclercq2015DT} by introducing utility functions on the space of classifiers. We consider three classes of problem: parameter inference (section \ref{sec:Utility for parameter inference: cosmic web analysis}), model selection (section \ref{sec:Utility for model selection: dark energy equation of state}), and prediction of new observations (section \ref{sec:Utility for predictions: galaxy colors}). In each of these general situations, we propose a utility function based on the concept of mutual information and motivated in the general framework of Bayesian design. We summarize them below:
\begin{itemize}
\item for parameter inference: $U_1(\xi) = I[\mathrm{T}\!:\!d|\xi]$ (equation \eqref{eq:utility_inference}), the mutual information between the classification $\mathrm{T}$ and the data $d$, given the classifier $\xi$,
\item for model selection: $U_2(\xi) = I[ \mathcal{M}\!:\!\mathcal{R}(d) | \xi]$ (equation \eqref{eq:utility_model_selection}), the mutual information between the model indicator $\mathcal{M}$ and the mixture $\mathcal{R}(d)$ of posterior distributions, conditional on the different competing models and on the classifier $\xi$,
\item for predictions: $U_3(\xi) = I[\mathrm{T}\!:\!c|\xi]$ (equation \eqref{eq:utility_colors}), the mutual information between the classification $\mathrm{T}$ and the new observations $c$, given the classifier $\xi$.
\end{itemize}
In practice, due to the difficulty of combining competing goals, the decision maker may be unwilling or unable to specify a unique utility function. Given the set of possible utility functions for different situations, a target function for several design objectives can be written down, for example, as a weighted average of different utilities.

As an illustration of our methodological framework, we assessed the relative performance of the {\tweb}, {\diva}, {\origami} for different goals, one of each type mentioned above: optimization of the information content of cosmic web maps, comparison of dark energy models, and prediction of galaxy colors. Our physical findings can be summarized as follows. We found that the {\tweb} maximizes the information content of web-type maps (especially in the densest regions), but that {\diva} may have to be preferred due to its lower sensitivity to artifacts. Unsurprisingly, Lagrangian classifiers ({\diva} and {\origami}), which exploit the displacement field, excel at finding the regions of the cosmic web, such as the boundaries of voids, that are the most sensitive to the equation of state of dark energy. Finally, all classifiers agree on the general trend of red galaxies in clusters, and blue galaxies in sheets and voids. The information gained on galaxy colors is the highest with the {\tweb}, slightly less with {\origami}, and lowest with {\diva}; but the absolute number stays rather low. Though investigation of this question should be made much more comprehensive, this result is indicative of the as-of-now limited understanding of the connection between galaxy properties and the cosmic web, which is essential to the development of a consistent cosmological theory of galaxy formation and evolution.

Numerical results given in table \ref{tb:utilities} depend on method-internal parameters, in particular a scale on the Eulerian or Lagrangian grid. Though we did not investigate this question, in addition to making comparisons between different classifiers, the same formalism can be used within each classification method to assign a utility for each parameter choice, then decide on which values to use. This allows to probe the hierarchical nature of the cosmic web quantitatively and to focus on the optimal filtering scale for the considered problem. At the largest filtering scales (for example, when studying the dark energy equation of state), we expect the results of the {\tweb} and {\diva} to converge, since the methods are equivalent at first order in Lagrangian perturbation theory; whereas {\origami} will miss all the phase-space foldings that happen below the considered scale. At the smallest filtering scales (for example, when studying galaxies), we expect an increase of the information gained on all features that are intrinsically local, such as many galaxy properties.

In order to facilitate the use of our methods and as a complement to our earlier release of density fields and cosmic web maps, we made publicly available the maps, analysis scripts and galaxy catalog used in this paper, which can be used to reproduce our results. These products are available from the first author's website, currently hosted at \href{http://icg.port.ac.uk/~leclercq/}{http://icg.port.ac.uk/$\sim$leclercq/}.

Though we used the common terms of voids, sheets, filaments, and clusters, this paper can be considered as a generic way to optimally define four summary statistics A, B, C, D of the large-scale structure, depending on the desired use. Therefore, beyond the specific application of cosmic web analysis, our methodology opens the way to the automatic design of summary statistics of the LSS that capture targeted parts of its information content. In the coming area of accurate cosmology with deep galaxy surveys, we expect that the optimal design of analysis procedures will play an ever-increasing role.

\appendix

\section{Information theory}
\label{apx:Information theory}

We review here some useful information-theoretic notions. For simplicity, we consider only discrete random variables, but the generalization to continuous variables is possible, by replacing discrete sums by integrals. Throughout this appendix, $X$ and $Y$ are two discrete random variables with respective possible values in $\mathcal{X} = \{x_0, ..., x_n\}$ and $\mathcal{Y} = \{y_0, ..., y_m\}$. We note their respective pmfs $\mathcal{P}(X)$ and $\mathcal{P}(Y)$. We denote by $\mathcal{Q}(X) \equiv \mathcal{P}(X')$ the pmf of another discrete random variable $X'$ with possible values in $\mathcal{X}$.

\subsection{Jensen's inequality}
\label{apx:Jensen's inequality}

An important result used in the following is Jensen's inequality \citep{Jensen1906} in a probabilistic setting. Let us consider a probability space, a random variable $X$ with probability distribution $\mathcal{P}(X)$ and a convex function $\varphi$. Then
\begin{equation}
\varphi(\left\langle X \right\rangle_{\mathcal{P}(X)}) \leq \left\langle \varphi(X) \right\rangle_{\mathcal{P}(X)} 
\end{equation}
where the brackets indicate the expectation value of the quantity inside, under the probability $\mathcal{P}(X)$.

\subsection{Information content and Shannon entropy}
\label{apx:Information content and Shannon entropy}

The \textit{information content} (or \textit{self-information}) of $X$ is defined by
\begin{equation}
I[X] \equiv - \sum_{x\in\mathcal{X}} \log_2 \mathcal{P}(x) ,
\end{equation}
where the $\mathcal{P}(x)$ are the probabilities of possible events. The \textit{entropy} \citep{Shannon1948} is the expectation of the information content under the probability itself, i.e.
\begin{equation}
H[X] \equiv \left\langle I[X] \right\rangle_{\mathcal{P}(X)} = \left\langle -\log_2 \mathcal{P}(X) \right\rangle_{\mathcal{P}(X)} .
\end{equation}
It can be written explicitly
\begin{equation}
H[X] = - \sum_{x\in\mathcal{X}} \mathcal{P}(x) \log_2 \mathcal{P}(x) .
\label{eq:def_entropy}
\end{equation}
Information content and Shannon entropy are non-negative quantities. Furthermore, Jensen's inequality (section \ref{apx:Jensen's inequality}) implies that:
\begin{eqnarray}
H[X] & = & \sum_{x\in\mathcal{X}} \mathcal{P}(x) \log_2\left(\frac{1}{\mathcal{P}(x)}\right) \nonumber\\
&\leq & \log_2\left( \sum_{x\in\mathcal{X}} \mathcal{P}(x) \frac{1}{\mathcal{P}(x)} \right) = \log_2 |\mathcal{X}| ,
\end{eqnarray}
since $-\log_2$ is a convex function. This maximal entropy is effectively attained in the case of a uniform pmf: uncertainty is maximal when all possible events are equiprobable. 

The \textit{joint entropy} of two random variables $X$ and $Y$ is defined as
\begin{eqnarray}
H[X,Y] & \equiv & \left\langle I[X,Y] \right\rangle_{\mathcal{P}(X,Y)} \nonumber\\
& = & - \sum_{x\in\mathcal{X},y\in\mathcal{Y}} \mathcal{P}(x,y) \log_2 \mathcal{P}(x,y) .
\label{eq:def_joint_entropy}
\end{eqnarray}

One may also define the \textit{conditional entropy} of $X$ given $Y$ as
\begin{equation}
H[X|Y] \equiv \left\langle H[X|Y=y] \right\rangle_{\mathcal{P}(Y)} = \sum_{y\in\mathcal{Y}} \mathcal{P}(y) \, H[X|Y=y] .
\end{equation}
Using
\begin{equation}
H[X|Y=y] = - \sum_{x\in\mathcal{X}} \mathcal{P}(x|y) \log_2 \mathcal{P}(x|y)
\label{eq:def_conditional_entropy}
\end{equation}
and $\mathcal{P}(x,y)=\mathcal{P}(x|y)\mathcal{P}(y)$, it is easy to show that the conditional entropy verifies
\begin{equation}
H[X|Y] = \sum_{x\in\mathcal{X},y\in\mathcal{Y}} \mathcal{P}(x,y) \log_2 \left( \frac{\mathcal{P}(y)}{\mathcal{P}(x,y)}\right) .
\label{eq:prop_conditional_entropy}
\end{equation}
From equations \eqref{eq:def_entropy}, \eqref{eq:def_joint_entropy} and \eqref{eq:prop_conditional_entropy}, one can derive the \textit{chain rule} of conditional entropy:
\begin{equation}
H[X|Y] = H[X,Y] - H[Y],
\end{equation}
from which follows straightforwardly an equivalent of Bayes' theorem for entropies,
\begin{equation}
H[X|Y]+H[Y] = H[Y|X]+H[X] .
\end{equation}

Finally, the \textit{cross entropy} between two random variables $X$ and $X'$ with possible values in the same set $\mathcal{X}$ and respective pmfs $\mathcal{P}(X)$ and $\mathcal{Q}(X)$ is
\begin{equation}
H[X||X'] \equiv - \sum_{x\in\mathcal{X}} \mathcal{P}(x) \log_2 \mathcal{Q}(x) .
\label{eq:def_cross_entropy}
\end{equation}

\subsection{Mutual information}
\label{apx:Mutual information}

The \textit{mutual information} of two variables $X$ and $Y$ is defined as
\begin{equation}
I[X\!:\!Y] \equiv \sum_{x\in\mathcal{X},y\in\mathcal{Y}} \mathcal{P}(x,y) \log_2\left( \frac{\mathcal{P}(x,y)}{\mathcal{P}(x)\mathcal{P}(y)} \right).
\label{eq:def_mutual_information}
\end{equation}
It is a symmetric measure of inherent dependence of $X$ and $Y$. Jensen's inequality (section \ref{apx:Jensen's inequality}) implies that it is non-negative:
\begin{eqnarray}
-I[X\!:\!Y] & = & \sum_{x\in\mathcal{X},y\in\mathcal{Y}} \mathcal{P}(x,y) \log_2\left( \frac{\mathcal{P}(x)\mathcal{P}(y)}{\mathcal{P}(x,y)} \right) \nonumber\\
& \leq & \log_2\left( \sum_{x\in\mathcal{X},y\in\mathcal{Y}} \mathcal{P}(x,y) \frac{\mathcal{P}(x)\mathcal{P}(y)}{\mathcal{P}(x,y)} \right) \nonumber\\
& = & \log_2\left(\sum_{x\in\mathcal{X},y\in\mathcal{Y}}\mathcal{P}(x)\mathcal{P}(y) \right) = 0.
\end{eqnarray}

A remarkable property is that the entropy satisfies $H[X] = I[X\!:\!X]$, the mutual information of $X$ and itself.

Using the definition of conditional probabilities, one can also show that mutual information can be equivalently expressed as:
\begin{eqnarray}
I[X\!:\!Y] & = & H[X] - H[X|Y] \label{eq:mutual_info_entropy}\\
& = & H[Y] - H[Y|X] \\
& = & H[X]+H[Y]-H[X,Y] \\
& = & H[X,Y]-H[X|Y]-H[Y|X].
\end{eqnarray}
It follows that for any $X$ and $Y$, $H[X|Y] \leq H[X]$. Therefore, conditional entropy should be understood as the amount of randomness remaining in $X$ once $Y$ is known. $H[X|Y]=0$ if and only if the value of $X$ is completely determined by the value of $Y$. Conversely, $H[X|Y]=H[X]$ if and only if $X$ and $Y$ are independent random variables.

\subsection{Kullback-Leibler divergence}
\label{apx:Kullback-Leibler divergence}

In this section and in the following, we consider two discrete random variables $X$ and $X'$ with possible values in $\mathcal{X}$ and respective pmfs $\mathcal{P}(X)$ and $\mathcal{Q}(X) \equiv \mathcal{P}(X')$. When there is no ambiguity, we simplify the formalism and note $I[\mathcal{P}] \equiv I[X]$, $H[\mathcal{P}]~\equiv~H[X]$, $H[\mathcal{P}||\mathcal{Q}]~\equiv~H[X||X']$, etc.

The Kullback-Leibler divergence \citep{Kullback1951} is a non-symmetric measure of the difference between two probability distributions. It is defined as 
\begin{equation}
D_\mathrm{KL}\left[ \mathcal{P}||\mathcal{Q} \right] \equiv \sum_{x\in\mathcal{X}} \mathcal{P}(x) \log_2\left( \frac{\mathcal{P}(x)}{\mathcal{Q}(x)} \right) .
\label{eq:def_DKL}
\end{equation}
It can also be expressed in terms of the entropy of $\mathcal{P}$ and the cross entropy between $\mathcal{P}$ and $\mathcal{Q}$ (see equations \eqref{eq:def_entropy} and \eqref{eq:def_cross_entropy}):
\begin{equation}
D_\mathrm{KL}\left[ \mathcal{P}||\mathcal{Q} \right] = H\left[ \mathcal{P}||\mathcal{Q} \right] - H\left[ \mathcal{P} \right].
\end{equation}
An important result, known as Gibbs' inequality, states that the Kullback-Leibler divergence is non-negative, reaching zero if and only if $\mathcal{P}=\mathcal{Q}$. Equivalently, for two pmfs $\mathcal{P}$ and $\mathcal{Q}$, $H[\mathcal{P}] \leq H\left[ \mathcal{P}||\mathcal{Q} \right]$, i.e. the (self) entropy of $\mathcal{P}$ is always smaller than the cross entropy of $\mathcal{P}$ with any other pmf $\mathcal{Q}$. The proof uses the inequality $\ln(x) \leq x -1$ for all $x>0$, with equality if and only if $x=1$. Denoting by $\widetilde{\mathcal{X}}$ the subset of $\mathcal{X}$ for which $\mathcal{P}(x)$ is non-zero, we have:
\begin{eqnarray}
\sum_{x \in \widetilde{\mathcal{X}}} \mathcal{P}(x) \ln \left(\frac{\mathcal{Q}(x)}{\mathcal{P}(x)}\right) & \leq & \sum_{x \in \widetilde{\mathcal{X}}} \mathcal{P}(x) \left( \frac{\mathcal{Q}(x)}{\mathcal{P}(x)}-1 \right) \nonumber\\
& = & \sum_{x \in \widetilde{\mathcal{X}}} \mathcal{Q}(x) - \sum_{x \in \widetilde{\mathcal{X}}} \mathcal{P}(x) \nonumber\\
& = & \sum_{x \in \widetilde{\mathcal{X}}} \mathcal{Q}(x) - 1 \leq 0 .
\end{eqnarray}
The property follows trivially.

As can be seen from its definition, equation \eqref{eq:def_mutual_information}, mutual information is related to the Kullback-Leibler divergence:
\begin{equation}
I\left[X\!:\!Y \right] = D_\mathrm{KL}\left[ \mathcal{P}(x,y) || \mathcal{P}(x) \mathcal{P}(y) \right] . 
\end{equation}
Furthermore, using $\mathcal{P}(x,y)=\mathcal{P}(x|y)\mathcal{P}(y)$, we obtain
\begin{eqnarray}
I\left[X\!:\!Y \right] & = & \sum_{y\in\mathcal{Y}} \mathcal{P}(y) D_\mathrm{KL}\left[\mathcal{P}(x|y) || \mathcal{P}(x) \right] \nonumber\\
& = & \left\langle D_\mathrm{KL}\left[\mathcal{P}(x|y) || \mathcal{P}(x) \right] \right\rangle_{\mathcal{P}(Y)} .
\label{eq:mutual_info_DKL}
\end{eqnarray}
Mutual information can thus be understood as the expectation of the Kullback-Leibler divergence $D_\mathrm{KL}\left[\mathcal{P}(x|y) || \mathcal{P}(x) \right]$ (seen as a random variable in $y$) of the conditional distribution $\mathcal{P}(x|y)$ from the unconditional distribution $\mathcal{P}(x)$.

A convenient way to think of $D_\mathrm{KL}\left[ \mathcal{P}||\mathcal{Q} \right]$ is as of a quantification of the information lost when $\mathcal{Q}$ is used to approximate $\mathcal{P}$. In Bayesian statistics, the Kullback-Leibler divergence can be used as a measure of the information gained in moving from the prior distribution of some quantity $x$, $\mathcal{P}(x)$ to the posterior distribution of $x$ given the data $d$, $\mathcal{P}(x|d)$. According to Bayes' theorem,
\begin{equation}
\mathcal{P}(x|d) = \frac{\mathcal{P}(d|x)}{\mathcal{P}(d)} \mathcal{P}(x) .
\end{equation}
In this context, we refer to the Kullback-Leibler divergence of the posterior from the prior,
\begin{equation}
D_\mathrm{KL}\left[ \mathcal{P}(x|d) || \mathcal{P}(x) \right] = \sum_{x\in\mathcal{X}} \mathcal{P}(x|d) \log_2\left( \frac{\mathcal{P}(x|d)}{\mathcal{P}(x)} \right) ,
\end{equation}
as the \textit{information gain}.

\subsection{Jensen-Shannon divergence}
\label{apx:Jensen-Shannon divergence}

The Jensen-Shannon divergence \citep{Lin1991} is a symmetrized version of the Kullback-Leibler divergence. It is defined as
\begin{equation}
D_\mathrm{JS}\left[ \mathcal{P}\!:\!\mathcal{Q} \right] \equiv \frac{1}{2} D_\mathrm{KL}\left[ \mathcal{P} \middle\| \mathcal{R} \right] + \frac{1}{2} D_\mathrm{KL}\left[ \mathcal{Q} \middle\| \mathcal{R} \right] ,
\label{eq:def_DJS} 
\end{equation}
where $\mathcal{R} \equiv \left( \mathcal{P}+\mathcal{Q} \right)/2$. The above definition can be generalized to more than two distributions, by noting
\begin{equation}
D_\mathrm{JS}\left[ \mathcal{P}_1\!:\!\mathcal{P}_2\!:\! ... \!:\!\mathcal{P}_n\right] \equiv \frac{1}{n} \sum_{i=1}^n D_\mathrm{KL}\left[ \mathcal{P}_i || \frac{\sum_{j=1}^n \mathcal{P}_j}{n} \right] .
\label{eq:def_DJS_generalized}
\end{equation}
The square root of the Jensen-Shannon divergence is a metric and is often referred to as Jensen-Shannon distance \citep{Endres2003}.

Using equation \eqref{eq:def_DKL}, one can check explicitly that
\begin{equation}
D_\mathrm{JS}\left[ \mathcal{P}\!:\!\mathcal{Q} \right] = H[\mathcal{R}] - \frac{1}{2} H[\mathcal{P}] - \frac{1}{2} H[\mathcal{Q}] .
\label{eq:DJS_property}
\end{equation}

The Jensen-Shannon divergence can also be related to the concept of mutual information. Let us consider the random variable $Z$ that takes the values $0$ and $1$ with probabilities $1/2$ and $1/2$, and the random variable $R~\equiv~ZX'+(1-Z)X$. The pmf of $R$ is $\mathcal{R}=(\mathcal{P}+\mathcal{Q})/2$. Using equation \eqref{eq:mutual_info_entropy}, the mutual information between $R$ and $Z$ is
\begin{equation}
I[R\!:\!Z] = H[R] - H[R|Z] ,
\end{equation}
where $H[R]$ is $H[\mathcal{R}]$, and (using equation \eqref{eq:def_conditional_entropy})
\begin{eqnarray}
H[R|Z] & = & \frac{1}{2} H[R|Z=0] + \frac{1}{2} H[R|Z=1] \nonumber\\
& = & \frac{1}{2} H[X] + \frac{1}{2} H[X'] \nonumber\\
& = & \frac{1}{2} H[\mathcal{P}] + \frac{1}{2} H[\mathcal{Q}] .
\end{eqnarray}
Thus, given equation \eqref{eq:DJS_property}, we have
\begin{equation}
D_\mathrm{JS}\left[ \mathcal{P}\!:\!\mathcal{Q}\right] = I[R\!:\!Z],
\label{eq:DJS_mutual_info}
\end{equation}
the mutual information between the mixture variable $R$ and the indicator $Z$ used to produce the mixture. It follows from this result that the Jensen-Shannon divergence is always between $0$ and $1$~Sh, since mutual information is non-negative and since
\begin{equation}
I[R\!:\!Z] = H[Z] - H[Z|X] \leq H[Z] = 1.
\end{equation}

\section{Bayesian experimental design}
\label{apx:Bayesian experimental design}

In Bayesian experimental design \citep[e.g.][]{Chaloner1995}, the expected utility of an experiment with design $\xi$ can be defined as
\begin{equation}
U(\xi) \equiv \left\langle U(d,\xi) \right\rangle_{\mathcal{P}(d|\xi)} = \int \mathcal{P}(d|\xi) \, U(d,\xi) \: \mathrm{d}d .
\label{eq:utility_xi}
\end{equation}
It should incorporate experimental aims and be specific to the targeted application. In this appendix, we review some of the more commonly used utility functions. 

\subsection{Parameter inference utility functions}
\label{apx:Parameter inference utility functions}

Precise parameter inference is an frequent goal for experimental design. In this case, $U(d,\xi)$ is a function of the posterior probability distribution function $\mathcal{P}(\theta|d,\xi)$ of the inferred parameters $\theta$. It is often chosen as the gain in Shannon information, i.e.
\begin{eqnarray}
U(d,\xi) & \equiv & D_\mathrm{KL}\left[\mathcal{P}(\theta|d,\xi) || \mathcal{P}(\theta|\xi)\right] \nonumber \\
& = & \int \mathcal{P}(\theta|d,\xi) \log_2\left( \frac{\mathcal{P}(\theta|d,\xi)}{\mathcal{P}(\theta|\xi)}\right) \: \mathrm{d}\theta ,
\label{eq:utility_d_xi}
\end{eqnarray}
where $D_\mathrm{KL}$ is the Kullback-Leibler divergence of the posterior from the prior (see section \ref{apx:Kullback-Leibler divergence}). In this fashion, the utility $U(\xi)$ is the expected information gain under the experimental design $\xi$. Given equations \eqref{eq:utility_xi} and \eqref{eq:utility_d_xi}, it can also be written as
\begin{eqnarray}
U(\xi) & = & \int \mathcal{P}(d|\xi) \, D_\mathrm{KL}\left[\mathcal{P}(\theta|d,\xi) || \mathcal{P}(\theta|\xi)\right] \: \mathrm{d}d \nonumber \\
& = & \iint \mathcal{P}(\theta,d|\xi) \log_2\left( \frac{\mathcal{P}(\theta,d|\xi)}{\mathcal{P}(\theta|\xi)\mathcal{P}(d|\xi)}\right) \: \mathrm{d}\theta \, \mathrm{d}d \nonumber \\
& \equiv & I\left[\theta\!:\!d|\xi\right],
\label{eq:utility_inference_mutual_info}
\end{eqnarray}
i.e. the mutual information between $\theta$ and $d$, conditional on the experimental design $\xi$ (see also equations \eqref{eq:def_mutual_information} and \eqref{eq:mutual_info_DKL}).

In some situations, instead of using the expected information gain, it is acceptable to simplify the problem by considering scalar functions of the posterior covariance matrix $\mathrm{cov}(\theta|d,\xi)$. For example, a design maximizing the expected value of
\begin{equation}
U_\mathrm{A}(d,\xi) \equiv \frac{1}{\mathrm{tr}(\mathrm{cov}(\theta|d,\xi)^{-1})}
\end{equation}
is called A-optimal, and a design maximizing the expected value of
\begin{equation}
U_\mathrm{D}(d,\xi) \equiv \mathrm{det}(\mathrm{cov}(\theta|d,\xi))
\end{equation}
is called D-optimal \citep[see][for an application to the design of cosmological surveys]{Bassett2005}.

\subsection{Model selection utility functions}
\label{apx:Model selection utility functions}

In model selection questions, the quantity of interest is not the posterior distributions of different models, but the \textit{Bayes factor}, defined as
\begin{equation}
\mathcal{B}_{12}(\xi) \equiv \frac{\mathcal{P}(d|\xi,\mathcal{M}_1)}{\mathcal{P}(d|\xi,\mathcal{M}_2)},
\end{equation}
i.e. the ratio of \textit{evidences}, where for $i=1,2$,
\begin{equation}
\mathcal{P}(d|\xi,\mathcal{M}_i) = \int \mathcal{P}(d|\theta,\xi,\mathcal{M}_i) \, \mathcal{P}(\theta|\xi,\mathcal{M}_i) \: \mathrm{d} \theta .
\end{equation}
Naively, optimizing experimental design to effectively allow model selection suggests to simply use as utility
\begin{equation}
U(\xi) = \mathcal{B}_{12}(\xi).
\end{equation}
Forecasting a predictive distribution of Bayes factors has been studied in the case of nested models using the Savage-Dickey ratio \citep{Trotta2007}, but it is computationally intractable for most non-linear models that are not nested.

Another common idea in the Bayesian design literature for model selection is to use mutual information: the optimal design $\xi$ is the one that maximizes the mutual information between the model indicator $\mathcal{M}$ (a random variable on the space of possible models, that takes the value $\mathcal{M}_i$ with the meta-prior probability $\mathcal{P}(\mathcal{M}_i)$) and the future observation $d$:
\begin{equation}
U(\xi) = I\left[ \mathcal{M}\!:\!d |\xi \right] 
\end{equation}
\citep[see, for example,][]{Cavagnaro2010}.

The model selection utility known as ``total separation'' \citep{Roth1965} rather aims at finding designs that yield the largest difference between the means of posterior predictive distributions of rival models.

In the context of involved non-linear models of biochemical reaction networks, \citet{Vanlier2014} proposed a model selection utility based on the Jensen-Shannon divergence between posterior predictive distributions. Let us denote by $\mathcal{P}_i \equiv \mathcal{P}(d|\tilde{d},\mathcal{M}_i,\xi)$ the probability of a new measurement $d$ given the available data set $\tilde{d}$, under model $\mathcal{M}_i$ ($i=1,2$) and experimental design $\xi$. Using property \eqref{eq:DJS_mutual_info}, the utility is
\begin{equation}
U(\xi) = D_\mathrm{JS}\left[ \mathcal{P}_1\!:\!\mathcal{P}_2 | \xi \right] = I[\mathcal{M}\!:\!\mathcal{R} | \xi],
\label{eq:utility_DJS_mixture}
\end{equation}
where $\mathcal{R}$ is the mixture of predictive densities and $\mathcal{M}$ is the model indicator. Informally, maximizing the Jensen-Shannon divergence between predictive distributions is maximizing the reduction of uncertainty in the determination of which of the two distributions the new measurement comes from. For analytically tractable models, \citet{Vanlier2014} showed that the Jensen-Shannon divergence of predictive distributions is approximately monotonically related to the expected change in the Bayes factor, in favor of the model that generates new data. This supports the Jensen-Shannon divergence as a useful quantity for model discrimination.

\subsection{Utilities for prediction of future observations}
\label{apx:Utilities for prediction of future observations}

The design $\xi$ that optimizes the predictive power of the experiment, once data $d$ are acquired, is the one which allows the best prediction of future observations $t$, conditional on $d$ and $\xi$ \citep[e.g.][]{Vanlier2012}. In this situation, the Kullback-Leibler divergence of the posterior predictive distribution from the prior predictive distribution can be used as joint utility:
\begin{equation}
U(d,\xi) = D_\mathrm{KL}\left[ \mathcal{P}(t|d,\xi) || \mathcal{P}(t|\xi) \right] ,
\end{equation}
and the utility of $\xi$ is the expected information gain for the predicted observation:
\begin{equation}
U(\xi) = \left\langle D_\mathrm{KL}\left[ \mathcal{P}(t|d,\xi) || \mathcal{P}(t|\xi) \right] \right\rangle_{\mathcal{P}(d)} .
\end{equation}
Using property \eqref{eq:mutual_info_DKL}, this is equivalent to the mutual information between the predicted observation $t$ and the upcoming observation $d$, conditional on the design $\xi$:
\begin{equation}
U(\xi) = I\left[ t\!:\!d |\xi \right] .
\end{equation}
It is also the change of entropy from the prior predictive distribution to the posterior predictive distribution (see equation \eqref{eq:mutual_info_entropy}),
\begin{equation}
U(\xi) = H\left[ \mathcal{P}(t|\xi) \right] - H\left[ \mathcal{P}(t|d,\xi) \right] .
\end{equation}

The above formalism is directly analog to that of supervised machine learning, where $T$ denotes a set of training examples, each of the form $(\textbf{t},\ell) = (t_1,t_2,...,t_n,\ell)$, where $t_a \in \mathrm{Vals}(a)$ is the value of the $a$-th attribute of example $\textbf{t}$ and $\ell$ is its class label. The goal is to determine which of the attributes are the most informative, if one wants to predict the class label of future elements. The utility of an attribute $a$ is the expected information gain:
\begin{eqnarray}
U(a) & = & H\left[ T \right] - H\left[ T|a \right] \label{eq:machine_learning}\\
& = & H\left[ T \right] - \sum_{\tilde{a} \in \mathrm{Vals}(a)} \frac{| \left\lbrace \ell | t_a = \tilde{a} \right\rbrace |}{|T|} H\left[ \left\lbrace \ell | t_a = \tilde{a} \right\rbrace \right] \nonumber .
\end{eqnarray}

\section*{Statement of contribution}

\small{FL conceived the project, performed the study and wrote the paper. GL developed {\diva}; JJ developed {\borg} and lead the SDSS analysis. GL and JJ contributed to the construction of utility functions. BW was involved in the conception and design of Bayesian large-scale structure inference and contributed to the interpretation of results. All authors read and approved the final manuscript.}

\acknowledgments

\small{We are grateful to Bridget Falck and Mark Neyrinck for useful comments on the draft of this paper. FL thanks Will Percival for useful discussions and Thomas Tram for help with the {\class} code. This work has made use of the publicly available \href{http://class-code.net/}{\class} and \href{http://folk.uio.no/bridgetf/origami.html}{\origami} codes. Numerical computations were done on the Sciama High Performance Compute (HPC) cluster which is supported by the ICG, SEPNet and the University of Portsmouth. This work has also made use of the Horizon Cluster hosted by Institut d'Astrophysique de Paris; we thank St\'ephane Rouberol for running smoothly this cluster for us.}

\small{FL acknowledges funding from European Research Council through grant 614030, Darksurvey. JJ is partially supported by a Feodor Lynen Fellowship by the Alexander von Humboldt foundation. BW acknowledges funding from an ANR Chaire d'Excellence (ANR-10-CEXC-004-01) and the UPMC Chaire Internationale in Theoretical Cosmology. This work has been done within the Labex \href{http://ilp.upmc.fr/}{Institut Lagrange de Paris} (reference ANR-10-LABX-63) part of the Idex SUPER, and received financial state aid managed by the Agence Nationale de la Recherche, as part of the programme Investissements d'avenir under the reference ANR-11-IDEX-0004-02. This research was supported by the DFG cluster of excellence ``\href{www.universe-cluster.de}{Origin and Structure of the Universe}''.}

\section*{References}
\bibliography{/home/leclercq/workspace/biblio}

\begin{thebibliography}{63}%
\makeatletter
\providecommand \@ifxundefined [1]{%
 \@ifx{#1\undefined}
}%
\providecommand \@ifnum [1]{%
 \ifnum #1\expandafter \@firstoftwo
 \else \expandafter \@secondoftwo
 \fi
}%
\providecommand \@ifx [1]{%
 \ifx #1\expandafter \@firstoftwo
 \else \expandafter \@secondoftwo
 \fi
}%
\providecommand \natexlab [1]{#1}%
\providecommand \enquote  [1]{``#1''}%
\providecommand \bibnamefont  [1]{#1}%
\providecommand \bibfnamefont [1]{#1}%
\providecommand \citenamefont [1]{#1}%
\providecommand \href@noop [0]{\@secondoftwo}%
\providecommand \href [0]{\begingroup \@sanitize@url \@href}%
\providecommand \@href[1]{\@@startlink{#1}\@@href}%
\providecommand \@@href[1]{\endgroup#1\@@endlink}%
\providecommand \@sanitize@url [0]{\catcode `\\12\catcode `\$12\catcode
  `\&12\catcode `\#12\catcode `\^12\catcode `\_12\catcode `\%12\relax}%
\providecommand \@@startlink[1]{}%
\providecommand \@@endlink[0]{}%
\newcommand{\PineGreen}[1]{\textcolor{PineGreen}{#1}}%
\providecommand \url  [0]{\begingroup\@sanitize@url \@url }%
\providecommand \@url [1]{\endgroup\@href {#1}{\urlprefix }}%
\providecommand \urlprefix  [0]{URL }%
\providecommand \Eprint [0]{\href }%
\providecommand \doibase [0]{http://dx.doi.org/}%
\providecommand \selectlanguage [0]{\@gobble}%
\providecommand \bibinfo  [0]{\@secondoftwo}%
\providecommand \bibfield  [0]{\@secondoftwo}%
\providecommand \translation [1]{[#1]}%
\providecommand \BibitemOpen [0]{}%
\providecommand \bibitemStop [0]{}%
\providecommand \bibitemNoStop [0]{.\EOS\space}%
\providecommand \EOS [0]{\spacefactor3000\relax}%
\providecommand \BibitemShut  [1]{\csname bibitem#1\endcsname}%
\let\auto@bib@innerbib\@empty
\bibitem [{{Alpaslan} {\textit{et~al}}\mbox{.}(2015)\citenamefont {{Alpaslan}},
  \citenamefont {{Driver}}, \citenamefont {{Robotham}}, \citenamefont
  {{Obreschkow}}, \citenamefont {{Andrae}}, \citenamefont {{Cluver}},
  \citenamefont {{Kelvin}}, \citenamefont {{Lange}}, \citenamefont {{Owers}},
  \citenamefont {{Taylor}}, \citenamefont {{Andrews}}, \citenamefont
  {{Bamford}}, \citenamefont {{Bland-Hawthorn}}, \citenamefont {{Brough}},
  \citenamefont {{Brown}}, \citenamefont {{Colless}}, \citenamefont {{Davies}},
  \citenamefont {{Eardley}}, \citenamefont {{Grootes}}, \citenamefont
  {{Hopkins}}, \citenamefont {{Kennedy}}, \citenamefont {{Liske}},
  \citenamefont {{Lara-L{\'o}pez}}, \citenamefont {{L{\'o}pez-S{\'a}nchez}},
  \citenamefont {{Loveday}}, \citenamefont {{Madore}}, \citenamefont
  {{Mahajan}}, \citenamefont {{Meyer}}, \citenamefont {{Moffett}},
  \citenamefont {{Norberg}}, \citenamefont {{Penny}}, \citenamefont
  {{Pimbblet}}, \citenamefont {{Popescu}}, \citenamefont {{Seibert}},\ \&\
  \citenamefont {{Tuffs}}}]{Alpaslan2015}%
{(\PineGreen{{Alpaslan} {\textit{et~al}}\mbox{.}}, \PineGreen{2015})}
  \BibitemOpen
  \bibfield  {author} {\bibinfo {author} {\bibfnamefont {M.}~\bibnamefont
  {{Alpaslan}}}, \bibinfo {author} {\bibfnamefont {S.}~\bibnamefont
  {{Driver}}}, \bibinfo {author} {\bibfnamefont {A.~S.~G.}\ \bibnamefont
  {{Robotham}}}, \bibinfo {author} {\bibfnamefont {D.}~\bibnamefont
  {{Obreschkow}}}, \bibinfo {author} {\bibfnamefont {E.}~\bibnamefont
  {{Andrae}}}, \bibinfo {author} {\bibfnamefont {M.}~\bibnamefont {{Cluver}}},
  \bibinfo {author} {\bibfnamefont {L.~S.}\ \bibnamefont {{Kelvin}}}, \bibinfo
  {author} {\bibfnamefont {R.}~\bibnamefont {{Lange}}}, \bibinfo {author}
  {\bibfnamefont {M.}~\bibnamefont {{Owers}}}, \bibinfo {author} {\bibfnamefont
  {E.~N.}\ \bibnamefont {{Taylor}}}, \bibinfo {author} {\bibfnamefont {S.~K.}\
  \bibnamefont {{Andrews}}}, \bibinfo {author} {\bibfnamefont {S.}~\bibnamefont
  {{Bamford}}}, \bibinfo {author} {\bibfnamefont {J.}~\bibnamefont
  {{Bland-Hawthorn}}}, \bibinfo {author} {\bibfnamefont {S.}~\bibnamefont
  {{Brough}}}, \bibinfo {author} {\bibfnamefont {M.~J.~I.}\ \bibnamefont
  {{Brown}}}, \bibinfo {author} {\bibfnamefont {M.}~\bibnamefont {{Colless}}},
  \bibinfo {author} {\bibfnamefont {L.~J.~M.}\ \bibnamefont {{Davies}}},
  \bibinfo {author} {\bibfnamefont {E.}~\bibnamefont {{Eardley}}}, \bibinfo
  {author} {\bibfnamefont {M.~W.}\ \bibnamefont {{Grootes}}}, \bibinfo {author}
  {\bibfnamefont {A.~M.}\ \bibnamefont {{Hopkins}}}, \bibinfo {author}
  {\bibfnamefont {R.}~\bibnamefont {{Kennedy}}}, \bibinfo {author}
  {\bibfnamefont {J.}~\bibnamefont {{Liske}}}, \bibinfo {author} {\bibfnamefont
  {M.~A.}\ \bibnamefont {{Lara-L{\'o}pez}}}, \bibinfo {author} {\bibfnamefont
  {{\'A}.~R.}\ \bibnamefont {{L{\'o}pez-S{\'a}nchez}}}, \bibinfo {author}
  {\bibfnamefont {J.}~\bibnamefont {{Loveday}}}, \bibinfo {author}
  {\bibfnamefont {B.~F.}\ \bibnamefont {{Madore}}}, \bibinfo {author}
  {\bibfnamefont {S.}~\bibnamefont {{Mahajan}}}, \bibinfo {author}
  {\bibfnamefont {M.}~\bibnamefont {{Meyer}}}, \bibinfo {author} {\bibfnamefont
  {A.}~\bibnamefont {{Moffett}}}, \bibinfo {author} {\bibfnamefont
  {P.}~\bibnamefont {{Norberg}}}, \bibinfo {author} {\bibfnamefont
  {S.}~\bibnamefont {{Penny}}}, \bibinfo {author} {\bibfnamefont {K.~A.}\
  \bibnamefont {{Pimbblet}}}, \bibinfo {author} {\bibfnamefont {C.~C.}\
  \bibnamefont {{Popescu}}}, \bibinfo {author} {\bibfnamefont {M.}~\bibnamefont
  {{Seibert}}}, \bibinfo {author} {\bibfnamefont {R.}~\bibnamefont {{Tuffs}}},\
  }\emph {{Galaxy And Mass Assembly (GAMA): trends in galaxy colours,
  morphology, and stellar populations with large-scale structure, group, and
  pair environments}},\ \href {\doibase 10.1093/mnras/stv1176} {\bibfield
  {journal} {\bibinfo  {journal} {\mnras}\ }\textbf {\bibinfo {volume} {451}},\
  \bibinfo {pages} {3249} (\bibinfo {year} {2015})},\ \Eprint
  {http://arxiv.org/abs/1505.05518} {arXiv:1505.05518
  [astro-ph.GA]}\BibitemShut {NoStop}%
\bibitem [{{Bassett}(2005)\citenamefont {{Bassett}}}]{Bassett2005}%
{(\PineGreen{{Bassett}}, \PineGreen{2005})}  \BibitemOpen
  \bibfield  {author} {\bibinfo {author} {\bibfnamefont {B.~A.}\ \bibnamefont
  {{Bassett}}},\ }\emph {{Optimizing cosmological surveys in a crowded
  market}},\ \href {\doibase 10.1103/PhysRevD.71.083517} {\bibfield  {journal}
  {\bibinfo  {journal} {\prd}\ }\textbf {\bibinfo {volume} {71}},\ \bibinfo
  {eid} {083517} (\bibinfo {year} {2005})},\ \Eprint
  {http://arxiv.org/abs/astro-ph/0407201} {astro-ph/0407201}\BibitemShut
  {NoStop}%
\bibitem [{{Bernardeau} {\textit{et~al}}\mbox{.}(2002)\citenamefont
  {{Bernardeau}}, \citenamefont {{Colombi}}, \citenamefont {{Gazta{\~n}aga}},\
  \&\ \citenamefont {{Scoccimarro}}}]{Bernardeau2002}%
{(\PineGreen{{Bernardeau} {\textit{et~al}}\mbox{.}}, \PineGreen{2002})}
  \BibitemOpen
  \bibfield  {author} {\bibinfo {author} {\bibfnamefont {F.}~\bibnamefont
  {{Bernardeau}}}, \bibinfo {author} {\bibfnamefont {S.}~\bibnamefont
  {{Colombi}}}, \bibinfo {author} {\bibfnamefont {E.}~\bibnamefont
  {{Gazta{\~n}aga}}}, \bibinfo {author} {\bibfnamefont {R.}~\bibnamefont
  {{Scoccimarro}}},\ }\emph {{Large-scale structure of the Universe and
  cosmological perturbation theory}},\ \href {\doibase
  10.1016/S0370-1573(02)00135-7} {\bibfield  {journal} {\bibinfo  {journal}
  {\physrep}\ }\textbf {\bibinfo {volume} {367}},\ \bibinfo {pages} {1}
  (\bibinfo {year} {2002})},\ \Eprint {http://arxiv.org/abs/astro-ph/0112551}
  {astro-ph/0112551}\BibitemShut {NoStop}%
\bibitem [{{Blanton} \& {Roweis}(2007)\citenamefont {{Blanton}}\ \&\
  \citenamefont {{Roweis}}}]{Blanton2007}%
{(\PineGreen{{Blanton} \& {Roweis}}, \PineGreen{2007})}  \BibitemOpen
  \bibfield  {author} {\bibinfo {author} {\bibfnamefont {M.~R.}\ \bibnamefont
  {{Blanton}}}, \bibinfo {author} {\bibfnamefont {S.}~\bibnamefont
  {{Roweis}}},\ }\emph {{K-Corrections and Filter Transformations in the
  Ultraviolet, Optical, and Near-Infrared}},\ \href {\doibase 10.1086/510127}
  {\bibfield  {journal} {\bibinfo  {journal} {\aj}\ }\textbf {\bibinfo {volume}
  {133}},\ \bibinfo {pages} {734} (\bibinfo {year} {2007})},\ \Eprint
  {http://arxiv.org/abs/astro-ph/0606170} {astro-ph/0606170}\BibitemShut
  {NoStop}%
\bibitem [{{Blanton} {\textit{et~al}}\mbox{.}(2005)\citenamefont {{Blanton}},
  \citenamefont {{Eisenstein}}, \citenamefont {{Hogg}}, \citenamefont
  {{Schlegel}},\ \&\ \citenamefont {{Brinkmann}}}]{Blanton2005a}%
{(\PineGreen{{Blanton} {\textit{et~al}}\mbox{.}}, \PineGreen{2005})}
  \BibitemOpen
  \bibfield  {author} {\bibinfo {author} {\bibfnamefont {M.~R.}\ \bibnamefont
  {{Blanton}}}, \bibinfo {author} {\bibfnamefont {D.}~\bibnamefont
  {{Eisenstein}}}, \bibinfo {author} {\bibfnamefont {D.~W.}\ \bibnamefont
  {{Hogg}}}, \bibinfo {author} {\bibfnamefont {D.~J.}\ \bibnamefont
  {{Schlegel}}}, \bibinfo {author} {\bibfnamefont {J.}~\bibnamefont
  {{Brinkmann}}},\ }\emph {{Relationship between Environment and the Broadband
  Optical Properties of Galaxies in the Sloan Digital Sky Survey}},\ \href
  {\doibase 10.1086/422897} {\bibfield  {journal} {\bibinfo  {journal} {\apj}\
  }\textbf {\bibinfo {volume} {629}},\ \bibinfo {pages} {143} (\bibinfo {year}
  {2005})},\ \Eprint {http://arxiv.org/abs/astro-ph/0310453}
  {astro-ph/0310453}\BibitemShut {NoStop}%
\bibitem [{{Blanton} {\textit{et~al}}\mbox{.}(2003{\natexlab{a}})\citenamefont
  {{Blanton}}, \citenamefont {{Lin}}, \citenamefont {{Lupton}}, \citenamefont
  {{Maley}}, \citenamefont {{Young}}, \citenamefont {{Zehavi}},\ \&\
  \citenamefont {{Loveday}}}]{Blanton2003a}%
{(\PineGreen{{Blanton} {\textit{et~al}}\mbox{.}},
  \PineGreen{2003{\natexlab{a}}})}  \BibitemOpen
  \bibfield  {author} {\bibinfo {author} {\bibfnamefont {M.~R.}\ \bibnamefont
  {{Blanton}}}, \bibinfo {author} {\bibfnamefont {H.}~\bibnamefont {{Lin}}},
  \bibinfo {author} {\bibfnamefont {R.~H.}\ \bibnamefont {{Lupton}}}, \bibinfo
  {author} {\bibfnamefont {F.~M.}\ \bibnamefont {{Maley}}}, \bibinfo {author}
  {\bibfnamefont {N.}~\bibnamefont {{Young}}}, \bibinfo {author} {\bibfnamefont
  {I.}~\bibnamefont {{Zehavi}}}, \bibinfo {author} {\bibfnamefont
  {J.}~\bibnamefont {{Loveday}}},\ }\emph {{An Efficient Targeting Strategy for
  Multiobject Spectrograph Surveys: the Sloan Digital Sky Survey ``Tiling''
  Algorithm}},\ \href {\doibase 10.1086/344761} {\bibfield  {journal} {\bibinfo
   {journal} {\aj}\ }\textbf {\bibinfo {volume} {125}},\ \bibinfo {pages}
  {2276} (\bibinfo {year} {2003}{\natexlab{a}})},\ \Eprint
  {http://arxiv.org/abs/astro-ph/0105535} {astro-ph/0105535}\BibitemShut
  {NoStop}%
\bibitem [{{Blanton} {\textit{et~al}}\mbox{.}(2003{\natexlab{b}})\citenamefont
  {{Blanton}}, \citenamefont {{Brinkmann}}, \citenamefont {{Csabai}},
  \citenamefont {{Doi}}, \citenamefont {{Eisenstein}}, \citenamefont
  {{Fukugita}}, \citenamefont {{Gunn}}, \citenamefont {{Hogg}},\ \&\
  \citenamefont {{Schlegel}}}]{Blanton2003}%
{(\PineGreen{{Blanton} {\textit{et~al}}\mbox{.}},
  \PineGreen{2003{\natexlab{b}}})}  \BibitemOpen
  \bibfield  {author} {\bibinfo {author} {\bibfnamefont {M.~R.}\ \bibnamefont
  {{Blanton}}}, \bibinfo {author} {\bibfnamefont {J.}~\bibnamefont
  {{Brinkmann}}}, \bibinfo {author} {\bibfnamefont {I.}~\bibnamefont
  {{Csabai}}}, \bibinfo {author} {\bibfnamefont {M.}~\bibnamefont {{Doi}}},
  \bibinfo {author} {\bibfnamefont {D.}~\bibnamefont {{Eisenstein}}}, \bibinfo
  {author} {\bibfnamefont {M.}~\bibnamefont {{Fukugita}}}, \bibinfo {author}
  {\bibfnamefont {J.~E.}\ \bibnamefont {{Gunn}}}, \bibinfo {author}
  {\bibfnamefont {D.~W.}\ \bibnamefont {{Hogg}}}, \bibinfo {author}
  {\bibfnamefont {D.~J.}\ \bibnamefont {{Schlegel}}},\ }\emph {{Estimating
  Fixed-Frame Galaxy Magnitudes in the Sloan Digital Sky Survey}},\ \href
  {\doibase 10.1086/342935} {\bibfield  {journal} {\bibinfo  {journal} {\aj}\
  }\textbf {\bibinfo {volume} {125}},\ \bibinfo {pages} {2348} (\bibinfo {year}
  {2003}{\natexlab{b}})},\ \Eprint {http://arxiv.org/abs/astro-ph/0205243}
  {astro-ph/0205243}\BibitemShut {NoStop}%
\bibitem [{{Blas}, {Lesgourgues} \& {Tram}(2011)\citenamefont {{Blas}},
  \citenamefont {{Lesgourgues}},\ \&\ \citenamefont {{Tram}}}]{Blas2011}%
{(\PineGreen{{Blas}, {Lesgourgues} \& {Tram}}, \PineGreen{2011})}  \BibitemOpen
  \bibfield  {author} {\bibinfo {author} {\bibfnamefont {D.}~\bibnamefont
  {{Blas}}}, \bibinfo {author} {\bibfnamefont {J.}~\bibnamefont
  {{Lesgourgues}}}, \bibinfo {author} {\bibfnamefont {T.}~\bibnamefont
  {{Tram}}},\ }\emph {{The Cosmic Linear Anisotropy Solving System (CLASS).
  Part II: Approximation schemes}},\ \href {\doibase
  10.1088/1475-7516/2011/07/034} {\bibfield  {journal} {\bibinfo  {journal}
  {\jcap}\ }\textbf {\bibinfo {volume} {7}},\ \bibinfo {eid} {034} (\bibinfo
  {year} {2011})},\ \Eprint {http://arxiv.org/abs/1104.2933} {arXiv:1104.2933
  [astro-ph.CO]}\BibitemShut {NoStop}%
\bibitem [{{Bond}, {Kofman} \& {Pogosyan}(1996)\citenamefont {{Bond}},
  \citenamefont {{Kofman}},\ \&\ \citenamefont {{Pogosyan}}}]{Bond1996}%
{(\PineGreen{{Bond}, {Kofman} \& {Pogosyan}}, \PineGreen{1996})}  \BibitemOpen
  \bibfield  {author} {\bibinfo {author} {\bibfnamefont {J.~R.}\ \bibnamefont
  {{Bond}}}, \bibinfo {author} {\bibfnamefont {L.}~\bibnamefont {{Kofman}}},
  \bibinfo {author} {\bibfnamefont {D.}~\bibnamefont {{Pogosyan}}},\ }\emph
  {{How filaments of galaxies are woven into the cosmic web}},\ \href {\doibase
  10.1038/380603a0} {\bibfield  {journal} {\bibinfo  {journal} {\nat}\ }\textbf
  {\bibinfo {volume} {380}},\ \bibinfo {pages} {603} (\bibinfo {year}
  {1996})},\ \Eprint {http://arxiv.org/abs/astro-ph/9512141}
  {astro-ph/9512141}\BibitemShut {NoStop}%
\bibitem [{{Cavagnaro} {\textit{et~al}}\mbox{.}(2010)\citenamefont
  {{Cavagnaro}}, \citenamefont {{Myung}}, \citenamefont {{Pitt}},\ \&\
  \citenamefont {{Kujala}}}]{Cavagnaro2010}%
{(\PineGreen{{Cavagnaro} {\textit{et~al}}\mbox{.}}, \PineGreen{2010})}
  \BibitemOpen
  \bibfield  {author} {\bibinfo {author} {\bibfnamefont {D.~R.}\ \bibnamefont
  {{Cavagnaro}}}, \bibinfo {author} {\bibfnamefont {J.~I.}\ \bibnamefont
  {{Myung}}}, \bibinfo {author} {\bibfnamefont {M.~A.}\ \bibnamefont {{Pitt}}},
  \bibinfo {author} {\bibfnamefont {J.~V.}\ \bibnamefont {{Kujala}}},\ }\emph
  {{Adaptive Design Optimization: A Mutual Information-Based Approach to Model
  Discrimination in Cognitive Science}},\ \href {\doibase
  10.1162/neco.2009.02-09-959} {\bibfield  {journal} {\bibinfo  {journal}
  {Neural Computation}\ }\textbf {\bibinfo {volume} {22}},\ \bibinfo {pages}
  {887} (\bibinfo {year} {2010})}\BibitemShut {NoStop}%
\bibitem [{{Chaloner} \& {Verdinelli}(1995)\citenamefont {{Chaloner}}\ \&\
  \citenamefont {{Verdinelli}}}]{Chaloner1995}%
{(\PineGreen{{Chaloner} \& {Verdinelli}}, \PineGreen{1995})}  \BibitemOpen
  \bibfield  {author} {\bibinfo {author} {\bibfnamefont {K.}~\bibnamefont
  {{Chaloner}}}, \bibinfo {author} {\bibfnamefont {I.}~\bibnamefont
  {{Verdinelli}}},\ }\emph {{Bayesian Experimental Design: A Review}},\ \href
  {\doibase 10.1214/ss/1177009939} {\bibfield  {journal} {\bibinfo  {journal}
  {Statistical Science}\ }\textbf {\bibinfo {volume} {10}},\ \bibinfo {pages}
  {273} (\bibinfo {year} {1995})}\BibitemShut {NoStop}%
\bibitem [{{de Haan} {\textit{et~al}}\mbox{.}(2016)\citenamefont {{de Haan}},
  \citenamefont {{Benson}}, \citenamefont {{Bleem}}, \citenamefont {{Allen}},
  \citenamefont {{Applegate}}, \citenamefont {{Ashby}}, \citenamefont
  {{Bautz}}, \citenamefont {{Bayliss}}, \citenamefont {{Bocquet}},
  \citenamefont {{Brodwin}}, \citenamefont {{Carlstrom}}, \citenamefont
  {{Chang}}, \citenamefont {{Chiu}}, \citenamefont {{Cho}}, \citenamefont
  {{Clocchiatti}}, \citenamefont {{Crawford}}, \citenamefont {{Crites}},
  \citenamefont {{Desai}}, \citenamefont {{Dietrich}}, \citenamefont {{Dobbs}},
  \citenamefont {{Doucouliagos}}, \citenamefont {{Foley}}, \citenamefont
  {{Forman}}, \citenamefont {{Garmire}}, \citenamefont {{George}},
  \citenamefont {{Gladders}}, \citenamefont {{Gonzalez}}, \citenamefont
  {{Gupta}}, \citenamefont {{Halverson}}, \citenamefont {{Hlavacek-Larrondo}},
  \citenamefont {{Hoekstra}}, \citenamefont {{Holder}}, \citenamefont
  {{Holzapfel}}, \citenamefont {{Hou}}, \citenamefont {{Hrubes}}, \citenamefont
  {{Huang}}, \citenamefont {{Jones}}, \citenamefont {{Keisler}}, \citenamefont
  {{Knox}}, \citenamefont {{Lee}}, \citenamefont {{Leitch}}, \citenamefont
  {{von der Linden}}, \citenamefont {{Luong-Van}}, \citenamefont {{Mantz}},
  \citenamefont {{Marrone}}, \citenamefont {{McDonald}}, \citenamefont
  {{McMahon}}, \citenamefont {{Meyer}}, \citenamefont {{Mocanu}}, \citenamefont
  {{Mohr}}, \citenamefont {{Murray}}, \citenamefont {{Padin}}, \citenamefont
  {{Pryke}}, \citenamefont {{Rapetti}}, \citenamefont {{Reichardt}},
  \citenamefont {{Rest}}, \citenamefont {{Ruel}}, \citenamefont {{Ruhl}},
  \citenamefont {{Saliwanchik}}, \citenamefont {{Saro}}, \citenamefont
  {{Sayre}}, \citenamefont {{Schaffer}}, \citenamefont {{Schrabback}},
  \citenamefont {{Shirokoff}}, \citenamefont {{Song}}, \citenamefont
  {{Spieler}}, \citenamefont {{Stalder}}, \citenamefont {{Stanford}},
  \citenamefont {{Staniszewski}}, \citenamefont {{Stark}}, \citenamefont
  {{Story}}, \citenamefont {{Stubbs}}, \citenamefont {{Vanderlinde}},
  \citenamefont {{Vieira}}, \citenamefont {{Vikhlinin}}, \citenamefont
  {{Williamson}},\ \&\ \citenamefont {{Zenteno}}}]{deHaan2016}%
{(\PineGreen{{de Haan} {\textit{et~al}}\mbox{.}}, \PineGreen{2016})}
  \BibitemOpen
  \bibfield  {author} {\bibinfo {author} {\bibfnamefont {T.}~\bibnamefont {{de
  Haan}}}, \bibinfo {author} {\bibfnamefont {B.~A.}\ \bibnamefont {{Benson}}},
  \bibinfo {author} {\bibfnamefont {L.~E.}\ \bibnamefont {{Bleem}}}, \bibinfo
  {author} {\bibfnamefont {S.~W.}\ \bibnamefont {{Allen}}}, \bibinfo {author}
  {\bibfnamefont {D.~E.}\ \bibnamefont {{Applegate}}}, \bibinfo {author}
  {\bibfnamefont {M.~L.~N.}\ \bibnamefont {{Ashby}}}, \bibinfo {author}
  {\bibfnamefont {M.}~\bibnamefont {{Bautz}}}, \bibinfo {author} {\bibfnamefont
  {M.}~\bibnamefont {{Bayliss}}}, \bibinfo {author} {\bibfnamefont
  {S.}~\bibnamefont {{Bocquet}}}, \bibinfo {author} {\bibfnamefont
  {M.}~\bibnamefont {{Brodwin}}}, \bibinfo {author} {\bibfnamefont {J.~E.}\
  \bibnamefont {{Carlstrom}}}, \bibinfo {author} {\bibfnamefont {C.~L.}\
  \bibnamefont {{Chang}}}, \bibinfo {author} {\bibfnamefont {I.}~\bibnamefont
  {{Chiu}}}, \bibinfo {author} {\bibfnamefont {H.}~\bibnamefont {{Cho}}},
  \bibinfo {author} {\bibfnamefont {A.}~\bibnamefont {{Clocchiatti}}}, \bibinfo
  {author} {\bibfnamefont {T.~M.}\ \bibnamefont {{Crawford}}}, \bibinfo
  {author} {\bibfnamefont {A.~T.}\ \bibnamefont {{Crites}}}, \bibinfo {author}
  {\bibfnamefont {S.}~\bibnamefont {{Desai}}}, \bibinfo {author} {\bibfnamefont
  {J.~P.}\ \bibnamefont {{Dietrich}}}, \bibinfo {author} {\bibfnamefont
  {M.~A.}\ \bibnamefont {{Dobbs}}}, \bibinfo {author} {\bibfnamefont {A.~N.}\
  \bibnamefont {{Doucouliagos}}}, \bibinfo {author} {\bibfnamefont {R.~J.}\
  \bibnamefont {{Foley}}}, \bibinfo {author} {\bibfnamefont {W.~R.}\
  \bibnamefont {{Forman}}}, \bibinfo {author} {\bibfnamefont {G.~P.}\
  \bibnamefont {{Garmire}}}, \bibinfo {author} {\bibfnamefont {E.~M.}\
  \bibnamefont {{George}}}, \bibinfo {author} {\bibfnamefont {M.~D.}\
  \bibnamefont {{Gladders}}}, \bibinfo {author} {\bibfnamefont {A.~H.}\
  \bibnamefont {{Gonzalez}}}, \bibinfo {author} {\bibfnamefont
  {N.}~\bibnamefont {{Gupta}}}, \bibinfo {author} {\bibfnamefont {N.~W.}\
  \bibnamefont {{Halverson}}}, \bibinfo {author} {\bibfnamefont
  {J.}~\bibnamefont {{Hlavacek-Larrondo}}}, \bibinfo {author} {\bibfnamefont
  {H.}~\bibnamefont {{Hoekstra}}}, \bibinfo {author} {\bibfnamefont {G.~P.}\
  \bibnamefont {{Holder}}}, \bibinfo {author} {\bibfnamefont {W.~L.}\
  \bibnamefont {{Holzapfel}}}, \bibinfo {author} {\bibfnamefont
  {Z.}~\bibnamefont {{Hou}}}, \bibinfo {author} {\bibfnamefont {J.~D.}\
  \bibnamefont {{Hrubes}}}, \bibinfo {author} {\bibfnamefont {N.}~\bibnamefont
  {{Huang}}}, \bibinfo {author} {\bibfnamefont {C.}~\bibnamefont {{Jones}}},
  \bibinfo {author} {\bibfnamefont {R.}~\bibnamefont {{Keisler}}}, \bibinfo
  {author} {\bibfnamefont {L.}~\bibnamefont {{Knox}}}, \bibinfo {author}
  {\bibfnamefont {A.~T.}\ \bibnamefont {{Lee}}}, \bibinfo {author}
  {\bibfnamefont {E.~M.}\ \bibnamefont {{Leitch}}}, \bibinfo {author}
  {\bibfnamefont {A.}~\bibnamefont {{von der Linden}}}, \bibinfo {author}
  {\bibfnamefont {D.}~\bibnamefont {{Luong-Van}}}, \bibinfo {author}
  {\bibfnamefont {A.}~\bibnamefont {{Mantz}}}, \bibinfo {author} {\bibfnamefont
  {D.~P.}\ \bibnamefont {{Marrone}}}, \bibinfo {author} {\bibfnamefont
  {M.}~\bibnamefont {{McDonald}}}, \bibinfo {author} {\bibfnamefont {J.~J.}\
  \bibnamefont {{McMahon}}}, \bibinfo {author} {\bibfnamefont {S.~S.}\
  \bibnamefont {{Meyer}}}, \bibinfo {author} {\bibfnamefont {L.~M.}\
  \bibnamefont {{Mocanu}}}, \bibinfo {author} {\bibfnamefont {J.~J.}\
  \bibnamefont {{Mohr}}}, \bibinfo {author} {\bibfnamefont {S.~S.}\
  \bibnamefont {{Murray}}}, \bibinfo {author} {\bibfnamefont {S.}~\bibnamefont
  {{Padin}}}, \bibinfo {author} {\bibfnamefont {C.}~\bibnamefont {{Pryke}}},
  \bibinfo {author} {\bibfnamefont {D.}~\bibnamefont {{Rapetti}}}, \bibinfo
  {author} {\bibfnamefont {C.~L.}\ \bibnamefont {{Reichardt}}}, \bibinfo
  {author} {\bibfnamefont {A.}~\bibnamefont {{Rest}}}, \bibinfo {author}
  {\bibfnamefont {J.}~\bibnamefont {{Ruel}}}, \bibinfo {author} {\bibfnamefont
  {J.~E.}\ \bibnamefont {{Ruhl}}}, \bibinfo {author} {\bibfnamefont {B.~R.}\
  \bibnamefont {{Saliwanchik}}}, \bibinfo {author} {\bibfnamefont
  {A.}~\bibnamefont {{Saro}}}, \bibinfo {author} {\bibfnamefont {J.~T.}\
  \bibnamefont {{Sayre}}}, \bibinfo {author} {\bibfnamefont {K.~K.}\
  \bibnamefont {{Schaffer}}}, \bibinfo {author} {\bibfnamefont
  {T.}~\bibnamefont {{Schrabback}}}, \bibinfo {author} {\bibfnamefont
  {E.}~\bibnamefont {{Shirokoff}}}, \bibinfo {author} {\bibfnamefont
  {J.}~\bibnamefont {{Song}}}, \bibinfo {author} {\bibfnamefont {H.~G.}\
  \bibnamefont {{Spieler}}}, \bibinfo {author} {\bibfnamefont {B.}~\bibnamefont
  {{Stalder}}}, \bibinfo {author} {\bibfnamefont {S.~A.}\ \bibnamefont
  {{Stanford}}}, \bibinfo {author} {\bibfnamefont {Z.}~\bibnamefont
  {{Staniszewski}}}, \bibinfo {author} {\bibfnamefont {A.~A.}\ \bibnamefont
  {{Stark}}}, \bibinfo {author} {\bibfnamefont {K.~T.}\ \bibnamefont
  {{Story}}}, \bibinfo {author} {\bibfnamefont {C.~W.}\ \bibnamefont
  {{Stubbs}}}, \bibinfo {author} {\bibfnamefont {K.}~\bibnamefont
  {{Vanderlinde}}}, \bibinfo {author} {\bibfnamefont {J.~D.}\ \bibnamefont
  {{Vieira}}}, \bibinfo {author} {\bibfnamefont {A.}~\bibnamefont
  {{Vikhlinin}}}, \bibinfo {author} {\bibfnamefont {R.}~\bibnamefont
  {{Williamson}}}, \bibinfo {author} {\bibfnamefont {A.}~\bibnamefont
  {{Zenteno}}},\ }\emph {{Cosmological Constraints from Galaxy Clusters in the
  2500 square-degree SPT-SZ Survey}},\ \href@noop {} {\bibfield  {journal}
  {\bibinfo  {journal} {ArXiv e-prints}\ } (\bibinfo {year} {2016})},\ \Eprint
  {http://arxiv.org/abs/1603.06522} {arXiv:1603.06522
  [astro-ph.CO]}\BibitemShut {NoStop}%
\bibitem [{{Duane} {\textit{et~al}}\mbox{.}(1987)\citenamefont {{Duane}},
  \citenamefont {{Kennedy}}, \citenamefont {{Pendleton}},\ \&\ \citenamefont
  {{Roweth}}}]{Duane1987}%
{(\PineGreen{{Duane} {\textit{et~al}}\mbox{.}}, \PineGreen{1987})}
  \BibitemOpen
  \bibfield  {author} {\bibinfo {author} {\bibfnamefont {S.}~\bibnamefont
  {{Duane}}}, \bibinfo {author} {\bibfnamefont {A.~D.}\ \bibnamefont
  {{Kennedy}}}, \bibinfo {author} {\bibfnamefont {B.~J.}\ \bibnamefont
  {{Pendleton}}}, \bibinfo {author} {\bibfnamefont {D.}~\bibnamefont
  {{Roweth}}},\ }\emph {{Hybrid Monte Carlo}},\ \href {\doibase
  10.1016/0370-2693(87)91197-X} {\bibfield  {journal} {\bibinfo  {journal}
  {Physics Letters B}\ }\textbf {\bibinfo {volume} {195}},\ \bibinfo {pages}
  {216} (\bibinfo {year} {1987})}\BibitemShut {NoStop}%
\bibitem [{{Endres} \& {Schindelin}(2003)\citenamefont {{Endres}}\ \&\
  \citenamefont {{Schindelin}}}]{Endres2003}%
{(\PineGreen{{Endres} \& {Schindelin}}, \PineGreen{2003})}  \BibitemOpen
  \bibfield  {author} {\bibinfo {author} {\bibfnamefont {D.~M.}\ \bibnamefont
  {{Endres}}}, \bibinfo {author} {\bibfnamefont {J.~E.}\ \bibnamefont
  {{Schindelin}}},\ }\emph {{A new metric for probability distributions}},\
  \href {\doibase 10.1109/tit.2003.813506} {\bibfield  {journal} {\bibinfo
  {journal} {IEEE Transactions on Information Theory}\ }\textbf {\bibinfo
  {volume} {49}},\ \bibinfo {pages} {1858} (\bibinfo {year}
  {2003})}\BibitemShut {NoStop}%
\bibitem [{{Erdo{\v g}du} {\textit{et~al}}\mbox{.}(2004)\citenamefont {{Erdo{\v
  g}du}}, \citenamefont {{Lahav}}, \citenamefont {{Zaroubi}}, \citenamefont
  {{Efstathiou}}, \citenamefont {{Moody}}, \citenamefont {{Peacock}},
  \citenamefont {{Colless}}, \citenamefont {{Baldry}}, \citenamefont {{Baugh}},
  \citenamefont {{Bland-Hawthorn}}, \citenamefont {{Bridges}}, \citenamefont
  {{Cannon}}, \citenamefont {{Cole}}, \citenamefont {{Collins}}, \citenamefont
  {{Couch}}, \citenamefont {{Dalton}}, \citenamefont {{De Propris}},
  \citenamefont {{Driver}}, \citenamefont {{Ellis}}, \citenamefont {{Frenk}},
  \citenamefont {{Glazebrook}}, \citenamefont {{Jackson}}, \citenamefont
  {{Lewis}}, \citenamefont {{Lumsden}}, \citenamefont {{Maddox}}, \citenamefont
  {{Madgwick}}, \citenamefont {{Norberg}}, \citenamefont {{Peterson}},
  \citenamefont {{Sutherland}},\ \&\ \citenamefont {{Taylor}}}]{Erdovgdu2004}%
{(\PineGreen{{Erdo{\v g}du} {\textit{et~al}}\mbox{.}}, \PineGreen{2004})}
  \BibitemOpen
  \bibfield  {author} {\bibinfo {author} {\bibfnamefont {P.}~\bibnamefont
  {{Erdo{\v g}du}}}, \bibinfo {author} {\bibfnamefont {O.}~\bibnamefont
  {{Lahav}}}, \bibinfo {author} {\bibfnamefont {S.}~\bibnamefont {{Zaroubi}}},
  \bibinfo {author} {\bibfnamefont {G.}~\bibnamefont {{Efstathiou}}}, \bibinfo
  {author} {\bibfnamefont {S.}~\bibnamefont {{Moody}}}, \bibinfo {author}
  {\bibfnamefont {J.~A.}\ \bibnamefont {{Peacock}}}, \bibinfo {author}
  {\bibfnamefont {M.}~\bibnamefont {{Colless}}}, \bibinfo {author}
  {\bibfnamefont {I.~K.}\ \bibnamefont {{Baldry}}}, \bibinfo {author}
  {\bibfnamefont {C.~M.}\ \bibnamefont {{Baugh}}}, \bibinfo {author}
  {\bibfnamefont {J.}~\bibnamefont {{Bland-Hawthorn}}}, \bibinfo {author}
  {\bibfnamefont {T.}~\bibnamefont {{Bridges}}}, \bibinfo {author}
  {\bibfnamefont {R.}~\bibnamefont {{Cannon}}}, \bibinfo {author}
  {\bibfnamefont {S.}~\bibnamefont {{Cole}}}, \bibinfo {author} {\bibfnamefont
  {C.}~\bibnamefont {{Collins}}}, \bibinfo {author} {\bibfnamefont
  {W.}~\bibnamefont {{Couch}}}, \bibinfo {author} {\bibfnamefont
  {G.}~\bibnamefont {{Dalton}}}, \bibinfo {author} {\bibfnamefont
  {R.}~\bibnamefont {{De Propris}}}, \bibinfo {author} {\bibfnamefont {S.~P.}\
  \bibnamefont {{Driver}}}, \bibinfo {author} {\bibfnamefont {R.~S.}\
  \bibnamefont {{Ellis}}}, \bibinfo {author} {\bibfnamefont {C.~S.}\
  \bibnamefont {{Frenk}}}, \bibinfo {author} {\bibfnamefont {K.}~\bibnamefont
  {{Glazebrook}}}, \bibinfo {author} {\bibfnamefont {C.}~\bibnamefont
  {{Jackson}}}, \bibinfo {author} {\bibfnamefont {I.}~\bibnamefont {{Lewis}}},
  \bibinfo {author} {\bibfnamefont {S.}~\bibnamefont {{Lumsden}}}, \bibinfo
  {author} {\bibfnamefont {S.}~\bibnamefont {{Maddox}}}, \bibinfo {author}
  {\bibfnamefont {D.}~\bibnamefont {{Madgwick}}}, \bibinfo {author}
  {\bibfnamefont {P.}~\bibnamefont {{Norberg}}}, \bibinfo {author}
  {\bibfnamefont {B.~A.}\ \bibnamefont {{Peterson}}}, \bibinfo {author}
  {\bibfnamefont {W.}~\bibnamefont {{Sutherland}}}, \bibinfo {author}
  {\bibfnamefont {K.}~\bibnamefont {{Taylor}}},\ }\emph {{The 2dF Galaxy
  Redshift Survey: Wiener reconstruction of the cosmic web}},\ \href {\doibase
  10.1111/j.1365-2966.2004.07984.x} {\bibfield  {journal} {\bibinfo  {journal}
  {\mnras}\ }\textbf {\bibinfo {volume} {352}},\ \bibinfo {pages} {939}
  (\bibinfo {year} {2004})},\ \Eprint {http://arxiv.org/abs/astro-ph/0312546}
  {astro-ph/0312546}\BibitemShut {NoStop}%
\bibitem [{{Falck}, {Neyrinck} \& {Szalay}(2012)\citenamefont {{Falck}},
  \citenamefont {{Neyrinck}},\ \&\ \citenamefont {{Szalay}}}]{Falck2012}%
{(\PineGreen{{Falck}, {Neyrinck} \& {Szalay}}, \PineGreen{2012})}  \BibitemOpen
  \bibfield  {author} {\bibinfo {author} {\bibfnamefont {B.~L.}\ \bibnamefont
  {{Falck}}}, \bibinfo {author} {\bibfnamefont {M.~C.}\ \bibnamefont
  {{Neyrinck}}}, \bibinfo {author} {\bibfnamefont {A.~S.}\ \bibnamefont
  {{Szalay}}},\ }\emph {{ORIGAMI: Delineating Halos Using Phase-space Folds}},\
  \href {\doibase 10.1088/0004-637X/754/2/126} {\bibfield  {journal} {\bibinfo
  {journal} {\apj}\ }\textbf {\bibinfo {volume} {754}},\ \bibinfo {eid} {126}
  (\bibinfo {year} {2012})},\ \Eprint {http://arxiv.org/abs/1201.2353}
  {arXiv:1201.2353 [astro-ph.CO]}\BibitemShut {NoStop}%
\bibitem [{{Falck}, {Koyama} \& {Zhao}(2015)\citenamefont {{Falck}},
  \citenamefont {{Koyama}},\ \&\ \citenamefont {{Zhao}}}]{Falck2015a}%
{(\PineGreen{{Falck}, {Koyama} \& {Zhao}}, \PineGreen{2015})}  \BibitemOpen
  \bibfield  {author} {\bibinfo {author} {\bibfnamefont {B.}~\bibnamefont
  {{Falck}}}, \bibinfo {author} {\bibfnamefont {K.}~\bibnamefont {{Koyama}}},
  \bibinfo {author} {\bibfnamefont {G.-B.}\ \bibnamefont {{Zhao}}},\ }\emph
  {{Cosmic web and environmental dependence of screening: Vainshtein vs.
  chameleon}},\ \href {\doibase 10.1088/1475-7516/2015/07/049} {\bibfield
  {journal} {\bibinfo  {journal} {\jcap}\ }\textbf {\bibinfo {volume} {7}},\
  \bibinfo {eid} {049} (\bibinfo {year} {2015})},\ \Eprint
  {http://arxiv.org/abs/1503.06673} {arXiv:1503.06673
  [astro-ph.CO]}\BibitemShut {NoStop}%
\bibitem [{{Fisher}, {Faltenbacher} \& {Johnson}(2016)\citenamefont {{Fisher}},
  \citenamefont {{Faltenbacher}},\ \&\ \citenamefont
  {{Johnson}}}]{FisherFaltenbacherJohnson2016}%
{(\PineGreen{{Fisher}, {Faltenbacher} \& {Johnson}}, \PineGreen{2016})}
  \BibitemOpen
  \bibfield  {author} {\bibinfo {author} {\bibfnamefont {J.~D.}\ \bibnamefont
  {{Fisher}}}, \bibinfo {author} {\bibfnamefont {A.}~\bibnamefont
  {{Faltenbacher}}}, \bibinfo {author} {\bibfnamefont {M.~S.~T.}\ \bibnamefont
  {{Johnson}}},\ }\emph {{Lagrangian methods of cosmic web classification}},\
  \href {\doibase 10.1093/mnras/stw370} {\bibfield  {journal} {\bibinfo
  {journal} {\mnras}\ }\textbf {\bibinfo {volume} {458}},\ \bibinfo {pages}
  {1517} (\bibinfo {year} {2016})},\ \Eprint {http://arxiv.org/abs/1511.05971}
  {arXiv:1511.05971 [astro-ph.CO]}\BibitemShut {NoStop}%
\bibitem [{{Forero-Romero} {\textit{et~al}}\mbox{.}(2009)\citenamefont
  {{Forero-Romero}}, \citenamefont {{Hoffman}}, \citenamefont
  {{Gottl{\"o}ber}}, \citenamefont {{Klypin}},\ \&\ \citenamefont
  {{Yepes}}}]{Forero-Romero2009}%
{(\PineGreen{{Forero-Romero} {\textit{et~al}}\mbox{.}}, \PineGreen{2009})}
  \BibitemOpen
  \bibfield  {author} {\bibinfo {author} {\bibfnamefont {J.~E.}\ \bibnamefont
  {{Forero-Romero}}}, \bibinfo {author} {\bibfnamefont {Y.}~\bibnamefont
  {{Hoffman}}}, \bibinfo {author} {\bibfnamefont {S.}~\bibnamefont
  {{Gottl{\"o}ber}}}, \bibinfo {author} {\bibfnamefont {A.}~\bibnamefont
  {{Klypin}}}, \bibinfo {author} {\bibfnamefont {G.}~\bibnamefont {{Yepes}}},\
  }\emph {{A dynamical classification of the cosmic web}},\ \href {\doibase
  10.1111/j.1365-2966.2009.14885.x} {\bibfield  {journal} {\bibinfo  {journal}
  {\mnras}\ }\textbf {\bibinfo {volume} {396}},\ \bibinfo {pages} {1815}
  (\bibinfo {year} {2009})},\ \Eprint {http://arxiv.org/abs/0809.4135}
  {arXiv:0809.4135 [astro-ph]}\BibitemShut {NoStop}%
\bibitem [{{Frank}, {Jasche} \& {En{\ss}lin}(2016)\citenamefont {{Frank}},
  \citenamefont {{Jasche}},\ \&\ \citenamefont
  {{En{\ss}lin}}}]{FrankJascheEnslin2016}%
{(\PineGreen{{Frank}, {Jasche} \& {En{\ss}lin}}, \PineGreen{2016})}
  \BibitemOpen
  \bibfield  {author} {\bibinfo {author} {\bibfnamefont {P.}~\bibnamefont
  {{Frank}}}, \bibinfo {author} {\bibfnamefont {J.}~\bibnamefont {{Jasche}}},
  \bibinfo {author} {\bibfnamefont {T.~A.}\ \bibnamefont {{En{\ss}lin}}},\
  }\emph {{SOMBI: Bayesian identification of parameter relations in
  unstructured cosmological data}},\ \href@noop {} {\bibfield  {journal}
  {\bibinfo  {journal} {ArXiv e-prints}\ } (\bibinfo {year} {2016})},\ \Eprint
  {http://arxiv.org/abs/1602.08497} {arXiv:1602.08497
  [astro-ph.CO]}\BibitemShut {NoStop}%
\bibitem [{{Hahn} {\textit{et~al}}\mbox{.}(2007)\citenamefont {{Hahn}},
  \citenamefont {{Porciani}}, \citenamefont {{Carollo}},\ \&\ \citenamefont
  {{Dekel}}}]{Hahn2007a}%
{(\PineGreen{{Hahn} {\textit{et~al}}\mbox{.}}, \PineGreen{2007})}  \BibitemOpen
  \bibfield  {author} {\bibinfo {author} {\bibfnamefont {O.}~\bibnamefont
  {{Hahn}}}, \bibinfo {author} {\bibfnamefont {C.}~\bibnamefont {{Porciani}}},
  \bibinfo {author} {\bibfnamefont {C.~M.}\ \bibnamefont {{Carollo}}}, \bibinfo
  {author} {\bibfnamefont {A.}~\bibnamefont {{Dekel}}},\ }\emph {{Properties of
  dark matter haloes in clusters, filaments, sheets and voids}},\ \href
  {\doibase 10.1111/j.1365-2966.2006.11318.x} {\bibfield  {journal} {\bibinfo
  {journal} {\mnras}\ }\textbf {\bibinfo {volume} {375}},\ \bibinfo {pages}
  {489} (\bibinfo {year} {2007})},\ \Eprint
  {http://arxiv.org/abs/astro-ph/0610280} {astro-ph/0610280}\BibitemShut
  {NoStop}%
\bibitem [{{Hamaus} {\textit{et~al}}\mbox{.}(2016)\citenamefont {{Hamaus}},
  \citenamefont {{Pisani}}, \citenamefont {{Sutter}}, \citenamefont {{Lavaux}},
  \citenamefont {{Escoffier}}, \citenamefont {{Wandelt}},\ \&\ \citenamefont
  {{Weller}}}]{Hamaus2016}%
{(\PineGreen{{Hamaus} {\textit{et~al}}\mbox{.}}, \PineGreen{2016})}
  \BibitemOpen
  \bibfield  {author} {\bibinfo {author} {\bibfnamefont {N.}~\bibnamefont
  {{Hamaus}}}, \bibinfo {author} {\bibfnamefont {A.}~\bibnamefont {{Pisani}}},
  \bibinfo {author} {\bibfnamefont {P.~M.}\ \bibnamefont {{Sutter}}}, \bibinfo
  {author} {\bibfnamefont {G.}~\bibnamefont {{Lavaux}}}, \bibinfo {author}
  {\bibfnamefont {S.}~\bibnamefont {{Escoffier}}}, \bibinfo {author}
  {\bibfnamefont {B.~D.}\ \bibnamefont {{Wandelt}}}, \bibinfo {author}
  {\bibfnamefont {J.}~\bibnamefont {{Weller}}},\ }\emph {{Constraints on
  cosmology and gravity from the dynamics of voids}},\ \href@noop {} {\bibfield
   {journal} {\bibinfo  {journal} {ArXiv e-prints}\ } (\bibinfo {year}
  {2016})},\ \Eprint {http://arxiv.org/abs/1602.01784} {arXiv:1602.01784
  [astro-ph.CO]}\BibitemShut {NoStop}%
\bibitem [{{Harvey} {\textit{et~al}}\mbox{.}(2015)\citenamefont {{Harvey}},
  \citenamefont {{Massey}}, \citenamefont {{Kitching}}, \citenamefont
  {{Taylor}},\ \&\ \citenamefont {{Tittley}}}]{Harvey2015}%
{(\PineGreen{{Harvey} {\textit{et~al}}\mbox{.}}, \PineGreen{2015})}
  \BibitemOpen
  \bibfield  {author} {\bibinfo {author} {\bibfnamefont {D.}~\bibnamefont
  {{Harvey}}}, \bibinfo {author} {\bibfnamefont {R.}~\bibnamefont {{Massey}}},
  \bibinfo {author} {\bibfnamefont {T.}~\bibnamefont {{Kitching}}}, \bibinfo
  {author} {\bibfnamefont {A.}~\bibnamefont {{Taylor}}}, \bibinfo {author}
  {\bibfnamefont {E.}~\bibnamefont {{Tittley}}},\ }\emph {{The nongravitational
  interactions of dark matter in colliding galaxy clusters}},\ \href {\doibase
  10.1126/science.1261381} {\bibfield  {journal} {\bibinfo  {journal}
  {Science}\ }\textbf {\bibinfo {volume} {347}},\ \bibinfo {pages} {1462}
  (\bibinfo {year} {2015})},\ \Eprint {http://arxiv.org/abs/1503.07675}
  {arXiv:1503.07675 [astro-ph.CO]}\BibitemShut {NoStop}%
\bibitem [{{Hoffman} {\textit{et~al}}\mbox{.}(2012)\citenamefont {{Hoffman}},
  \citenamefont {{Metuki}}, \citenamefont {{Yepes}}, \citenamefont
  {{Gottl{\"o}ber}}, \citenamefont {{Forero-Romero}}, \citenamefont
  {{Libeskind}},\ \&\ \citenamefont {{Knebe}}}]{Hoffman2012}%
{(\PineGreen{{Hoffman} {\textit{et~al}}\mbox{.}}, \PineGreen{2012})}
  \BibitemOpen
  \bibfield  {author} {\bibinfo {author} {\bibfnamefont {Y.}~\bibnamefont
  {{Hoffman}}}, \bibinfo {author} {\bibfnamefont {O.}~\bibnamefont {{Metuki}}},
  \bibinfo {author} {\bibfnamefont {G.}~\bibnamefont {{Yepes}}}, \bibinfo
  {author} {\bibfnamefont {S.}~\bibnamefont {{Gottl{\"o}ber}}}, \bibinfo
  {author} {\bibfnamefont {J.~E.}\ \bibnamefont {{Forero-Romero}}}, \bibinfo
  {author} {\bibfnamefont {N.~I.}\ \bibnamefont {{Libeskind}}}, \bibinfo
  {author} {\bibfnamefont {A.}~\bibnamefont {{Knebe}}},\ }\emph {{A kinematic
  classification of the cosmic web}},\ \href {\doibase
  10.1111/j.1365-2966.2012.21553.x} {\bibfield  {journal} {\bibinfo  {journal}
  {\mnras}\ }\textbf {\bibinfo {volume} {425}},\ \bibinfo {pages} {2049}
  (\bibinfo {year} {2012})},\ \Eprint {http://arxiv.org/abs/1201.3367}
  {arXiv:1201.3367 [astro-ph.CO]}\BibitemShut {NoStop}%
\bibitem [{{Hogg} {\textit{et~al}}\mbox{.}(2003)\citenamefont {{Hogg}},
  \citenamefont {{Blanton}}, \citenamefont {{Eisenstein}}, \citenamefont
  {{Gunn}}, \citenamefont {{Schlegel}}, \citenamefont {{Zehavi}}, \citenamefont
  {{Bahcall}}, \citenamefont {{Brinkmann}}, \citenamefont {{Csabai}},
  \citenamefont {{Schneider}}, \citenamefont {{Weinberg}},\ \&\ \citenamefont
  {{York}}}]{Hogg2003}%
{(\PineGreen{{Hogg} {\textit{et~al}}\mbox{.}}, \PineGreen{2003})}  \BibitemOpen
  \bibfield  {author} {\bibinfo {author} {\bibfnamefont {D.~W.}\ \bibnamefont
  {{Hogg}}}, \bibinfo {author} {\bibfnamefont {M.~R.}\ \bibnamefont
  {{Blanton}}}, \bibinfo {author} {\bibfnamefont {D.~J.}\ \bibnamefont
  {{Eisenstein}}}, \bibinfo {author} {\bibfnamefont {J.~E.}\ \bibnamefont
  {{Gunn}}}, \bibinfo {author} {\bibfnamefont {D.~J.}\ \bibnamefont
  {{Schlegel}}}, \bibinfo {author} {\bibfnamefont {I.}~\bibnamefont
  {{Zehavi}}}, \bibinfo {author} {\bibfnamefont {N.~A.}\ \bibnamefont
  {{Bahcall}}}, \bibinfo {author} {\bibfnamefont {J.}~\bibnamefont
  {{Brinkmann}}}, \bibinfo {author} {\bibfnamefont {I.}~\bibnamefont
  {{Csabai}}}, \bibinfo {author} {\bibfnamefont {D.~P.}\ \bibnamefont
  {{Schneider}}}, \bibinfo {author} {\bibfnamefont {D.~H.}\ \bibnamefont
  {{Weinberg}}}, \bibinfo {author} {\bibfnamefont {D.~G.}\ \bibnamefont
  {{York}}},\ }\emph {{The Overdensities of Galaxy Environments as a Function
  of Luminosity and Color}},\ \href {\doibase 10.1086/374238} {\bibfield
  {journal} {\bibinfo  {journal} {\apjl}\ }\textbf {\bibinfo {volume} {585}},\
  \bibinfo {pages} {L5} (\bibinfo {year} {2003})},\ \Eprint
  {http://arxiv.org/abs/astro-ph/0212085} {astro-ph/0212085}\BibitemShut
  {NoStop}%
\bibitem [{{Hoyle} {\textit{et~al}}\mbox{.}(2015)\citenamefont {{Hoyle}},
  \citenamefont {{Rau}}, \citenamefont {{Zitlau}}, \citenamefont {{Seitz}},\
  \&\ \citenamefont {{Weller}}}]{Hoyle2015}%
{(\PineGreen{{Hoyle} {\textit{et~al}}\mbox{.}}, \PineGreen{2015})}
  \BibitemOpen
  \bibfield  {author} {\bibinfo {author} {\bibfnamefont {B.}~\bibnamefont
  {{Hoyle}}}, \bibinfo {author} {\bibfnamefont {M.~M.}\ \bibnamefont {{Rau}}},
  \bibinfo {author} {\bibfnamefont {R.}~\bibnamefont {{Zitlau}}}, \bibinfo
  {author} {\bibfnamefont {S.}~\bibnamefont {{Seitz}}}, \bibinfo {author}
  {\bibfnamefont {J.}~\bibnamefont {{Weller}}},\ }\emph {{Feature importance
  for machine learning redshifts applied to SDSS galaxies}},\ \href {\doibase
  10.1093/mnras/stv373} {\bibfield  {journal} {\bibinfo  {journal} {\mnras}\
  }\textbf {\bibinfo {volume} {449}},\ \bibinfo {pages} {1275} (\bibinfo {year}
  {2015})},\ \Eprint {http://arxiv.org/abs/1410.4696} {arXiv:1410.4696
  [astro-ph.IM]}\BibitemShut {NoStop}%
\bibitem [{{Jasche} \& {Wandelt}(2013)\citenamefont {{Jasche}}\ \&\
  \citenamefont {{Wandelt}}}]{Jasche2013BORG}%
{(\PineGreen{{Jasche} \& {Wandelt}}, \PineGreen{2013})}  \BibitemOpen
  \bibfield  {author} {\bibinfo {author} {\bibfnamefont {J.}~\bibnamefont
  {{Jasche}}}, \bibinfo {author} {\bibfnamefont {B.~D.}\ \bibnamefont
  {{Wandelt}}},\ }\emph {{Bayesian physical reconstruction of initial
  conditions from large-scale structure surveys}},\ \href {\doibase
  10.1093/mnras/stt449} {\bibfield  {journal} {\bibinfo  {journal} {\mnras}\
  }\textbf {\bibinfo {volume} {432}},\ \bibinfo {pages} {894} (\bibinfo {year}
  {2013})},\ \Eprint {http://arxiv.org/abs/1203.3639} {arXiv:1203.3639
  [astro-ph.CO]}\BibitemShut {NoStop}%
\bibitem [{{Jasche} \& {Kitaura}(2010)\citenamefont {{Jasche}}\ \&\
  \citenamefont {{Kitaura}}}]{JascheKitaura2010}%
{(\PineGreen{{Jasche} \& {Kitaura}}, \PineGreen{2010})}  \BibitemOpen
  \bibfield  {author} {\bibinfo {author} {\bibfnamefont {J.}~\bibnamefont
  {{Jasche}}}, \bibinfo {author} {\bibfnamefont {F.~S.}\ \bibnamefont
  {{Kitaura}}},\ }\emph {{Fast Hamiltonian sampling for large-scale structure
  inference}},\ \href {\doibase 10.1111/j.1365-2966.2010.16897.x} {\bibfield
  {journal} {\bibinfo  {journal} {\mnras}\ }\textbf {\bibinfo {volume} {407}},\
  \bibinfo {pages} {29} (\bibinfo {year} {2010})},\ \Eprint
  {http://arxiv.org/abs/0911.2496} {arXiv:0911.2496 [astro-ph.CO]}\BibitemShut
  {NoStop}%
\bibitem [{{Jasche}, {Leclercq} \& {Wandelt}(2015)\citenamefont {{Jasche}},
  \citenamefont {{Leclercq}},\ \&\ \citenamefont
  {{Wandelt}}}]{Jasche2015BORGSDSS}%
{(\PineGreen{{Jasche}, {Leclercq} \& {Wandelt}}, \PineGreen{2015})}
  \BibitemOpen
  \bibfield  {author} {\bibinfo {author} {\bibfnamefont {J.}~\bibnamefont
  {{Jasche}}}, \bibinfo {author} {\bibfnamefont {F.}~\bibnamefont
  {{Leclercq}}}, \bibinfo {author} {\bibfnamefont {B.~D.}\ \bibnamefont
  {{Wandelt}}},\ }\emph {{Past and present cosmic structure in the SDSS DR7
  main sample}},\ \href {\doibase 10.1088/1475-7516/2015/01/036} {\bibfield
  {journal} {\bibinfo  {journal} {\jcap}\ }\textbf {\bibinfo {volume} {1}},\
  \bibinfo {eid} {036} (\bibinfo {year} {2015})},\ \Eprint
  {http://arxiv.org/abs/1409.6308} {arXiv:1409.6308 [astro-ph.CO]}\BibitemShut
  {NoStop}%
\bibitem [{{Jasche} {\textit{et~al}}\mbox{.}(2010{\natexlab{a}})\citenamefont
  {{Jasche}}, \citenamefont {{Kitaura}}, \citenamefont {{Wandelt}},\ \&\
  \citenamefont {{En{\ss}lin}}}]{Jasche2010b}%
{(\PineGreen{{Jasche} {\textit{et~al}}\mbox{.}},
  \PineGreen{2010{\natexlab{a}}})}  \BibitemOpen
  \bibfield  {author} {\bibinfo {author} {\bibfnamefont {J.}~\bibnamefont
  {{Jasche}}}, \bibinfo {author} {\bibfnamefont {F.~S.}\ \bibnamefont
  {{Kitaura}}}, \bibinfo {author} {\bibfnamefont {B.~D.}\ \bibnamefont
  {{Wandelt}}}, \bibinfo {author} {\bibfnamefont {T.~A.}\ \bibnamefont
  {{En{\ss}lin}}},\ }\emph {{Bayesian power-spectrum inference for large-scale
  structure data}},\ \href {\doibase 10.1111/j.1365-2966.2010.16610.x}
  {\bibfield  {journal} {\bibinfo  {journal} {\mnras}\ }\textbf {\bibinfo
  {volume} {406}},\ \bibinfo {pages} {60} (\bibinfo {year}
  {2010}{\natexlab{a}})},\ \Eprint {http://arxiv.org/abs/0911.2493}
  {arXiv:0911.2493 [astro-ph.CO]}\BibitemShut {NoStop}%
\bibitem [{{Jasche} {\textit{et~al}}\mbox{.}(2010{\natexlab{b}})\citenamefont
  {{Jasche}}, \citenamefont {{Kitaura}}, \citenamefont {{Li}},\ \&\
  \citenamefont {{En{\ss}lin}}}]{Jasche2010a}%
{(\PineGreen{{Jasche} {\textit{et~al}}\mbox{.}},
  \PineGreen{2010{\natexlab{b}}})}  \BibitemOpen
  \bibfield  {author} {\bibinfo {author} {\bibfnamefont {J.}~\bibnamefont
  {{Jasche}}}, \bibinfo {author} {\bibfnamefont {F.~S.}\ \bibnamefont
  {{Kitaura}}}, \bibinfo {author} {\bibfnamefont {C.}~\bibnamefont {{Li}}},
  \bibinfo {author} {\bibfnamefont {T.~A.}\ \bibnamefont {{En{\ss}lin}}},\
  }\emph {{Bayesian non-linear large-scale structure inference of the Sloan
  Digital Sky Survey Data Release 7}},\ \href {\doibase
  10.1111/j.1365-2966.2010.17313.x} {\bibfield  {journal} {\bibinfo  {journal}
  {\mnras}\ }\textbf {\bibinfo {volume} {409}},\ \bibinfo {pages} {355}
  (\bibinfo {year} {2010}{\natexlab{b}})},\ \Eprint
  {http://arxiv.org/abs/0911.2498} {arXiv:0911.2498 [astro-ph.CO]}\BibitemShut
  {NoStop}%
\bibitem [{{Jensen}(1906)\citenamefont {{Jensen}}}]{Jensen1906}%
{(\PineGreen{{Jensen}}, \PineGreen{1906})}  \BibitemOpen
  \bibfield  {author} {\bibinfo {author} {\bibfnamefont {J.~L.~W.~V.}\
  \bibnamefont {{Jensen}}},\ }\emph {{Sur les fonctions convexes et les
  in{\'e}galit{\'e}s entre les valeurs moyennes}},\ \href {\doibase
  10.1007/bf02418571} {\bibfield  {journal} {\bibinfo  {journal} {Acta Math.}\
  }\textbf {\bibinfo {volume} {30}},\ \bibinfo {pages} {175} (\bibinfo {year}
  {1906})}\BibitemShut {NoStop}%
\bibitem [{{Kitaura}(2013)\citenamefont {{Kitaura}}}]{Kitaura2013}%
{(\PineGreen{{Kitaura}}, \PineGreen{2013})}  \BibitemOpen
  \bibfield  {author} {\bibinfo {author} {\bibfnamefont {F.-S.}\ \bibnamefont
  {{Kitaura}}},\ }\emph {{The initial conditions of the Universe from
  constrained simulations}},\ \href {\doibase 10.1093/mnrasl/sls029} {\bibfield
   {journal} {\bibinfo  {journal} {\mnras}\ }\textbf {\bibinfo {volume}
  {429}},\ \bibinfo {pages} {L84} (\bibinfo {year} {2013})},\ \Eprint
  {http://arxiv.org/abs/1203.4184} {arXiv:1203.4184 [astro-ph.CO]}\BibitemShut
  {NoStop}%
\bibitem [{{Kitaura} \& {En{\ss}lin}(2008)\citenamefont {{Kitaura}}\ \&\
  \citenamefont {{En{\ss}lin}}}]{Kitaura2008}%
{(\PineGreen{{Kitaura} \& {En{\ss}lin}}, \PineGreen{2008})}  \BibitemOpen
  \bibfield  {author} {\bibinfo {author} {\bibfnamefont {F.~S.}\ \bibnamefont
  {{Kitaura}}}, \bibinfo {author} {\bibfnamefont {T.~A.}\ \bibnamefont
  {{En{\ss}lin}}},\ }\emph {{Bayesian reconstruction of the cosmological
  large-scale structure: methodology, inverse algorithms and numerical
  optimization}},\ \href {\doibase 10.1111/j.1365-2966.2008.13341.x} {\bibfield
   {journal} {\bibinfo  {journal} {\mnras}\ }\textbf {\bibinfo {volume}
  {389}},\ \bibinfo {pages} {497} (\bibinfo {year} {2008})},\ \Eprint
  {http://arxiv.org/abs/0705.0429} {arXiv:0705.0429 [astro-ph]}\BibitemShut
  {NoStop}%
\bibitem [{{Kullback} \& {Leibler}(1951)\citenamefont {{Kullback}}\ \&\
  \citenamefont {{Leibler}}}]{Kullback1951}%
{(\PineGreen{{Kullback} \& {Leibler}}, \PineGreen{1951})}  \BibitemOpen
  \bibfield  {author} {\bibinfo {author} {\bibfnamefont {S.}~\bibnamefont
  {{Kullback}}}, \bibinfo {author} {\bibfnamefont {R.~A.}\ \bibnamefont
  {{Leibler}}},\ }\emph {{On Information and Sufficiency}},\ \href {\doibase
  10.1214/aoms/1177729694} {\bibfield  {journal} {\bibinfo  {journal} {The
  Annals of Mathematical Statistics}\ }\textbf {\bibinfo {volume} {22}},\
  \bibinfo {pages} {79} (\bibinfo {year} {1951})}\BibitemShut {NoStop}%
\bibitem [{{Lahav} {\textit{et~al}}\mbox{.}(1994)\citenamefont {{Lahav}},
  \citenamefont {{Fisher}}, \citenamefont {{Hoffman}}, \citenamefont
  {{Scharf}},\ \&\ \citenamefont {{Zaroubi}}}]{Lahav1994}%
{(\PineGreen{{Lahav} {\textit{et~al}}\mbox{.}}, \PineGreen{1994})}
  \BibitemOpen
  \bibfield  {author} {\bibinfo {author} {\bibfnamefont {O.}~\bibnamefont
  {{Lahav}}}, \bibinfo {author} {\bibfnamefont {K.~B.}\ \bibnamefont
  {{Fisher}}}, \bibinfo {author} {\bibfnamefont {Y.}~\bibnamefont {{Hoffman}}},
  \bibinfo {author} {\bibfnamefont {C.~A.}\ \bibnamefont {{Scharf}}}, \bibinfo
  {author} {\bibfnamefont {S.}~\bibnamefont {{Zaroubi}}},\ }\emph {{Wiener
  Reconstruction of All-Sky Galaxy Surveys in Spherical Harmonics}},\ \href
  {\doibase 10.1086/187244} {\bibfield  {journal} {\bibinfo  {journal} {\apjl}\
  }\textbf {\bibinfo {volume} {423}},\ \bibinfo {pages} {L93} (\bibinfo {year}
  {1994})},\ \Eprint {http://arxiv.org/abs/astro-ph/9311059}
  {astro-ph/9311059}\BibitemShut {NoStop}%
\bibitem [{{Lavaux} \& {Jasche}(2016)\citenamefont {{Lavaux}}\ \&\
  \citenamefont {{Jasche}}}]{Lavaux2016BORG2MPP}%
{(\PineGreen{{Lavaux} \& {Jasche}}, \PineGreen{2016})}  \BibitemOpen
  \bibfield  {author} {\bibinfo {author} {\bibfnamefont {G.}~\bibnamefont
  {{Lavaux}}}, \bibinfo {author} {\bibfnamefont {J.}~\bibnamefont {{Jasche}}},\
  }\emph {{Unmasking the masked Universe: the 2M++ catalogue through Bayesian
  eyes}},\ \href {\doibase 10.1093/mnras/stv2499} {\bibfield  {journal}
  {\bibinfo  {journal} {\mnras}\ }\textbf {\bibinfo {volume} {455}},\ \bibinfo
  {pages} {3169} (\bibinfo {year} {2016})},\ \Eprint
  {http://arxiv.org/abs/1509.05040} {arXiv:1509.05040
  [astro-ph.CO]}\BibitemShut {NoStop}%
\bibitem [{{Lavaux} \& {Wandelt}(2010)\citenamefont {{Lavaux}}\ \&\
  \citenamefont {{Wandelt}}}]{Lavaux2010}%
{(\PineGreen{{Lavaux} \& {Wandelt}}, \PineGreen{2010})}  \BibitemOpen
  \bibfield  {author} {\bibinfo {author} {\bibfnamefont {G.}~\bibnamefont
  {{Lavaux}}}, \bibinfo {author} {\bibfnamefont {B.~D.}\ \bibnamefont
  {{Wandelt}}},\ }\emph {{Precision cosmology with voids: definition, methods,
  dynamics}},\ \href {\doibase 10.1111/j.1365-2966.2010.16197.x} {\bibfield
  {journal} {\bibinfo  {journal} {\mnras}\ }\textbf {\bibinfo {volume} {403}},\
  \bibinfo {pages} {1392} (\bibinfo {year} {2010})},\ \Eprint
  {http://arxiv.org/abs/0906.4101} {arXiv:0906.4101 [astro-ph.CO]}\BibitemShut
  {NoStop}%
\bibitem [{{Leclercq}(2015)\citenamefont {{Leclercq}}}]{LeclercqThesis}%
{(\PineGreen{{Leclercq}}, \PineGreen{2015})}  \BibitemOpen
  \bibfield  {author} {\bibinfo {author} {\bibfnamefont {F.}~\bibnamefont
  {{Leclercq}}},\ }\emph {\bibinfo {title} {{Bayesian large-scale structure
  inference and cosmic web analysis}}},\ \href@noop {} {Ph.D. thesis},\
  \bibinfo  {school} {{Institut d'Astrophysique de Paris}} (\bibinfo {year}
  {2015}),\ \Eprint {http://arxiv.org/abs/1512.04985} {arXiv:1512.04985
  [astro-ph.CO]}\BibitemShut {NoStop}%
\bibitem [{{Leclercq}, {Jasche} \& {Wandelt}(2015{\natexlab{a}})\citenamefont
  {{Leclercq}}, \citenamefont {{Jasche}},\ \&\ \citenamefont
  {{Wandelt}}}]{Leclercq2015DT}%
{(\PineGreen{{Leclercq}, {Jasche} \& {Wandelt}},
  \PineGreen{2015{\natexlab{a}}})}  \BibitemOpen
  \bibfield  {author} {\bibinfo {author} {\bibfnamefont {F.}~\bibnamefont
  {{Leclercq}}}, \bibinfo {author} {\bibfnamefont {J.}~\bibnamefont
  {{Jasche}}}, \bibinfo {author} {\bibfnamefont {B.}~\bibnamefont
  {{Wandelt}}},\ }\emph {{Cosmic web-type classification using decision
  theory}},\ \href {\doibase 10.1051/0004-6361/201526006} {\bibfield  {journal}
  {\bibinfo  {journal} {\aap}\ }\textbf {\bibinfo {volume} {576}},\ \bibinfo
  {eid} {L17} (\bibinfo {year} {2015}{\natexlab{a}})},\ \Eprint
  {http://arxiv.org/abs/1503.00730} {arXiv:1503.00730
  [astro-ph.CO]}\BibitemShut {NoStop}%
\bibitem [{{Leclercq}, {Jasche} \& {Wandelt}(2015{\natexlab{b}})\citenamefont
  {{Leclercq}}, \citenamefont {{Jasche}},\ \&\ \citenamefont
  {{Wandelt}}}]{Leclercq2015ST}%
{(\PineGreen{{Leclercq}, {Jasche} \& {Wandelt}},
  \PineGreen{2015{\natexlab{b}}})}  \BibitemOpen
  \bibfield  {author} {\bibinfo {author} {\bibfnamefont {F.}~\bibnamefont
  {{Leclercq}}}, \bibinfo {author} {\bibfnamefont {J.}~\bibnamefont
  {{Jasche}}}, \bibinfo {author} {\bibfnamefont {B.}~\bibnamefont
  {{Wandelt}}},\ }\emph {{Bayesian analysis of the dynamic cosmic web in the
  SDSS galaxy survey}},\ \href {\doibase 10.1088/1475-7516/2015/06/015}
  {\bibfield  {journal} {\bibinfo  {journal} {\jcap}\ }\textbf {\bibinfo
  {volume} {6}},\ \bibinfo {eid} {015} (\bibinfo {year}
  {2015}{\natexlab{b}})},\ \Eprint {http://arxiv.org/abs/1502.02690}
  {arXiv:1502.02690 [astro-ph.CO]}\BibitemShut {NoStop}%
\bibitem [{{Leclercq} {\textit{et~al}}\mbox{.}(2016)\citenamefont {{Leclercq}},
  \citenamefont {{Jasche}}, \citenamefont {{Lavaux}},\ \&\ \citenamefont
  {{Wandelt}}}]{Leclercq2016}%
{(\PineGreen{{Leclercq} {\textit{et~al}}\mbox{.}}, \PineGreen{2016})}
  \BibitemOpen
  \bibfield  {author} {\bibinfo {author} {\bibfnamefont {F.}~\bibnamefont
  {{Leclercq}}}, \bibinfo {author} {\bibfnamefont {J.}~\bibnamefont
  {{Jasche}}}, \bibinfo {author} {\bibfnamefont {G.}~\bibnamefont {{Lavaux}}},
  \bibinfo {author} {\bibfnamefont {B.}~\bibnamefont {{Wandelt}}},\ }\emph
  {{Inference and classifications of the Lagrangian dark matter sheet in the
  SDSS}},\ \href@noop {} {\bibfield  {journal} {\bibinfo  {journal} {ArXiv
  e-prints}\ } (\bibinfo {year} {2016})},\ \Eprint
  {http://arxiv.org/abs/1601.00093} {arXiv:1601.00093
  [astro-ph.CO]}\BibitemShut {NoStop}%
\bibitem [{{Leclercq} {\textit{et~al}}\mbox{.}(2015)\citenamefont {{Leclercq}},
  \citenamefont {{Jasche}}, \citenamefont {{Sutter}}, \citenamefont
  {{Hamaus}},\ \&\ \citenamefont {{Wandelt}}}]{Leclercq2015DMVOIDS}%
{(\PineGreen{{Leclercq} {\textit{et~al}}\mbox{.}}, \PineGreen{2015})}
  \BibitemOpen
  \bibfield  {author} {\bibinfo {author} {\bibfnamefont {F.}~\bibnamefont
  {{Leclercq}}}, \bibinfo {author} {\bibfnamefont {J.}~\bibnamefont
  {{Jasche}}}, \bibinfo {author} {\bibfnamefont {P.~M.}\ \bibnamefont
  {{Sutter}}}, \bibinfo {author} {\bibfnamefont {N.}~\bibnamefont {{Hamaus}}},
  \bibinfo {author} {\bibfnamefont {B.}~\bibnamefont {{Wandelt}}},\ }\emph
  {{Dark matter voids in the SDSS galaxy survey}},\ \href {\doibase
  10.1088/1475-7516/2015/03/047} {\bibfield  {journal} {\bibinfo  {journal}
  {\jcap}\ }\textbf {\bibinfo {volume} {3}},\ \bibinfo {eid} {047} (\bibinfo
  {year} {2015})},\ \Eprint {http://arxiv.org/abs/1410.0355} {arXiv:1410.0355
  [astro-ph.CO]}\BibitemShut {NoStop}%
\bibitem [{{Li} {\textit{et~al}}\mbox{.}(2006)\citenamefont {{Li}},
  \citenamefont {{Kauffmann}}, \citenamefont {{Jing}}, \citenamefont {{White}},
  \citenamefont {{B{\"o}rner}},\ \&\ \citenamefont {{Cheng}}}]{Li2006}%
{(\PineGreen{{Li} {\textit{et~al}}\mbox{.}}, \PineGreen{2006})}  \BibitemOpen
  \bibfield  {author} {\bibinfo {author} {\bibfnamefont {C.}~\bibnamefont
  {{Li}}}, \bibinfo {author} {\bibfnamefont {G.}~\bibnamefont {{Kauffmann}}},
  \bibinfo {author} {\bibfnamefont {Y.~P.}\ \bibnamefont {{Jing}}}, \bibinfo
  {author} {\bibfnamefont {S.~D.~M.}\ \bibnamefont {{White}}}, \bibinfo
  {author} {\bibfnamefont {G.}~\bibnamefont {{B{\"o}rner}}}, \bibinfo {author}
  {\bibfnamefont {F.~Z.}\ \bibnamefont {{Cheng}}},\ }\emph {{The dependence of
  clustering on galaxy properties}},\ \href {\doibase
  10.1111/j.1365-2966.2006.10066.x} {\bibfield  {journal} {\bibinfo  {journal}
  {\mnras}\ }\textbf {\bibinfo {volume} {368}},\ \bibinfo {pages} {21}
  (\bibinfo {year} {2006})},\ \Eprint {http://arxiv.org/abs/astro-ph/0509873}
  {astro-ph/0509873}\BibitemShut {NoStop}%
\bibitem [{{Lin}(1991)\citenamefont {{Lin}}}]{Lin1991}%
{(\PineGreen{{Lin}}, \PineGreen{1991})}  \BibitemOpen
  \bibfield  {author} {\bibinfo {author} {\bibfnamefont {J.}~\bibnamefont
  {{Lin}}},\ }\emph {{Divergence measures based on the Shannon entropy}},\
  \href {\doibase 10.1109/18.61115} {\bibfield  {journal} {\bibinfo  {journal}
  {IEEE Transactions on Information Theory}\ }\textbf {\bibinfo {volume}
  {37}},\ \bibinfo {pages} {145} (\bibinfo {year} {1991})}\BibitemShut
  {NoStop}%
\bibitem [{{Martin}, {Ringeval} \& {Vennin}(2016)\citenamefont {{Martin}},
  \citenamefont {{Ringeval}},\ \&\ \citenamefont {{Vennin}}}]{Martin2016}%
{(\PineGreen{{Martin}, {Ringeval} \& {Vennin}}, \PineGreen{2016})}
  \BibitemOpen
  \bibfield  {author} {\bibinfo {author} {\bibfnamefont {J.}~\bibnamefont
  {{Martin}}}, \bibinfo {author} {\bibfnamefont {C.}~\bibnamefont
  {{Ringeval}}}, \bibinfo {author} {\bibfnamefont {V.}~\bibnamefont
  {{Vennin}}},\ }\emph {{Information gain on reheating: The one bit
  milestone}},\ \href {\doibase 10.1103/PhysRevD.93.103532} {\bibfield
  {journal} {\bibinfo  {journal} {\prd}\ }\textbf {\bibinfo {volume} {93}},\
  \bibinfo {eid} {103532} (\bibinfo {year} {2016})},\ \Eprint
  {http://arxiv.org/abs/1603.02606} {arXiv:1603.02606}\BibitemShut {NoStop}%
\bibitem [{{Merson} {\textit{et~al}}\mbox{.}(2016)\citenamefont {{Merson}},
  \citenamefont {{Jasche}}, \citenamefont {{Abdalla}}, \citenamefont {{Lahav}},
  \citenamefont {{Wandelt}}, \citenamefont {{Jones}},\ \&\ \citenamefont
  {{Colless}}}]{MersonJascheAbdallaEtAl2016}%
{(\PineGreen{{Merson} {\textit{et~al}}\mbox{.}}, \PineGreen{2016})}
  \BibitemOpen
  \bibfield  {author} {\bibinfo {author} {\bibfnamefont {A.~I.}\ \bibnamefont
  {{Merson}}}, \bibinfo {author} {\bibfnamefont {J.}~\bibnamefont {{Jasche}}},
  \bibinfo {author} {\bibfnamefont {F.~B.}\ \bibnamefont {{Abdalla}}}, \bibinfo
  {author} {\bibfnamefont {O.}~\bibnamefont {{Lahav}}}, \bibinfo {author}
  {\bibfnamefont {B.}~\bibnamefont {{Wandelt}}}, \bibinfo {author}
  {\bibfnamefont {D.~H.}\ \bibnamefont {{Jones}}}, \bibinfo {author}
  {\bibfnamefont {M.}~\bibnamefont {{Colless}}},\ }\emph {{Halo detection via
  large-scale Bayesian inference}},\ \href {\doibase 10.1093/mnras/stw948}
  {\bibfield  {journal} {\bibinfo  {journal} {\mnras}\ }\textbf {\bibinfo
  {volume} {460}},\ \bibinfo {pages} {1340} (\bibinfo {year} {2016})},\ \Eprint
  {http://arxiv.org/abs/1505.03528} {arXiv:1505.03528
  [astro-ph.CO]}\BibitemShut {NoStop}%
\bibitem [{{Nuza} {\textit{et~al}}\mbox{.}(2014)\citenamefont {{Nuza}},
  \citenamefont {{Kitaura}}, \citenamefont {{He{\ss}}}, \citenamefont
  {{Libeskind}},\ \&\ \citenamefont {{M{\"u}ller}}}]{Nuza2014}%
{(\PineGreen{{Nuza} {\textit{et~al}}\mbox{.}}, \PineGreen{2014})}  \BibitemOpen
  \bibfield  {author} {\bibinfo {author} {\bibfnamefont {S.~E.}\ \bibnamefont
  {{Nuza}}}, \bibinfo {author} {\bibfnamefont {F.-S.}\ \bibnamefont
  {{Kitaura}}}, \bibinfo {author} {\bibfnamefont {S.}~\bibnamefont
  {{He{\ss}}}}, \bibinfo {author} {\bibfnamefont {N.~I.}\ \bibnamefont
  {{Libeskind}}}, \bibinfo {author} {\bibfnamefont {V.}~\bibnamefont
  {{M{\"u}ller}}},\ }\emph {{The cosmic web of the Local Universe: cosmic
  variance, matter content and its relation to galaxy morphology}},\ \href
  {\doibase 10.1093/mnras/stu1746} {\bibfield  {journal} {\bibinfo  {journal}
  {\mnras}\ }\textbf {\bibinfo {volume} {445}},\ \bibinfo {pages} {988}
  (\bibinfo {year} {2014})},\ \Eprint {http://arxiv.org/abs/1406.1004}
  {arXiv:1406.1004 [astro-ph.CO]}\BibitemShut {NoStop}%
\bibitem [{{Patiri} {\textit{et~al}}\mbox{.}(2006)\citenamefont {{Patiri}},
  \citenamefont {{Prada}}, \citenamefont {{Holtzman}}, \citenamefont
  {{Klypin}},\ \&\ \citenamefont {{Betancort-Rijo}}}]{Patiri2006}%
{(\PineGreen{{Patiri} {\textit{et~al}}\mbox{.}}, \PineGreen{2006})}
  \BibitemOpen
  \bibfield  {author} {\bibinfo {author} {\bibfnamefont {S.~G.}\ \bibnamefont
  {{Patiri}}}, \bibinfo {author} {\bibfnamefont {F.}~\bibnamefont {{Prada}}},
  \bibinfo {author} {\bibfnamefont {J.}~\bibnamefont {{Holtzman}}}, \bibinfo
  {author} {\bibfnamefont {A.}~\bibnamefont {{Klypin}}}, \bibinfo {author}
  {\bibfnamefont {J.}~\bibnamefont {{Betancort-Rijo}}},\ }\emph {{The
  properties of galaxies in voids}},\ \href {\doibase
  10.1111/j.1365-2966.2006.10975.x} {\bibfield  {journal} {\bibinfo  {journal}
  {\mnras}\ }\textbf {\bibinfo {volume} {372}},\ \bibinfo {pages} {1710}
  (\bibinfo {year} {2006})},\ \Eprint {http://arxiv.org/abs/astro-ph/0605703}
  {astro-ph/0605703}\BibitemShut {NoStop}%
\bibitem [{{Planck Collaboration}(2015)\citenamefont {{Planck
  Collaboration}}}]{Planck2015ISW}%
{(\PineGreen{{Planck Collaboration}}, \PineGreen{2015})}  \BibitemOpen
  \bibfield  {author} {\bibinfo {author} {\bibnamefont {{Planck
  Collaboration}}},\ }\emph {{Planck 2015 results. XXI. The integrated
  Sachs-Wolfe effect}},\ \href@noop {} {\bibfield  {journal} {\bibinfo
  {journal} {ArXiv e-prints}\ } (\bibinfo {year} {2015})},\ \Eprint
  {http://arxiv.org/abs/1502.01595} {arXiv:1502.01595
  [astro-ph.CO]}\BibitemShut {NoStop}%
\bibitem [{{Roth}(1965)\citenamefont {{Roth}}}]{Roth1965}%
{(\PineGreen{{Roth}}, \PineGreen{1965})}  \BibitemOpen
  \bibfield  {author} {\bibinfo {author} {\bibfnamefont {P.~M.}\ \bibnamefont
  {{Roth}}},\ }\emph {\bibinfo {title} {{Design of Experiments for
  Discrimination Among Rival Models}}},\ \href@noop {} {Ph.D. thesis},\
  \bibinfo  {school} {{Princeton University}} (\bibinfo {year}
  {1965})\BibitemShut {NoStop}%
\bibitem [{{Seehars} {\textit{et~al}}\mbox{.}(2014)\citenamefont {{Seehars}},
  \citenamefont {{Amara}}, \citenamefont {{Refregier}}, \citenamefont
  {{Paranjape}},\ \&\ \citenamefont {{Akeret}}}]{Seehars2014}%
{(\PineGreen{{Seehars} {\textit{et~al}}\mbox{.}}, \PineGreen{2014})}
  \BibitemOpen
  \bibfield  {author} {\bibinfo {author} {\bibfnamefont {S.}~\bibnamefont
  {{Seehars}}}, \bibinfo {author} {\bibfnamefont {A.}~\bibnamefont {{Amara}}},
  \bibinfo {author} {\bibfnamefont {A.}~\bibnamefont {{Refregier}}}, \bibinfo
  {author} {\bibfnamefont {A.}~\bibnamefont {{Paranjape}}}, \bibinfo {author}
  {\bibfnamefont {J.}~\bibnamefont {{Akeret}}},\ }\emph {{Information gains
  from cosmic microwave background experiments}},\ \href {\doibase
  10.1103/PhysRevD.90.023533} {\bibfield  {journal} {\bibinfo  {journal}
  {\prd}\ }\textbf {\bibinfo {volume} {90}},\ \bibinfo {eid} {023533} (\bibinfo
  {year} {2014})},\ \Eprint {http://arxiv.org/abs/1402.3593}
  {arXiv:1402.3593}\BibitemShut {NoStop}%
\bibitem [{{Shannon}(1948)\citenamefont {{Shannon}}}]{Shannon1948}%
{(\PineGreen{{Shannon}}, \PineGreen{1948})}  \BibitemOpen
  \bibfield  {author} {\bibinfo {author} {\bibfnamefont {C.~E.}\ \bibnamefont
  {{Shannon}}},\ }\emph {{A Mathematical Theory of Communication}},\ \href
  {\doibase 10.1002/j.1538-7305.1948.tb01338.x} {\bibfield  {journal} {\bibinfo
   {journal} {Bell System Technical Journal}\ }\textbf {\bibinfo {volume}
  {27}},\ \bibinfo {pages} {379} (\bibinfo {year} {1948})}\BibitemShut
  {NoStop}%
\bibitem [{{Tassev}, {Zaldarriaga} \& {Eisenstein}(2013)\citenamefont
  {{Tassev}}, \citenamefont {{Zaldarriaga}},\ \&\ \citenamefont
  {{Eisenstein}}}]{Tassev2013}%
{(\PineGreen{{Tassev}, {Zaldarriaga} \& {Eisenstein}}, \PineGreen{2013})}
  \BibitemOpen
  \bibfield  {author} {\bibinfo {author} {\bibfnamefont {S.}~\bibnamefont
  {{Tassev}}}, \bibinfo {author} {\bibfnamefont {M.}~\bibnamefont
  {{Zaldarriaga}}}, \bibinfo {author} {\bibfnamefont {D.~J.}\ \bibnamefont
  {{Eisenstein}}},\ }\emph {{Solving large scale structure in ten easy steps
  with COLA}},\ \href {\doibase 10.1088/1475-7516/2013/06/036} {\bibfield
  {journal} {\bibinfo  {journal} {\jcap}\ }\textbf {\bibinfo {volume} {6}},\
  \bibinfo {eid} {036} (\bibinfo {year} {2013})},\ \Eprint
  {http://arxiv.org/abs/1301.0322} {arXiv:1301.0322 [astro-ph.CO]}\BibitemShut
  {NoStop}%
\bibitem [{{Trotta}(2007)\citenamefont {{Trotta}}}]{Trotta2007}%
{(\PineGreen{{Trotta}}, \PineGreen{2007})}  \BibitemOpen
  \bibfield  {author} {\bibinfo {author} {\bibfnamefont {R.}~\bibnamefont
  {{Trotta}}},\ }\emph {{Forecasting the Bayes factor of a future
  observation}},\ \href {\doibase 10.1111/j.1365-2966.2007.11861.x} {\bibfield
  {journal} {\bibinfo  {journal} {\mnras}\ }\textbf {\bibinfo {volume} {378}},\
  \bibinfo {pages} {819} (\bibinfo {year} {2007})},\ \Eprint
  {http://arxiv.org/abs/astro-ph/0703063} {astro-ph/0703063}\BibitemShut
  {NoStop}%
\bibitem [{{van de Weygaert} \& {Bond}(2008)\citenamefont {{van de Weygaert}}\
  \&\ \citenamefont {{Bond}}}]{vandeWeygaertBond2008}%
{(\PineGreen{{van de Weygaert} \& {Bond}}, \PineGreen{2008})}  \BibitemOpen
  \bibfield  {author} {\bibinfo {author} {\bibfnamefont {R.}~\bibnamefont {{van
  de Weygaert}}}, \bibinfo {author} {\bibfnamefont {J.~R.}\ \bibnamefont
  {{Bond}}},\ }\emph {{Clusters and the Theory of the Cosmic Web}},\ in\
  \href@noop {} {\emph {\bibinfo {booktitle} {A Pan-Chromatic View of Clusters
  of Galaxies and the Large-Scale Structure}}},\ \bibinfo {series} {Lecture
  Notes in Physics, Berlin Springer Verlag}, Vol.\ \bibinfo {volume} {740},\
  \bibinfo {editor} {edited by\ \bibinfo {editor} {\bibfnamefont
  {M.}~\bibnamefont {{Plionis}}}, \bibinfo {editor} {\bibfnamefont
  {O.}~\bibnamefont {{L{\'o}pez-Cruz}}}, \bibinfo {editor} {\bibfnamefont
  {D.}~\bibnamefont {{Hughes}}}}\ (\bibinfo {year} {2008})\ p.~\bibinfo {pages}
  {24}\BibitemShut {NoStop}%
\bibitem [{{Vanlier} {\textit{et~al}}\mbox{.}(2014)\citenamefont {{Vanlier}},
  \citenamefont {{Tiemann}}, \citenamefont {{Hilbers}},\ \&\ \citenamefont
  {{van Riel}}}]{Vanlier2014}%
{(\PineGreen{{Vanlier} {\textit{et~al}}\mbox{.}}, \PineGreen{2014})}
  \BibitemOpen
  \bibfield  {author} {\bibinfo {author} {\bibfnamefont {J.}~\bibnamefont
  {{Vanlier}}}, \bibinfo {author} {\bibfnamefont {C.~A.}\ \bibnamefont
  {{Tiemann}}}, \bibinfo {author} {\bibfnamefont {P.~A.~J.}\ \bibnamefont
  {{Hilbers}}}, \bibinfo {author} {\bibfnamefont {N.~A.~W.}\ \bibnamefont {{van
  Riel}}},\ }\emph {{Optimal experiment design for model selection in
  biochemical networks}},\ \href {\doibase 10.1186/1752-0509-8-20} {\bibfield
  {journal} {\bibinfo  {journal} {BMC Syst Biol}\ }\textbf {\bibinfo {volume}
  {8}},\ \bibinfo {pages} {20} (\bibinfo {year} {2014})}\BibitemShut {NoStop}%
\bibitem [{{Vanlier} {\textit{et~al}}\mbox{.}(2012)\citenamefont {{Vanlier}},
  \citenamefont {{Tiemann}}, \citenamefont {{Hilbers}},\ \&\ \citenamefont
  {{van Riel}}}]{Vanlier2012}%
{(\PineGreen{{Vanlier} {\textit{et~al}}\mbox{.}}, \PineGreen{2012})}
  \BibitemOpen
  \bibfield  {author} {\bibinfo {author} {\bibfnamefont {J.}~\bibnamefont
  {{Vanlier}}}, \bibinfo {author} {\bibfnamefont {C.~A.}\ \bibnamefont
  {{Tiemann}}}, \bibinfo {author} {\bibfnamefont {P.~A.~J.}\ \bibnamefont
  {{Hilbers}}}, \bibinfo {author} {\bibfnamefont {N.~A.~W.}\ \bibnamefont {{van
  Riel}}},\ }\emph {{A Bayesian approach to targeted experiment design}},\
  \href {\doibase 10.1093/bioinformatics/bts092} {\bibfield  {journal}
  {\bibinfo  {journal} {Bioinformatics}\ }\textbf {\bibinfo {volume} {28}},\
  \bibinfo {pages} {1136} (\bibinfo {year} {2012})}\BibitemShut {NoStop}%
\bibitem [{{Vogelsberger} \& {White}(2011)\citenamefont {{Vogelsberger}}\ \&\
  \citenamefont {{White}}}]{VogelsbergerWhite2011}%
{(\PineGreen{{Vogelsberger} \& {White}}, \PineGreen{2011})}  \BibitemOpen
  \bibfield  {author} {\bibinfo {author} {\bibfnamefont {M.}~\bibnamefont
  {{Vogelsberger}}}, \bibinfo {author} {\bibfnamefont {S.~D.~M.}\ \bibnamefont
  {{White}}},\ }\emph {{Streams and caustics: the fine-grained structure of
  {$\Lambda$} cold dark matter haloes}},\ \href {\doibase
  10.1111/j.1365-2966.2011.18224.x} {\bibfield  {journal} {\bibinfo  {journal}
  {\mnras}\ }\textbf {\bibinfo {volume} {413}},\ \bibinfo {pages} {1419}
  (\bibinfo {year} {2011})},\ \Eprint {http://arxiv.org/abs/1002.3162}
  {arXiv:1002.3162 [astro-ph.CO]}\BibitemShut {NoStop}%
\bibitem [{{Wang} {\textit{et~al}}\mbox{.}(2013)\citenamefont {{Wang}},
  \citenamefont {{Mo}}, \citenamefont {{Yang}},\ \&\ \citenamefont {{van den
  Bosch}}}]{Wang2013}%
{(\PineGreen{{Wang} {\textit{et~al}}\mbox{.}}, \PineGreen{2013})}  \BibitemOpen
  \bibfield  {author} {\bibinfo {author} {\bibfnamefont {H.}~\bibnamefont
  {{Wang}}}, \bibinfo {author} {\bibfnamefont {H.~J.}\ \bibnamefont {{Mo}}},
  \bibinfo {author} {\bibfnamefont {X.}~\bibnamefont {{Yang}}}, \bibinfo
  {author} {\bibfnamefont {F.~C.}\ \bibnamefont {{van den Bosch}}},\ }\emph
  {{Reconstructing the Initial Density Field of the Local Universe: Methods and
  Tests with Mock Catalogs}},\ \href {\doibase 10.1088/0004-637X/772/1/63}
  {\bibfield  {journal} {\bibinfo  {journal} {\apj}\ }\textbf {\bibinfo
  {volume} {772}},\ \bibinfo {eid} {63} (\bibinfo {year} {2013})},\ \Eprint
  {http://arxiv.org/abs/1301.1348} {arXiv:1301.1348 [astro-ph.CO]}\BibitemShut
  {NoStop}%
\bibitem [{{Wang} {\textit{et~al}}\mbox{.}(2014)\citenamefont {{Wang}},
  \citenamefont {{Mo}}, \citenamefont {{Yang}}, \citenamefont {{Jing}},\ \&\
  \citenamefont {{Lin}}}]{Wang2014a}%
{(\PineGreen{{Wang} {\textit{et~al}}\mbox{.}}, \PineGreen{2014})}  \BibitemOpen
  \bibfield  {author} {\bibinfo {author} {\bibfnamefont {H.}~\bibnamefont
  {{Wang}}}, \bibinfo {author} {\bibfnamefont {H.~J.}\ \bibnamefont {{Mo}}},
  \bibinfo {author} {\bibfnamefont {X.}~\bibnamefont {{Yang}}}, \bibinfo
  {author} {\bibfnamefont {Y.~P.}\ \bibnamefont {{Jing}}}, \bibinfo {author}
  {\bibfnamefont {W.~P.}\ \bibnamefont {{Lin}}},\ }\emph {{ELUCID---Exploring
  the Local Universe with the Reconstructed Initial Density Field. I.
  Hamiltonian Markov Chain Monte Carlo Method with Particle Mesh Dynamics}},\
  \href {\doibase 10.1088/0004-637X/794/1/94} {\bibfield  {journal} {\bibinfo
  {journal} {\apj}\ }\textbf {\bibinfo {volume} {794}},\ \bibinfo {eid} {94}
  (\bibinfo {year} {2014})},\ \Eprint {http://arxiv.org/abs/1407.3451}
  {arXiv:1407.3451 [astro-ph.CO]}\BibitemShut {NoStop}%
\bibitem [{{Wolz} {\textit{et~al}}\mbox{.}(2012)\citenamefont {{Wolz}},
  \citenamefont {{Kilbinger}}, \citenamefont {{Weller}},\ \&\ \citenamefont
  {{Giannantonio}}}]{Wolz2012}%
{(\PineGreen{{Wolz} {\textit{et~al}}\mbox{.}}, \PineGreen{2012})}  \BibitemOpen
  \bibfield  {author} {\bibinfo {author} {\bibfnamefont {L.}~\bibnamefont
  {{Wolz}}}, \bibinfo {author} {\bibfnamefont {M.}~\bibnamefont {{Kilbinger}}},
  \bibinfo {author} {\bibfnamefont {J.}~\bibnamefont {{Weller}}}, \bibinfo
  {author} {\bibfnamefont {T.}~\bibnamefont {{Giannantonio}}},\ }\emph {{On the
  validity of cosmological Fisher matrix forecasts}},\ \href {\doibase
  10.1088/1475-7516/2012/09/009} {\bibfield  {journal} {\bibinfo  {journal}
  {\jcap}\ }\textbf {\bibinfo {volume} {9}},\ \bibinfo {eid} {009} (\bibinfo
  {year} {2012})},\ \Eprint {http://arxiv.org/abs/1205.3984} {arXiv:1205.3984
  [astro-ph.CO]}\BibitemShut {NoStop}%
\bibitem [{{Zaroubi}(2002)\citenamefont {{Zaroubi}}}]{Zaroubi2002}%
{(\PineGreen{{Zaroubi}}, \PineGreen{2002})}  \BibitemOpen
  \bibfield  {author} {\bibinfo {author} {\bibfnamefont {S.}~\bibnamefont
  {{Zaroubi}}},\ }\emph {{Unbiased reconstruction of the large-scale
  structure}},\ \href {\doibase 10.1046/j.1365-8711.2002.05229.x} {\bibfield
  {journal} {\bibinfo  {journal} {\mnras}\ }\textbf {\bibinfo {volume} {331}},\
  \bibinfo {pages} {901} (\bibinfo {year} {2002})},\ \Eprint
  {http://arxiv.org/abs/astro-ph/0010561} {astro-ph/0010561}\BibitemShut
  {NoStop}%
\end{thebibliography}%

\end{document}